\documentclass[12pt]{article}

\usepackage{amsfonts,graphicx,amsmath,epsf,array}
\newcommand\encadremath[1]{\vbox{\hrule\hbox{\vrule\kern8pt
\vbox{\kern8pt \hbox{$\displaystyle #1$}\kern8pt}
\kern8pt\vrule}\hrule}}
\def\enca#1{\vbox{\hrule\hbox{
\vrule\kern8pt\vbox{\kern8pt \hbox{$\displaystyle #1$}
\kern8pt} \kern8pt\vrule}\hrule}}

\newcommand\figureframex[3]{
\begin{figure}[bth]
\hrule\hbox{\vrule\kern8pt
\vbox{\kern8pt \vbox{
\begin{center}
{\mbox{\epsfxsize=#1.truecm\epsfbox{#2}}}
\end{center}
\caption{#3}
}\kern8pt}
\kern8pt\vrule}\hrule
\end{figure}
}
\newcommand\figureframey[3]{
\begin{figure}[bth]
\hrule\hbox{\vrule\kern8pt
\vbox{\kern8pt \vbox{
\begin{center}
{\mbox{\epsfysize=#1.truecm\epsfbox{#2}}}
\end{center}
\caption{#3}
}\kern8pt}
\kern8pt\vrule}\hrule
\end{figure}
}

\makeatletter
\@addtoreset{equation}{section}
\makeatother
\newtheorem{theorem}{Theorem}[section]

\newtheorem{remark}{Remark}[section]
\newtheorem{proposition}{Proposition}[section]
\newtheorem{lemma}{Lemma}[section]
\newtheorem{corollary}{Corollary}[section]
\newtheorem{definition}{Definition}[section]
\def\br{\begin{remark}\rm\small}
\def\er{\end{remark}}
\def\bt{\begin{theorem}}
\def\et{\end{theorem}}
\def\bd{\begin{definition}}
\def\ed{\end{definition}}
\def\bp{\begin{proposition}}
\def\ep{\end{proposition}}
\def\bl{\begin{lemma}}
\def\el{\end{lemma}}
\def\bc{\begin{corollary}}
\def\ec{\end{corollary}}
\def\beaq{\begin{eqnarray}}
\def\eeaq{\end{eqnarray}}
\newcommand{\proof}[1]{{\noindent \bf proof:}\par
{#1} {$\square$}}

\newcommand{\ch}{\,\mathrm{ch}}
\newcommand{\sh}{\,\mathrm{sh}}

\newcommand{\beq}{\begin{equation}}
\newcommand{\eeq}{\end{equation}}
\newcommand{\bea}{\begin{eqnarray}}
\newcommand{\eea}{\end{eqnarray}}

\renewcommand{\and}{{\qquad {\rm and} \qquad}}

 \newcommand{\Tr}{{\,\rm Tr}\:}

\newcommand{\Res}{\mathop{\,\rm Res\,}}

\newcommand{\td}[1]{{\tilde{#1}}}

\newcommand{\uu}{\mathrm{u}}

\newcommand{\f}{\mathrm{f}}

\newcommand{\e}{{\,\rm e}\,}
\newcommand{\ee}[1]{{{\rm e}^{#1}}}

\newcommand{\C}{{\mathbf C}}

\newcommand{\Pint}{{\int\kern -1.em -\kern-.25em}}

\newcommand{\x}{\mathrm{x}}

\newcommand{\nn}{{\mathfrak n}}

\usepackage[pdftex]{hyperref}
\hypersetup{colorlinks,urlcolor=magenta,citecolor=red,linkcolor=blue,filecolor=black}

\begin{document}
\sloppy

\pagestyle{empty}
\hfill IPhT-t09/160
\begin{flushright}October 2009 \\ v3, October 2010 \end{flushright}
\addtolength{\baselineskip}{0.20\baselineskip}
\begin{center}
\vspace{26pt}
{\large \textbf{Enumeration of maps with self avoiding loops and \\ the $\mathcal{O}(\nn)$ model on random lattices of all topologies}}
\vspace{26pt}
\end{center}

\begin{center}
{\sl G.~Borot}\hspace*{0.05cm}\footnote{ E-mail: \href{mailto:gaetan.borot@cea.fr}{gaetan.borot@cea.fr}},
{\sl B.~Eynard}\hspace*{0.05cm}\footnote{ E-mail: \href{bertrand.eynard@cea.fr}{bertrand.eynard@cea.fr}} \\
\vspace{6pt}
Institut de Physique Th\'eorique,\\
CEA, IPhT, F-91191 Gif-sur-Yvette, France,\\
CNRS, URA 2306, F-91191 Gif-sur-Yvette, France.\\
\end{center}

\vspace{100pt}

\begin{center}
{\bf Abstract}
\end{center}

\vspace{0.2cm}
\vspace{0.5cm}

We compute the generating functions of a $\mathcal{O}(\nn)$ model (loop gas model) on a random lattice of any topology. On the disc and the cylinder, they were already known, and here we compute all the other topologies. We find that they obey a slightly deformed version of the topological recursion valid for the 1-hermitian matrix models. The generating functions of genus $g$ maps without boundaries are given by the symplectic invariants $F^g$ of a spectral curve $\Sigma_{\mathcal{O}(\mathfrak{n})}$. This spectral curve was known before, and it is in general not algebraic.
\vspace{0.5cm}

\newpage
\tableofcontents
\newpage

\vspace{26pt}
\pagestyle{plain}
\setcounter{page}{1}


\section*{Introduction}

The problem consists in counting random discrete surface, carrying random, self-avoiding, non intersecting loops, which can have $\nn$ possible colors. In statistical physics, this is called the \emph{${\mathcal O}(\nn)$ model} (or \emph{loop gas model}) on a random lattice, and it plays a very important role. The $\mathcal{O}(\nn)$ model on a regular lattice is one of the exactly solvable models of Baxter \cite{Baxter}. Its limit $\nn \rightarrow 0$ (or more generally $\nn = \nn_0 + \delta\nn$ with $\delta\nn \rightarrow 0$) counts configurations of self-avoiding polymers in two dimensions \cite{DuK}. On the random lattice, it is one of the basic toy models to understand the random geometry of discrete maps carrying structure. The $\mathcal{O}(1)$ model on random triangulations is dual to the Ising model on a random triangulations. The fully packed case of the $\mathcal{O}(\mathfrak{n})$ model is dual to the $\mathfrak{q}$-Potts model on the dual of random triangulations, with $\mathfrak{q} = \mathfrak{n}^2$. The $\mathcal{O}(\mathfrak{n})$ model with $\mathfrak{n} = 2\cos\big(\pi/(m + 1)\big)$ correspond the RSOS models with $m$ states. In general, the continuum limit of the model is related to the $(p,q)$ minimal models when $\mathfrak{n} = 2\cos(\pi p/q)$.

Matrix integrals provide powerful techniques for the combinatorics of maps \cite{E2}.The problem of counting random discrete surfaces without loops can be rephrased as a 1-matrix integral \cite{BIPZ,E1}. Techniques to solve this $1$-matrix model beyond spherical limit first appeared in \cite{Amb}, and culminated in the full solution in \cite{E1MM}, by a "topological recursion" formula. This structure was later enhanced \cite{EOFg} to multi-matrix models, and it was used beyond matrix models to associate "symplectic invariants" to an arbitrary spectral curve.

The $\mathcal{O}(\nn)$ model was also rewritten as a matrix integral \cite{K,K1}. In \cite{KS}, then \cite{EZJ}, the phase diagram with respect to $\mu$ such that $\nn = -2\cos(\pi\mu)$ was established. Besides, the critical exponents for the geometry of large maps with the topology of a disc were found. Later, a closed set of loop equations for the generating function of the $\mathcal{O}(\nn)$ maps was found \cite{E1,EK1}, and they were solved at least for maps with the topology of a disc or a cylinder. Some sparse other cases were investigated before \cite{KM,K2}. So far, an efficient algorithm was lacking to compute the generating functions of the $\mathcal{O}(\nn)$ model in higher topologies. It was not clear to which extent the method of \cite{E1MM} could be generalized. Indeed, the disc amplitude $\mathrm{W}_1^{(0)}(\x)$, which ought to give the spectral curve, is not algebraic when $\mu$ is not rational, and the loop equations seem rather different from the 1-matrix model case.

\medskip

In other words, do some symplectic invariants of \cite{EOFg} give the solution of the ${\mathcal O}(\nn)$ model ? The answer is yes, with a slight deformation (of parameter $\nn)$ of the notion of spectral curve. Thus, we obtain all correlation functions of the random lattice for any topology and any number of boundaries.

\subsection*{Outline}

Apart from the presentation of new explicit results for the all genus solution, our work comes in continuity with the foundation articles \cite{K1,EK1,EK2} on the $\mathcal{O}(\mathfrak{n})$ model on random lattice despite the time gap. We also wish to make these techniques accessible to combinatorialists, who are increasingly interested in the $\mathcal{O}(\mathfrak{n})$ model and the Potts model. For these reasons, we take space:
\begin{itemize}
\item[$\bullet$] To introduce the model and its matrix integral representation (Section~\ref{sec:On}).
\item[$\bullet$] To give two derivations of the loop equations, one straightforward from the matrix integral, and the other being its combinatorial analog (Section~\ref{sec:loopeq}).
\item[$\bullet$] To review the solution of the master loop equation (Section~\ref{sec:lineq} and \ref{sec:W10W20}), which was first determined in \cite{EZJ} for $n = 2\cos(\pi p/q)$, in \cite{EK1} for $|\mathfrak{n}| \leq 2$, in \cite{EK2} for $|\mathfrak{n}| > 2$. Our presentation is maybe more "algebro-geometric" minded, and this step is essential for our solution to the full set of loop equations (Section~\ref{sec:Wkg}).
\end{itemize}
Secondly, we extend to the standard properties of the topological recursion to the spectral curves of the $\mathcal{O}(\mathfrak{n})$ model (Section~\ref{sec:prop}).
Eventually, we show how our approach allow to recover earlier results for the limit of large maps \cite{KS}, and we complete them for all topologies in the light of the double scaling limit and topological recursion (Section~\ref{sec:lmaps}).

\section{The $\mathcal{O}(\nn)$ model}
\label{sec:On}

\subsection{Definition: loop gas on a random surface}
\label{sec:11}
Roughly speaking, a random discrete surface\footnote{See \cite{Berge} for a precise definition} (also called "map" in combinatorics) is a graph drawn on an oriented connected surface, such that all faces are polygons. It is thus obtained by gluing polygons along their edges. In the ${\mathcal O}(\nn)$ model, we have at our disposal empty polygons of size $j \geq 3$ (weight $t_j$ for each), and triangles carrying a piece of path (weight $-c$ for each). A configuration correspond to a map of genus $g$, with $v$ vertices, to which we give a Boltzmann weight:
\beq
\label{eq:weight} t^{v}\,N^{2-2g}\,\,\frac{(-c)^{\ell}\,\nn^{\#{\rm loops}}}{\#{\rm Aut}}\,\, \hat{t}_3^{\,n_3}\,\hat{t}_4^{\,n_4}\,\cdots\,\hat{t}_{d_{\textrm{max}}}^{\,n_{d_{\textrm{max}}}}
\eeq
This weight is non local because $\mathfrak{n}$ is coupled to the number of loops (i.e. closed paths). The fully packed model is an interesting special case : if we set $\hat{t}_j = 0$ ($j \geq 3$), all unmarked polygons are triangles carrying a piece of path.

\begin{figure}[h!]
\label{fig:mapon}
\begin{center}
\includegraphics[width=\textwidth]{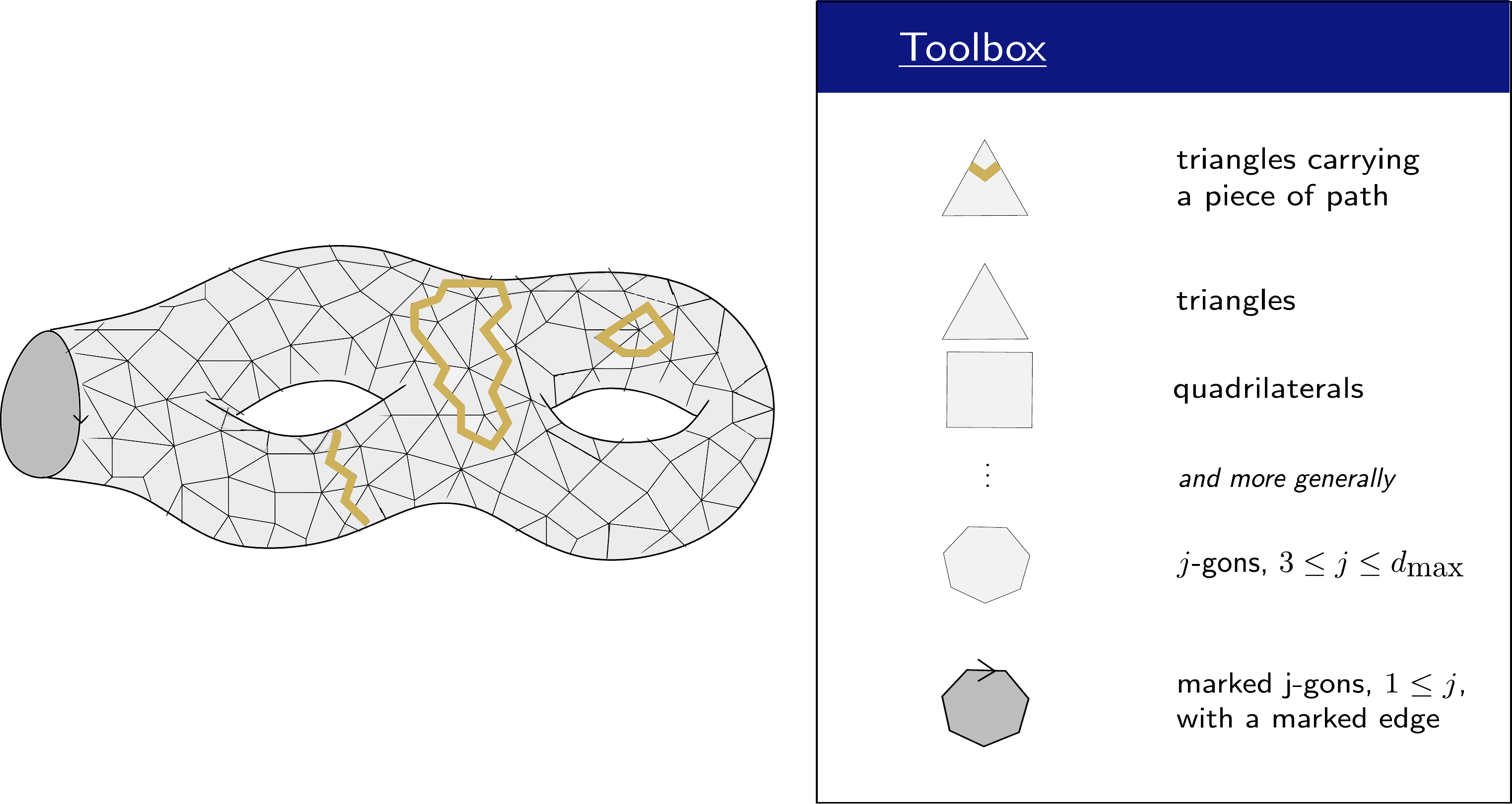}
\caption{\label{fig:poly}Example of a map in the $\mathcal{O}(\nn)$ model: $\mathcal{M} \in \mathbb{M}_1^{(2)}$.}
\end{center}
\end{figure}

In addition, we may consider maps with $k$ marked faces, with a marked edge on each marked face. We allow marked faces having more than one side, and we call them "boundaries". For $k = 1$, one also says that the maps are "rooted".

\bd
${\mathbb M}_k^{(g)}(v)$ is the set of connected oriented discrete surfaces of genus $g$, with $v$ vertices, obtained by gluing unmarked $j$-gons of degree $3 \leq j \leq d_{\textrm{max}}$, $k$ marked $j$-gons (of degree $1 \leq j$), and triangles carrying a piece of path, such that all the paths are loops (they are automatically self-avoiding).
\ed

\bp\label{propMkgfinite}
${\mathbb M}_k^{(g)}(v)$ is a finite set.
\ep

Indeed, in a configuration where marked faces have lengths $l_1,\dots,l_k$:
\beq
\#{\rm faces} = \sum_{j = 3}^{d_{\textrm{max}}} n_j + k + \ell \nonumber
\eeq
The total number of edges is half the number of half-edges i.e:
\beq
\#{\rm edges} = \frac{1}{2}\Big( \sum_{j = 3}^{d_{\textrm{max}}} j n_j + \sum_{i=1}^k l_i + 3\ell\Big) \nonumber
\eeq
The Euler characteristics is:
\beq
\chi = 2-2g = v-\#{\rm edges}+\#{\rm faces} \nonumber
\eeq
Hence:
\beq\label{eqchiMgkv}
2g-2+k+v = \frac{1}{2}\Big( \sum_{j = 3}^{d_{\textrm{max}}} (j-2)n_j + \sum_{i=1}^k l_i + \ell\Big)
\eeq
When $g, k, v$ are fixed, $n_j$, $l_i$ and $\ell$ are bounded, and there exists only a finite number of such surfaces.

\subsection{Generating functions}

As a convention, ${\mathbb M}_1^{(0)}(1)$ has one element, which is a map reduced to $1$ vertex.
\bd
We define a formal series in powers of $t$, counting genus $g$ maps with $k$ marked faces:
\beq
\mathrm{W}_k^{(g)} = \sum_{v=1}^\infty \,\, t^v \,\,\sum_{\mathcal{M}\in {\mathbb M}_k^{(g)}(v)}\,
\frac{(-c)^{\ell(\mathcal{M})}\, \nn^{\#{\rm loops}(\mathcal{M})}}{\#{\rm Aut}(\mathcal{M})}\, \frac{\hat{t}_3^{\,n_3(\mathcal{M})}\,\hat{t}_4^{\,n_4(\mathcal{M})}\,\cdots\, \hat{t}_{d_{\textrm{max}}}^{\,n_{d_{\textrm{max}}}(\mathcal{M})}}{(\x_1 - \frac{c}{2})^{l_1(\mathcal{M})+1}\cdots (\x_k - \frac{c}{2})^{l_k(\mathcal{M})+1}} \nonumber
\eeq
where the $i$-th marked face is a $l_i(\mathcal{M})$-gon. For closed maps, we call $\mathrm{W}_0^{(g)} = \mathrm{F}_g$.
\ed
This makes sense because the coefficient of $t^v$ is a finite sum. It is in fact polynomial in $\nn$ and $\hat{t}_j$'s, and a rational fraction in $\x_i$'s with poles only at $\x_i = \frac{c}{2}$. In particular for rooted maps, $\#{\rm Aut}(\mathcal{M})=1$. For instance:
\beq
\mathrm{W}_1^{(0)} = \frac{t}{\x - \frac{c}{2}} + \sum_{v \geq 2} t^{v} \sum_{\mathcal{M} \in \mathbb{M}_1^{(0)}(v)}\,(-c)^{\ell(\mathcal{M})}\,\nn^{\#{\rm loops}}\,\frac{\hat{t}_3^{\,n_3(\mathcal{M})}\,\hat{t}_4^{\,n_4(\mathcal{M})}\,\cdots\, \hat{t}_{d_{\textrm{max}}}^{\,n_{d_{\textrm{max}}}(\mathcal{M})}}{(\x - \frac{c}{2})^{l_1(\mathcal{M})+1}} \nonumber
\eeq
\begin{definition}
We may collect all genera:
\beq
\ln\mathrm{Z} = \sum_{g = 0}^{\infty} \left(\frac{N}{t}\right)^{2-2g} \mathrm{F}_g,\qquad
\mathrm{W}_k = \sum_{g = 0}^{\infty} \left(\frac{N}{t}\right)^{2 - 2g - k}\mathrm{W}_k^{(g)} \nonumber
\eeq
\end{definition}
Like in \cite{BIPZ}, all the relations between generating functions which appear in this article must be understood as equalities between formal power series of $t$. To each power $t^v$, the sum over $g$ in the right hand side ranges over a finite number of maps, with a maximal genus $g_{{\rm max}}(v)$. There is no problem of exchange of limits. Notice that, according to Eqn.~\ref{eqchiMgkv}, $2g - 2 + k + v \geq 0$, so that only nonnegative powers of $t$ appear in $\mathrm{W}_k$. Z is by construction the generating function of $\mathcal{O}(\nn)$ maps (possibly not connected) with weight given by Eqn.~\ref{eq:weight}.

\subsection{Matrix model}

\cite{BIPZ} opened the way to represent these partition functions as formal matrix integrals. The $\mathcal O(\nn)$ matrix model was first introduced by I. Kostov \cite{K1}:
\beq\label{defZOn}
\mathrm{Z} = \int_{\rm formal} \mathrm{d}\hat{M}\,\, \mathrm{d}A_1\cdots\,\mathrm{d}A_\nn\,\, \ee{-\frac{N}{t}\Tr \Big[ \hat{\mathrm{V}}(\hat{M}) + (\hat{M} + \frac{c}{2}) \sum_{i} A_i^2 \Big]}
\eeq
$\mathrm{d}\hat{M}$ and $\mathrm{d}A_i$ are the usual Lebesgue measures on the vector space $\mathcal{H}_N$ of $N\times N$ hermitian matrices. ${\mathrm{V}}(\hat{M})$ is the potential:
\bea
\hat{\mathrm{V}}(\hat{M}) & = & \frac{\hat{M}^2}{2} - \sum_{j = 3}^{d_{\mathrm{max}}} \frac{\hat{t}_j\,\hat{M}^j}{j} \equiv \frac{\hat{M}^2}{2} - \hat{\mathrm{V}}_{\geq 3}(\hat{M}) \nonumber
\eea
The notation $ \int_{\rm formal}$ actually means:
\begin{footnotesize}
\beq
\mathrm{Z} = \frac{1}{\mathrm{Z}_0}\sum_{j = 0}^{\infty} \frac{N^j t^{-j}}{j!}\,\,\int_{\mathbf{H}_N^{\nn+1}} \mathrm{d}\hat{M}\, \mathrm{d}A_1\cdots\,\mathrm{d}A_\nn\,e^{-\frac{N}{2t}\Tr\big(\hat{M}^2 + c\sum_{i = 1}^{\nn} A_i^2\Big)}\,\Big[\Tr \big(\hat{\mathrm{V}}_{\geq 3}(\hat{\mathrm{M}}) - \sum_{i = 1}^{\nn} \hat{M}\,A_i^2\big) \Big]^j \nonumber
\eeq
\end{footnotesize}
where $\mathrm{Z}_0$ is the Gaussian integral:
\beq
\mathrm{Z}_0 \equiv \int_{\mathbf{H}_N^{\nn + 1}} \mathrm{d}\hat{M}\,\mathrm{d}A_1\cdots\,\mathrm{d}A_\nn\,e^{-\frac{N}{2t}\Tr \Big(\hat{M}^2 + c\sum_{i = 1}^{\nn} A_i^2\Big)} = 2^{\frac{(\nn + 1)N}{2}}c^{-\frac{\nn\,N^2}{2}}\left(\frac{\pi{}t}{N}\right)^{\frac{(\nn + 1)N^2}{2}} \nonumber
\eeq
The coefficient of $t^v$ is a finite sum of moments of gaussian integrals over hermitian matrices of size $N\times N$. This model is defined for any $\nn \in \mathbb{C}$.
$\mathrm{Z}$, $\ln{\mathrm{Z}}$, $\ldots$ are well defined formal power series in $t$, and they coincide \cite{K} with the generating functions of maps in the $\mathcal{O}(\nn)$ model. For instance:
\beq
\mathrm{W}_k = \sum_{g = 0}^{\infty} \left(\frac{N}{t}\right)^{2-2g-k} \mathrm{W}_k^{(g)}  = \left< \prod_{i=1}^k \Tr \frac{1}{\x_i - \frac{c}{2} - \hat{M}}\right>_{C,\textrm{formal}} \nonumber
\eeq
$C$ stands for "cumulant", the expectation value is taken with respect to the formal measure of Eqn.~\ref{defZOn} and:
\beq
\left(\frac{1}{\x_i - \frac{c}{2} - \hat{M}}\right)_{\textrm{formal}} \equiv \sum_{j = 0}^{\infty} \frac{\hat{M}^j}{(\x_i - \frac{c}{2})^{j + 1}} \nonumber
\eeq
It is more convenient to study the following matrix model, where we shifted to $M = \hat{M} + \frac{c}{2}$:
\bea
\label{eq:mint} \mathrm{Z} & = &  \int_{\rm formal} \mathrm{d}M\,\mathrm{d}A_1\cdots\,\mathrm{d}A_\nn\,e^{-\frac{N}{t}\Tr \Big(\mathrm{V}(M) +  \sum_{i = 1}^{\nn} M A_i^2 \Big)} \\
\label{eq:WK}\mathrm{W}_k & = & \left< \prod_{i=1}^k \Tr \frac{1}{\x_i - M}\right>_{C}
\eea
where the expectation value is taken with respect to the measure of Eqn.~\ref{eq:mint}, and the potential is:
\beq
\mathrm{V}(M) = \hat{\mathrm{V}}(\hat{M}) \equiv t_0 \,\, + \,\, \sum_{j = 1}^{d_{\textrm{max}}}\,\frac{t_jM^j}{j} \nonumber
\eeq
To recover combinatorial quantities, one has to expand the correlation functions first as formal series in $t$, then as Laurent series in $\hat{\x_i} = \x_i - \frac{c}{2}$.
$\mathrm{W}_k$ and $\mathrm{W}_k^{(g)}$ are called correlation functions in the language of statistical physics. The fully packed case correspond to the quadratic potential
\beq
\mathrm{V}_{\mathrm{FPL}}(\x) = \frac{1}{2}\left(\x - \frac{c}{2}\right)^2 \nonumber
\eeq
The $A$ matrices are gaussian and can be integrated out. Then, $\mathrm{W}_k^{(g)}$'s appear as correlators of eigenvalues $(\lambda_i)$ of $\mathrm{M}$, with a weight on eigenvalues:
\beq
\label{pol}p(\lambda_1,\ldots,\lambda_N) \propto \frac{\prod_{1 \leq i < j \leq N} (\lambda_i - \lambda_j)^2}{\left[\prod_{1 \leq i,j \leq N} (\lambda_i + \lambda_j)\right]^{\mathfrak{n}/2}} \nonumber
\eeq

\subsection{Relation to conformal field theories}
\label{sec:CFTr}
Polygonal large maps constructed by matrix models provide a discretization of Riemann surfaces. In the continuum limit, physically speaking, it is thought to define a theory of 2D quantum gravity: observables should be weighted sums over all possible surfaces endowed with a metric. The self avoiding paths present in the $\mathcal{O}(\nn)$ model are in some sense matter fields added on the surface. The usual approach of 2D quantum gravity is Liouville field theory, which is a statistical model on the moduli space of Riemann surfaces endowed with a metric (see \cite{Liou} for a review). Liouville theory can also be coupled to matter fields. Both approaches are conjectured to coincide and to enjoy conformal invariance.

"Double scaling limits" of matrix models are conjectured\footnote{To our knowledge, this conjecture is up to now proved only for the 1-matrix model, near an edge point where the equilibrium density of eigenvalues y(x) behaves as $\x^{p/2}$ \cite{BE}} to be conformal field theories (CFT) coupled to gravity \cite{PagesJaunes}. This means in practice that the scaling exponents are given by Kac's table, that the double scaling limit (the definition is recalled in Section~\ref{sec:dbllimit}) $\mathrm{W}_k^*$ of the correlation functions satisfy \textsc{pde}'s on the spectral curve. These equations come from a representation of a Virasoro algebra with central charge c. In this correspondence, the double scaling limit of the $\mathcal{O}(\nn)$ matrix model should be a conformal theory with central charge:
\beq
\mathrm{c} = 1 - 6\left(\sqrt{\mathfrak{g}} - \frac{1}{\sqrt{\mathfrak{g}}}\right)^2
\eeq
where $\mathfrak{g}$ is such that $\nn = -2\cos(\pi\mathfrak{g})$. Various $\mathfrak{g}$ corresponding to the same $\nn$ define the various \emph{phases} of the $\mathcal{O}(\nn)$ model. Many studies on the $\mathcal{O}(\nn)$ model with boundary operators \cite{JS,KBLM,JEB,KS} support this proposal. See also the review \cite{DiFGZJ} and the article \cite{MSS}.

\begin{figure}[h!]
\begin{center}
\label{fig:known}
\includegraphics[width= 0.8\textwidth]{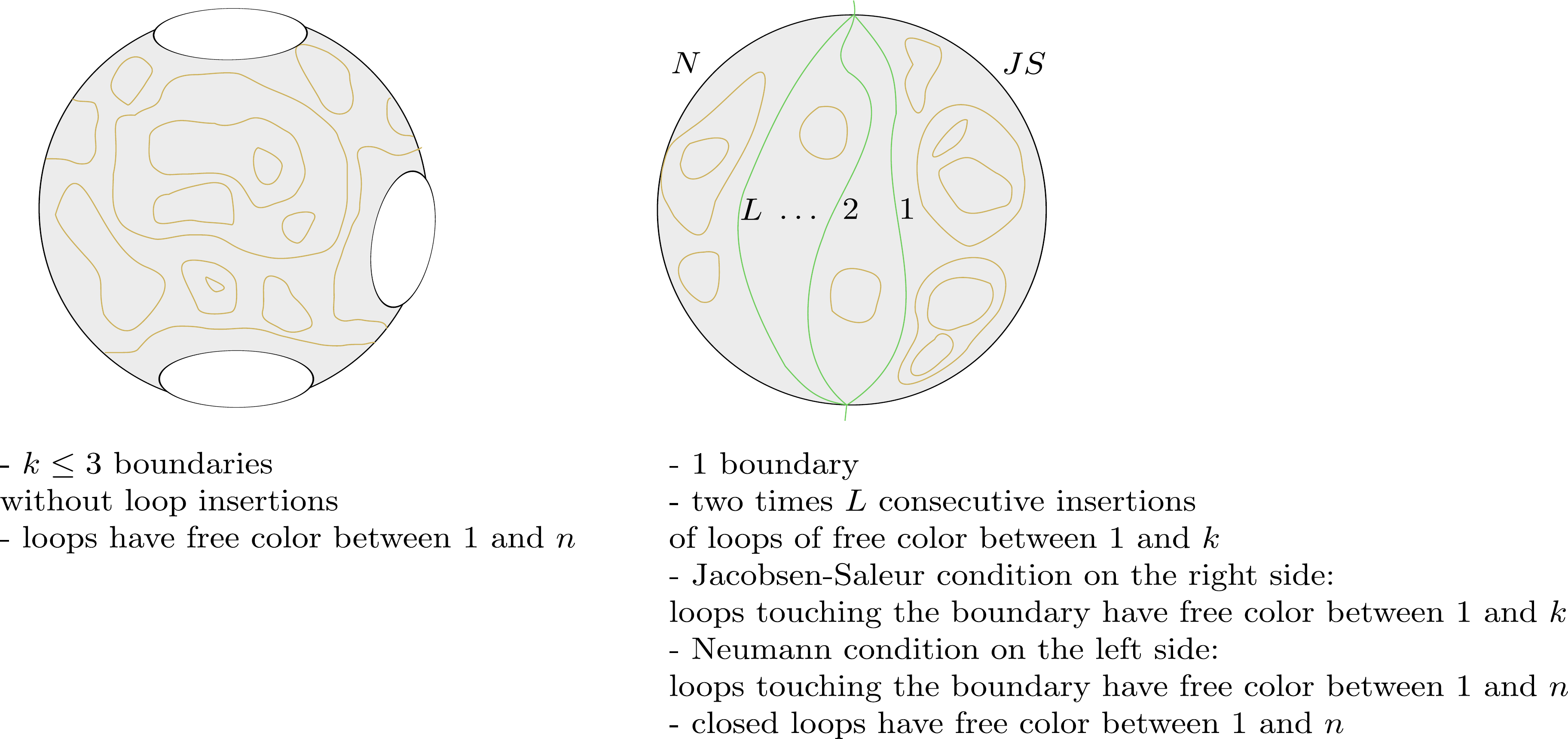}
\caption{\label{fig:g0}}\end{center}
\end{figure}

We bound ourselves to underline that, if those conjectures were correct, the critical exponents of the $\mathcal{O}(\nn)$ model would be known for any topology from the \textsc{kpz} relation \cite{KPZ,D,DK,Liou}. \textsc{kpz} is derived from Liouville conformal field theory and expresses how the critical exponents change when a CFT is coupled to gravity. Independently of CFT, rigorous results for the $\mathcal{O}(\nn)$ model were derived by sheer analysis of the loop equations, and they agree with the CFT predictions. In the literature, the exponents and the scaling form of the following \emph{genus $0$} functions of Fig.~\ref{fig:g0} were known.

\begin{figure}[h!]
\begin{center}
\label{fig:article}
\includegraphics[width=0.35\textwidth]{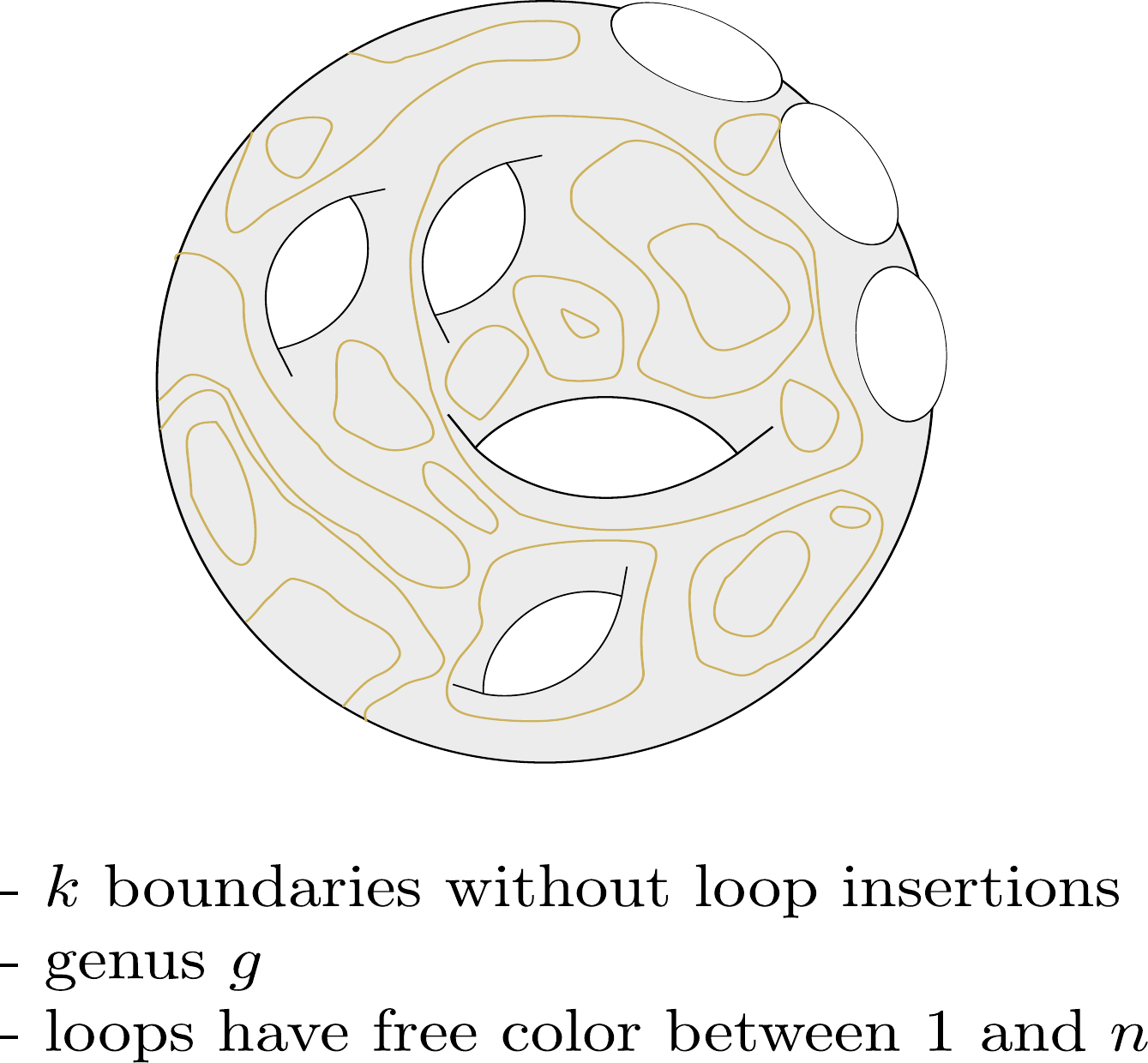}
\caption{\label{fig:g1}}\end{center}
\end{figure}

\noindent In this article, we obtain (Section~\ref{sec:lmaps}) the exponents and the scaling forms in all topologies of functions without loop insertion on the boundaries (Fig.~\ref{fig:g1}).

\subsection{Boundary insertion operator}

We can mark a face of size $j$, by taking a derivative with respect to $t_j$:
\beq
\mathrm{W}_{k+1}^{(g)}(\x_1,\dots,\x_k,\x_{k+1}) = \sum_{j \geq 0} \frac{1}{(\x_{k+1} - \frac{c}{2})^{j+1}}\,\, \frac{\partial \mathrm{W}_{k}^{(g)}(\x_1,\dots,\x_k)}{\partial \hat{t}_j} \nonumber
\eeq
Although we defined $\hat{t}_j$ only for $j\geq 3$ (this is the condition for Prop.~\ref{propMkgfinite} to hold), it is allowed to define $\hat{t}_2$, $\hat{t}_1$ and $\hat{t}_0$, and derivatives $\partial/\partial \hat{t}_0$ at $\hat{t}_0 = 0$, $\partial/\partial \hat{t}_1$ at $\hat{t}_1=0$, and $\partial/\partial \hat{t}_2$ at $\hat{t}_2=-1$. So, we define the "boundary insertion operator" as the formal series:
\bea
\frac{\partial}{\partial \hat{\mathrm{V}}(\hat{\x})} & \stackrel{{\rm def}}{=} & \sum_{j \geq 0}\,\frac{j}{\hat{\x}^{j + 1}}\,\, \frac{\partial}{\partial \hat{t}_j} \nonumber
\eea
In the new variable, $\x = \hat{\x} + \frac{c}{2}$, which is a formal series of the former one, we can do the resummation:
\beq
\frac{\partial}{\partial\hat{\mathrm{V}}(\hat{\x})} = \sum_{j \geq 0}\,\frac{j}{\x^{j + 1}}\frac{\partial}{\partial t_j} \equiv \frac{\partial}{\partial \mathrm{V}(\x)} \nonumber
\eeq
The reason for this definition is:
\beq
\mathrm{W}_{k+1}^{(g)}(\x_1,\ldots,\x_k,\x_{k+1}) = \frac{\partial}{\partial \mathrm{V}(\x_{k + 1})}\,\mathrm{W}_k^{(g)}(\x_1,\ldots,\x_k)
\eeq

\subsection{Loop equations}
\label{sec:loopeq}

The loop equations provide relationships between the generating functions $\mathrm{W}_k^{(g)}$'s. They can be derived either by integration by parts in the matrix integral, or by combinatorial manipulations. The first method is much faster, but the reader not familiar with matrix integrals may prefer the bijective proof. For completeness we recall the two possibilities.

\subsubsection{Derivation from the matrix integral}

Loop equations are the infinitesimal counterparts of the invariance of an integral under a continuous family of change of variable \cite{Mi83}. They are true both for formal integrals and convergent integrals, so we do not have to bother with variable $\hat{\x}$, and we can work directly with variable $\x$. One has to compute the jacobian of the change of variable and the variation of the exponential term. The loop equation merely states that the jacobian cancels the variation of the exponential, in expectation. The general method for computing jacobians of infinitesimal changes of matrix variables, is called "split and merge", and is exposed in many articles (for instance, \cite{E2}).

Let us write $I = \{\x_2,\ldots,\x_k\}$, and define the following auxiliary functions:
\beq
\mathrm{G}_k = \sum_{g = 0}^{\infty} \left(\frac{N}{t}\right)^{2-2g-k} \mathrm{G}_k^{(g)}  = \left< \Tr \frac{1}{\x-M}\,A_1^2\,\, \prod_{\x_i \in I} \Tr \frac{1}{\x_i-M}\right>_{C}\phantom{dodo} \nonumber
\eeq
\beq
\widetilde{\mathrm{G}}_k = \sum_{g = 0}^{\infty} \left(\frac{N}{t}\right)^{2-2g-k} \td{\mathrm{G}}^{(g)}_k  = \left< \Tr \frac{1}{\x-M}\,A_1\frac{1}{\x'-M}A_1\,\, \prod_{\x_i \in I} \Tr \frac{1}{\x_i-M}\right>_{C}\nonumber
\eeq
\beq
\mathrm{P}_k = \sum_{g = 0}^{\infty} \left(\frac{N}{t}\right)^{2-2g-k} \mathrm{P}_k^{(g)} = \left< \Tr \frac{\mathrm{V}'(\x)-\mathrm{V}'(M)}{\x-M}\,\,\, \prod_{\x_i \in I} \Tr \frac{1}{\x_i-M}\right>_{C}\phantom{dodo}\nonumber
\eeq
$\mathrm{P}_k^{(g)}$ is by construction the polynomial part of $\mathrm{V}'(\x)\, \mathrm{W}_k^{(g)}(\x,\x_2,\ldots,\x_k)$ at large x:
\beq
\mathrm{P}_k^{(g)}(\x,\x_2,\dots,\x_k) = \left(\mathrm{V}'(\x)\, \mathrm{W}_k^{(g)}(\x,\x_2,\dots,\x_k)\right)_+ \nonumber
\eeq

We indicate infinitesimal change of variable and the corresponding loop equations:
\begin{itemize}
\item[$\bullet$] With $M \to M + \epsilon\,\frac{1}{x-M}$, we obtain to first order in $\epsilon$:
\beq\label{loopeqW1G1}
\mathrm{W}_1(\x)^2 + \mathrm{W}_2(\x,\x) = \frac{N}{t}\,\Big[\mathrm{V}'(\x)\mathrm{W}_1(\x)-\mathrm{P}_1(\x) - \nn\,\mathrm{G}_1(\x) \Big]
\eeq
\item[$\bullet$] With $A_1\to A_1+\epsilon\,\frac{1}{\x-M}A_1\frac{1}{\x'-M}$, we obtain:
\beq\label{loopeqW1W1}
\mathrm{W}_1(\x)\mathrm{W}_1(\x')+\mathrm{W}_2(\x,\x') = \frac{N}{t}\left[(\x+\x') \widetilde{\mathrm{G}}_1(\x,\x') - \mathrm{G}_1(\x) - \mathrm{G}_1(\x')\right]
\eeq
\end{itemize}
Auxiliary functions can be eliminated by specializing to $\x = -\x'$ and combining the two equations:
\bea
&& \mathrm{W}_2(\x,\x) + \nn \mathrm{W}_2(\x,-\x) + \mathrm{W}_2(-\x,-\x) + \mathrm{W}_1(\x)^2 + \nn \mathrm{W}_1(\x)\mathrm{W}_1(-\x) + \mathrm{W}_1(-\x)^2 \nonumber\\
& = & \frac{N}{t}\left[\mathrm{V}'(x)\mathrm{W}_1(\x) + \mathrm{V}'(-\x)\mathrm{W}_1(\x) - \mathrm{P}_1(\x) - \mathrm{P}_1(-\x)\right] \nonumber
\eea
If we collect the highest power in $N$, which is $N^2$, we obtain:
\bt\label{MLP}
Master loop equation (quadratic in $\mathrm{W}_1^{(0)}$).
\small{\bea
\label{eq:mloopeq}
& & \mathrm{W}^{(0)}_1(\x)^2 + \nn \mathrm{W}^{(0)}_1(\x)\mathrm{W}^{(0)}_1(-\x) + \mathrm{W}^{(0)}_1(-\x)^2 \nonumber \\
& = & \mathrm{V}'(\x)\mathrm{W}^{(0)}_1(\x) + \mathrm{V}'(-\x)\mathrm{W}^{(0)}_1(\x) -\mathrm{P}^{(0)}_1(\x) - \mathrm{P}^{(0)}_1(-\x) \nonumber
\eea}
\et
We stress that Eqns.~\ref{loopeqW1G1} and \ref{loopeqW1W1} are true for any potential for which the quantities involved make sense. In particular, $(t_0,t_1,t_2)$ is not restricted to $(0,0,-1)$. Accordingly, we can obtain loop equations for all $\mathrm{W}_k$ by successive applications of $\frac{\partial}{\partial \mathrm{V}}$. Here is the general results for $k \geq 1$ and collecting the $N^{2 - 2g}$ terms ($g \geq 0$).
\bt
\label{MLPgk}Higher loop equations (linear in $\mathrm{W}_k^{(g)}$).
\begin{small}
\bea
\label{eq:loopeqkg}
& & \mathrm{W}^{(g-1)}_{k+1}(\x,\x,I) + \mathrm{W}^{(g-1)}_{k+1}(-\x,-\x,I) + \nn\,\mathrm{W}^{(g-1)}_{k+1}(\x,-\x,I) \nonumber  \\
& & + \sum_{J \subseteq I,\,0 \leq h \leq g}^{'} \left(\mathrm{W}^{(h)}_{|J| + 1}(\x,J)\,\mathrm{W}^{(g-h)}_{k-|J|}(\x,{}^{c}J) + \mathrm{W}^{(h)}_{|J| + 1}(-\x,J)\,\mathrm{W}^{(g-h)}_{k-|J|}(-\x,{}^{c}J)\right. \nonumber \\
& & \left.\phantom{dodododododo} + \nn\,\mathrm{W}^{(h)}_{|J| + 1}(\x,J)\,\mathrm{W}^{(g-h)}_{k-|J|}(-\x,{}^{c}J)\right) \nonumber \\
& & + \sum_{\x_i \in I}\frac{\partial}{\partial \x_i}\left(\frac{\mathrm{W}_{k - 1}^{(g)}(\x,I\setminus\{\x_i\}) - \mathrm{W}^{(g)}_{k - 1}(I)}{\x - \x_i} -  \frac{\mathrm{W}^{(g)}_{k - 1}(-\x,I\setminus\{\x_i\}) - \mathrm{W}^{(g)}_{k - 1}(I)}{\x + \x_i} \right) \nonumber \\
& = & \mathrm{V}'(\x)\mathrm{W}^{(g)}_{k}(\x,I) + \mathrm{V}'(-\x)\mathrm{W}^{(g)}_{k}(-\x,I) - \mathrm{P}^{(g)}_{k}(\x,I) - \mathrm{P}^{(g)}_{k}(-\x,I) \nonumber
\eea
\end{small}We take $\mathrm{W}_k^{(-1)} = 0$ as convention. $\sum^{'}$ means that we exclude the two terms where $W_k^{g}$ itself appears.
\et
Thm.~\ref{MLP} is just a specialization of Thm.~\ref{MLPgk} to $(k,g) = (1,0)$, but the master loop equations plays a special role. When $\nn = 0$, we recover the loop equations of maps \cite{Tutte1,Tutte2,E2}.

\subsubsection{Derivation \`a la Tutte}

Now, we give a bijective proof of Thm.~\ref{MLP}. We introduce the auxiliary functions:
\begin{itemize}
\item[$\bullet$] $\widetilde{\mathrm{G}}_k^{(g)}(\x,\x',I)$, which counts maps which carry a path of weight 1 (and not $\nn$) starting on a distinguished marked face (of length $l + l' + 2$), and ending on the same marked face. $l$ is the notation for the distance between the starting point and ending point, and the weight is $\propto \x^{-(l + 1)}\,\x'{}^{-(l' + 1)}$.
\begin{figure}[h!]
\begin{center}
\includegraphics[width=0.76\textwidth]{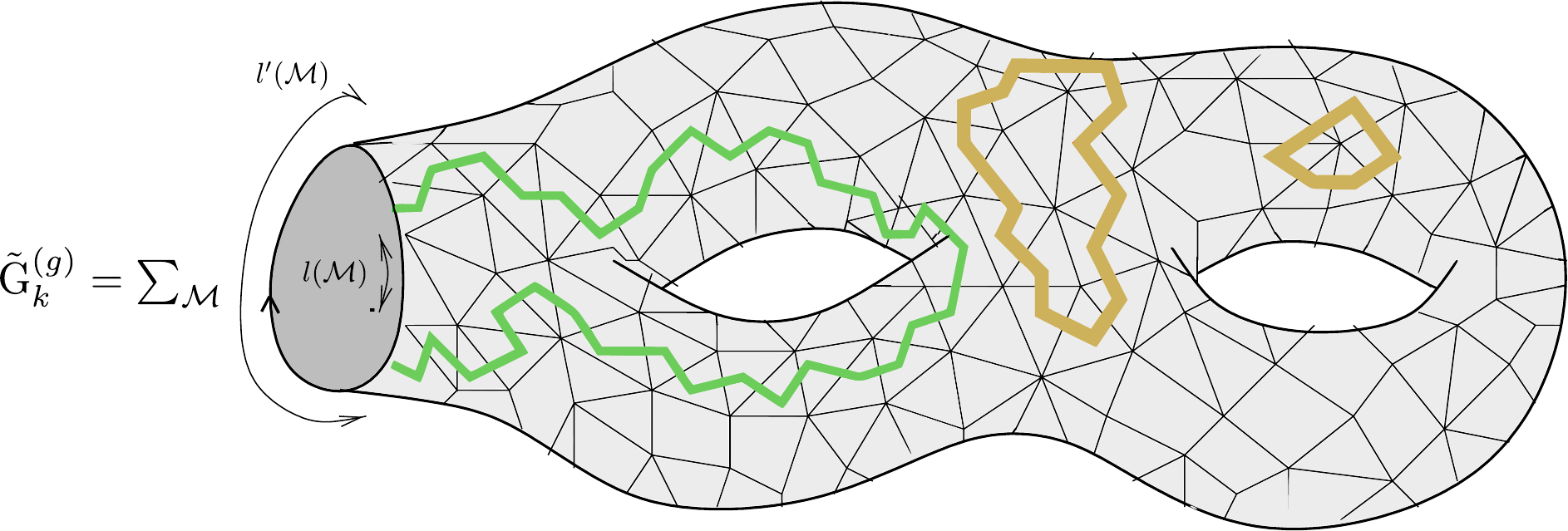}
\end{center}
\end{figure}
\item[$\bullet$] $\mathrm{G}_k^{(g)}(\x,I)$, defined by the same sum restricted to maps where the starting point is just next to the ending point ($l \equiv 1$). The weight is $\propto \x'{}^{-(l' + 1)}$. Actually:
    \beq
\mathrm{G}^{(g)}_k(\x') = \mathop{{\rm lim}}_{\x \to \infty}\,
\x\,\widetilde{\mathrm{G}}^{(g)}_k(\x,\x') \nonumber
    \eeq
\begin{figure}[h!]
\begin{center}
\includegraphics[width=0.76\textwidth]{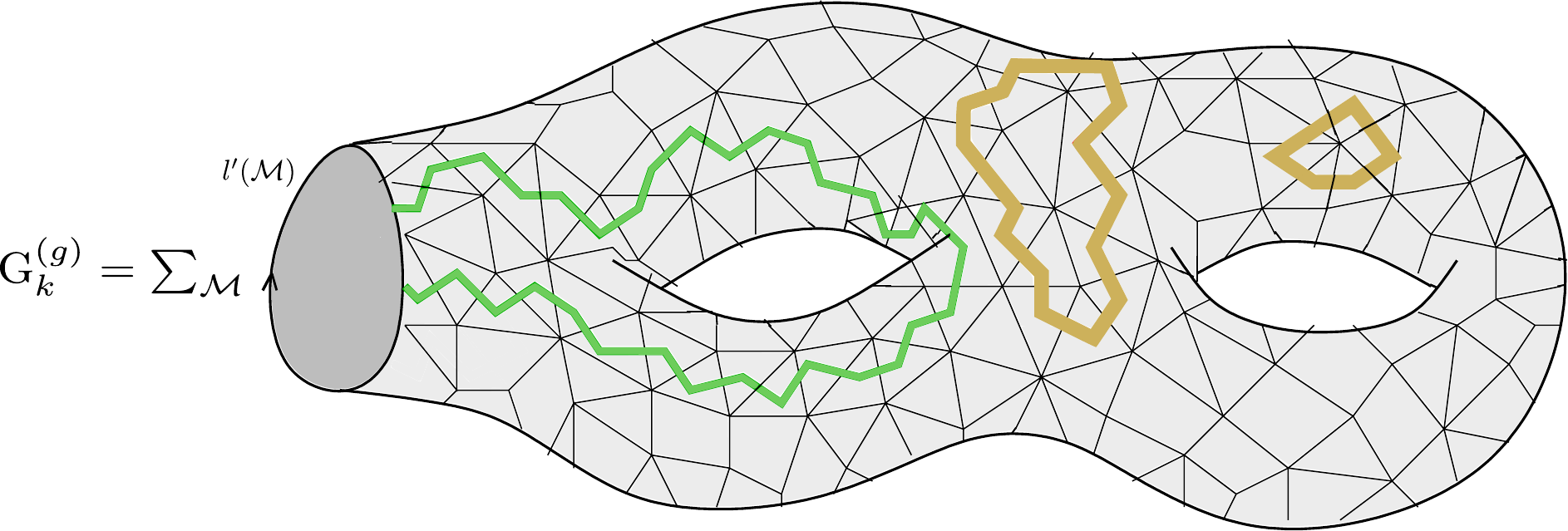}
\end{center}
\end{figure}
\end{itemize}
The combinatorial derivation follows the method of Tutte \cite{Tutte1,Tutte2}, i.e we look for a recursion on the number of edges.

\begin{itemize}
\item[$\bullet$] Consider $\mathrm{W}_1^{(0)}$, which counts rooted planar maps. If we remove the marked edge in a configuration counted by $\mathrm{W}_1^{(0)}$, we count $\left[\big(\x - \frac{c}{2}\big)\mathrm{W}_1^{(0)}(\x) - t\right]$ (taking care of the planar map with only 1 vertex) $\mathrm{W}_1^{(0)}$). Three cases occur when we remove the marked edge. the face on the other side of the marked edge could be an unmarked polygon, it could carry a piece of path, or it could be the marked face itself. Pictorially, if we fill the marked face in dark grey (the whole map should be a sphere):
    \vspace{0.1cm}
\begin{figure}[h!]
\begin{center}
\includegraphics[width=0.63\textwidth]{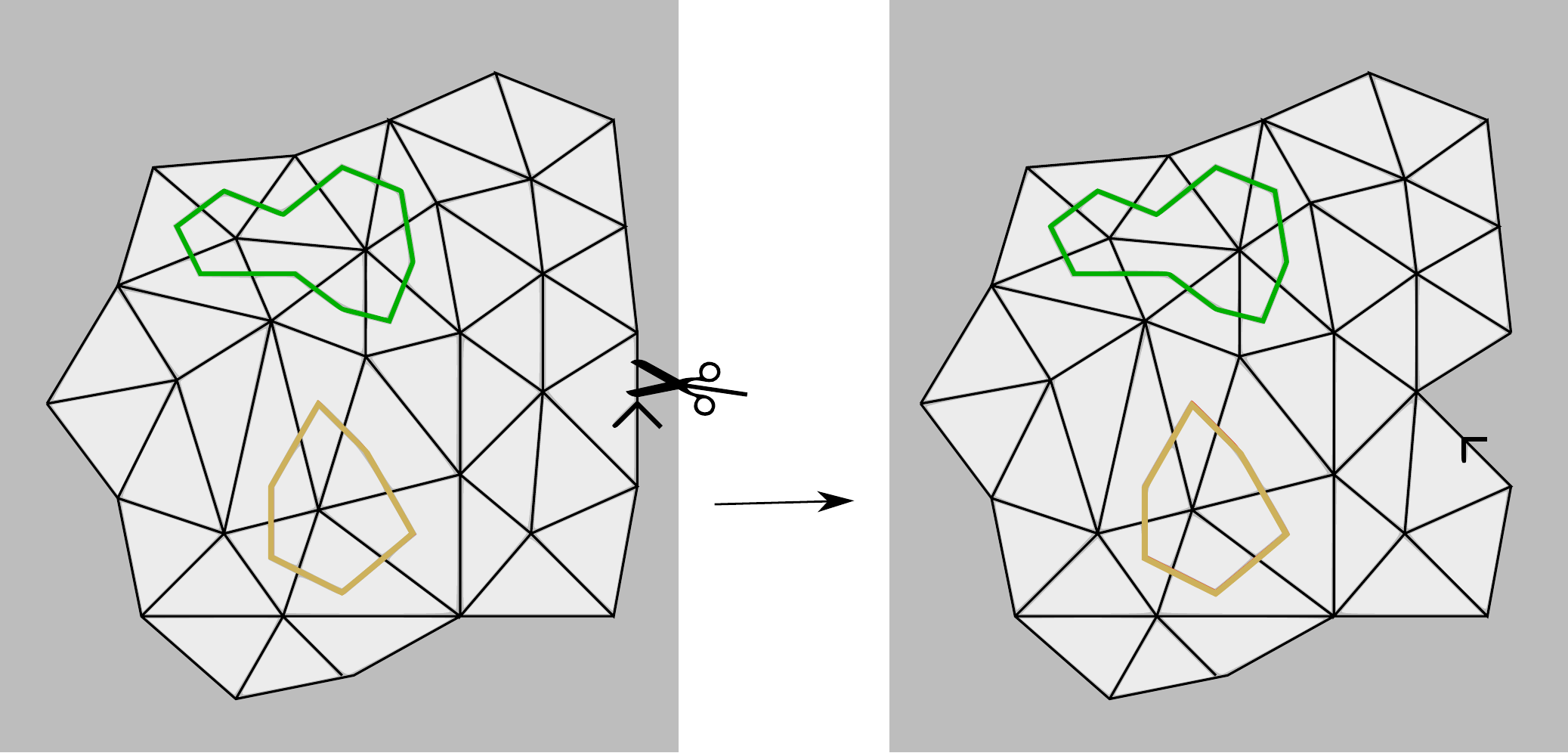}

\vspace{0.1cm}

\includegraphics[width=0.63\textwidth]{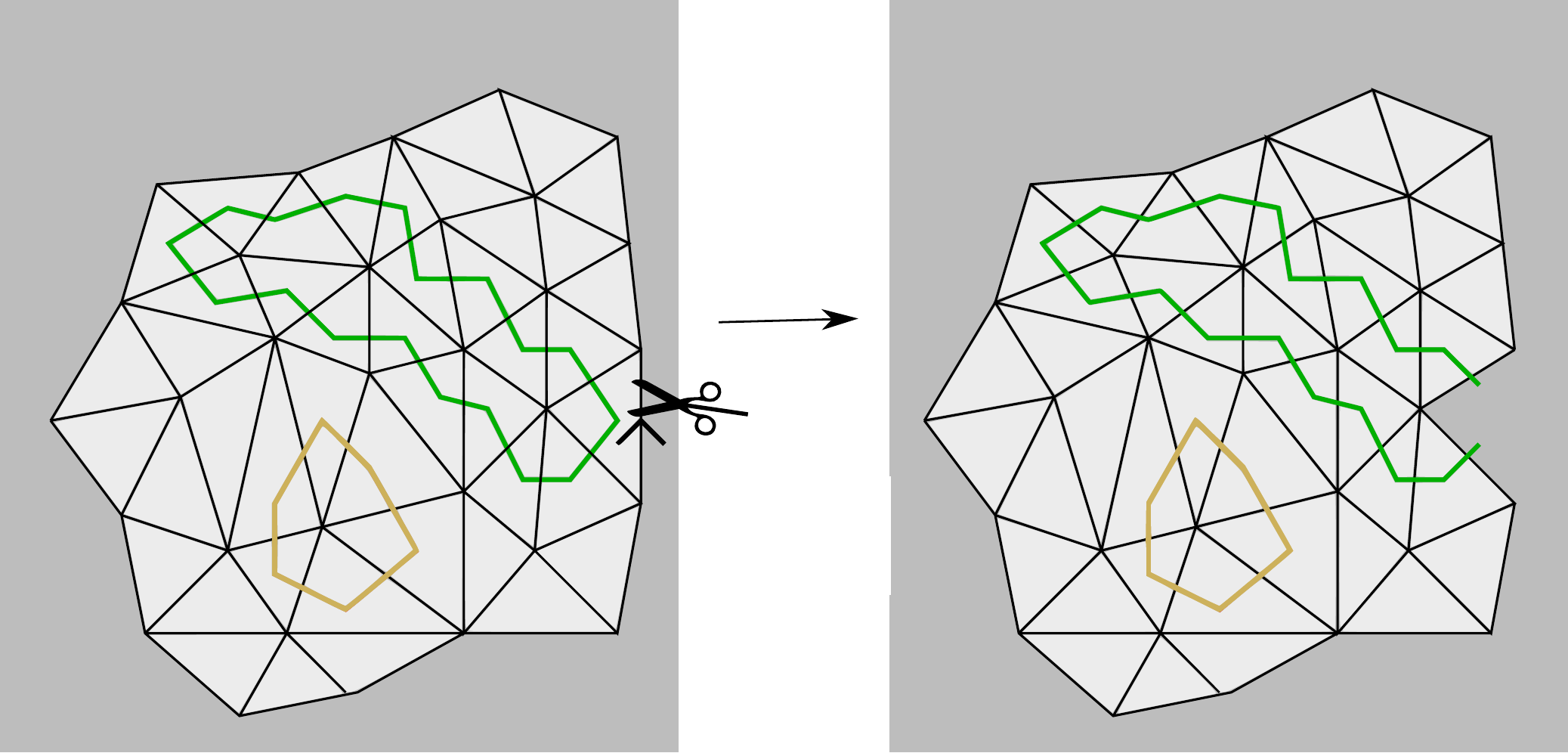}

\vspace{0.1cm}

\includegraphics[width=0.63\textwidth]{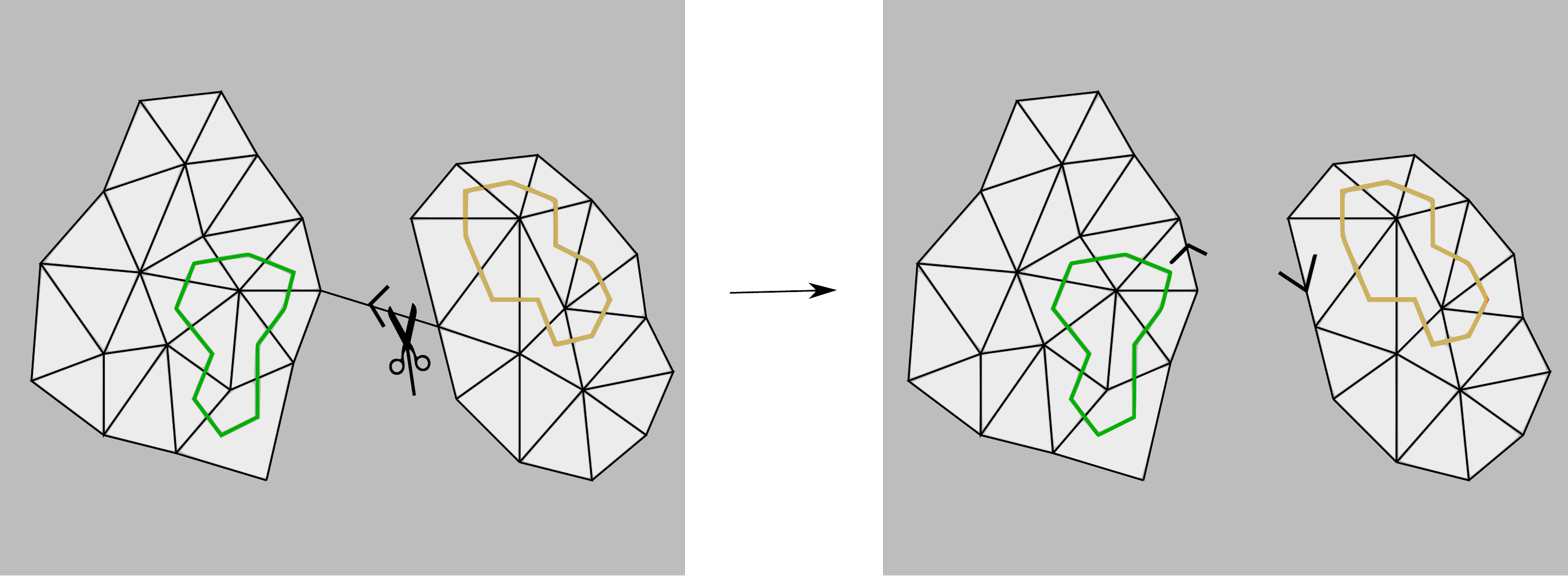}
\end{center}
\end{figure}
Thus:
\beq
\label{eq:eq1}\left(\x - \frac{c}{2}\right)\mathrm{W}_1^{(0)}(\x) - t = \left[\hat{\mathrm{V}}_{\geq 3}\left(\x - \frac{c}{2}\right)\mathrm{W}_1^{(0)}(\x)\right]_{-} + \nn \mathrm{G}_1^{(0)}(\x) + \mathrm{W}_1^{(0)}(\x)^2 \nonumber
\eeq
where $(\cdots)_{-}$ means the negative part of the Laurent expansion in $\x - \frac{c}{2}$. We see that the whole $\hat{\mathrm{V}}'\left(\x - \frac{c}{2}\right) = \mathrm{V}'(\x)$ is restored in this equation: nothing special happens with the quadratic part, and the loop equation takes a nice form because of the convention for the term $v = 1$ in $\mathrm{W}_1^{(0)}$. The polynomial (in the variable $\hat{\x}$ or $\x$):
\beq
\mathrm{P}_1^{(0)}(\x) = \left<\frac{\hat{\mathrm{V}}'(\hat{\x}) - \hat{\mathrm{V}}'(\hat{M})}{\hat{\x} - \hat{M}}\right>^{(0)} = \left<\frac{\mathrm{V}'(\x) - \mathrm{V}'(M)}{\x - M}\right>^{(0)} \nonumber
\eeq
is precisely the nonnegative part of the Laurent expansion of $\mathrm{V}'(\x)\mathrm{W}_1^{(0)}(\x)$. Hence Eqn.~\ref{loopeqW1G1}:
\beq
\mathrm{W}_1^{(0)}(\x)^2 = \mathrm{V}'(\x)\mathrm{W}_1^{(0)}(\x) - \mathrm{P}_1^{(0)}(\x) - \nn \mathrm{G}_1^{(0)}(\x) \nonumber
\eeq
\item[$\bullet$] Consider configurations counted by $\widetilde{\mathrm{G}}_1^{(0)}(\x,\x')$. If we remove the triangle of the marked face where the path starts, we are counting \mbox{$\left[\left(\x + \x' - c\right)\widetilde{\mathrm{G}}_1^{(0)}(\x,\x') - \mathrm{G}_1^{(0)}(\x) - \mathrm{G}_1^{(0)}(\x')\right]$}. Indeed, either $l$ or $l'$ is shortened by one, but we have to take care of the degenerate cases $\mathrm{G}_1^{(0)}(\x)$ and $\mathrm{G}_1^{(0)}(\x')$ where $l'$ (resp. $l$) shrinks to $0$.
There are two possibilities for this removal: either the path was of length 0 or not.
\vspace{0.1cm}
\begin{figure}[h!]
\begin{center}
\includegraphics[width=0.65\textwidth]{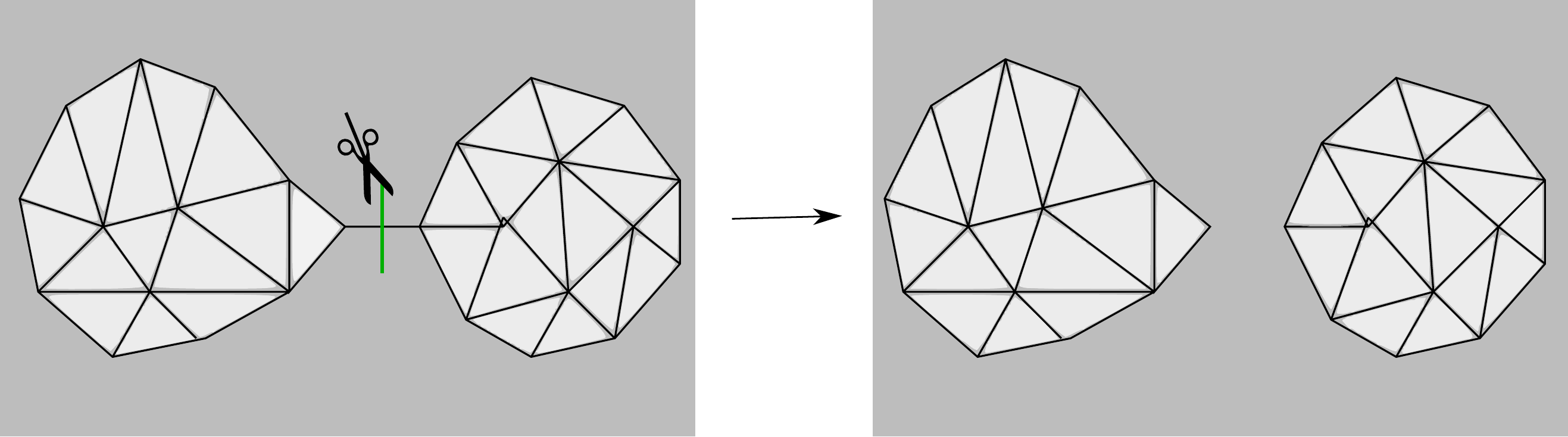}

\vspace{0.1cm}

\includegraphics[width=0.65\textwidth]{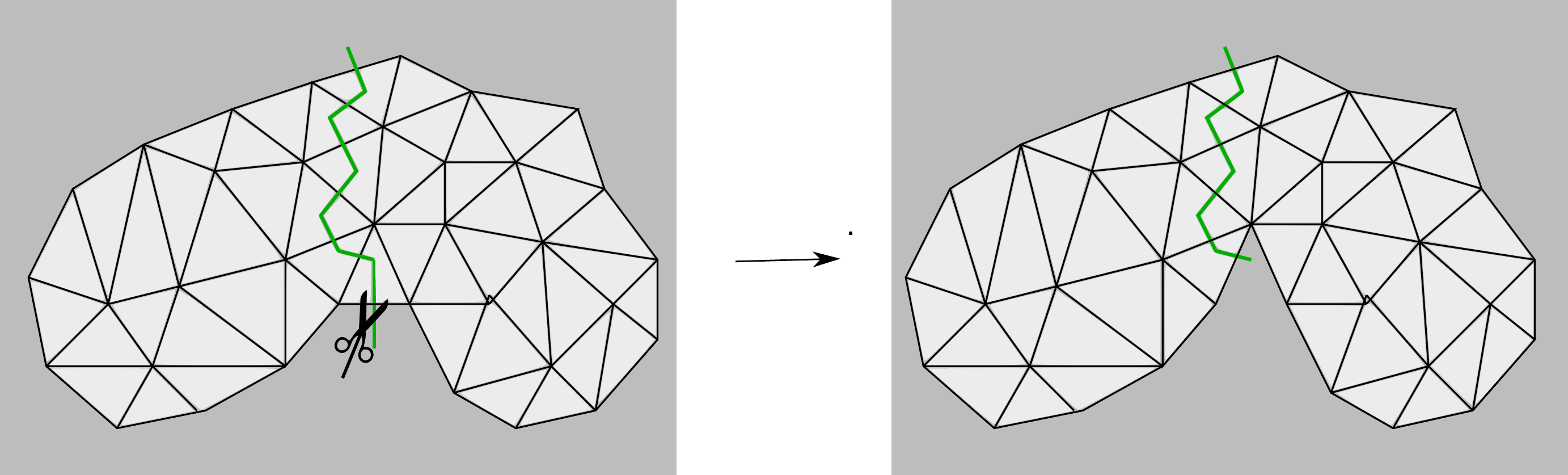}
\end{center}
\end{figure}
Thus, we find Eqn.~\ref{loopeqW1W1}:
\beq
\big((\x+\x' - c)\widetilde{\mathrm{G}}^{(0)}_1(\x,\x') - \mathrm{G}_1^{(0)}(\x) - \mathrm{G}_1^{(0)}(\x')\big) = \mathrm{W}_1^{(0)}(\x)\mathrm{W}_1^{(0)}(\x') - c\,\widetilde{\mathrm{G}}_1^{(0)}(\x,\x') \nonumber
\eeq
\end{itemize}

Higher loop equations (Thm.~\ref{MLPgk}) can be derived in a similar way.

\subsection{Analyticity properties}

To each order $t^v$, the sum over ${\mathbb M}_k^{(g)}(v)$ in $\mathrm{W}_k^{(g)}$ is a finite sum, and in particular, the coefficient of $t^v$ in $\mathrm{W}_k^{(g)}$ is a rational fraction of each $\x_i$, with poles only at $\x_i = \frac{c}{2}$, of maximal degree $4g-4+k+2v$. In Appendix \ref{app:1cut}, we prove that this implies the following analytical properties for the $\mathrm{W}_k^{(g)}$'s:

\begin{lemma}\label{Lemma1cut} 1-cut lemma. \\
There exists $\rho_0 > 0$ (depending only on the degree $d_{\textrm{max}}$ of the polynomial $\hat{\mathrm{V}}$, $\hat{t}_3,\dots, \hat{t}_{d_{\max}},  \nn$ and $c$, and a priori on $k,g$) such that, for all $r > 0$:
\begin{itemize}
\item[$\bullet$]  If $I = (\x_1,\dots,\x_k) \in \left(\C\setminus\mathcal{D}(\frac{c}{2}\,; r)\right)^k \;\Rightarrow$, then $\mathrm{W}_k^{(g)}(I)$, as a formal series in $t$, has a radius of convergence $\geq \frac{r}{\rho_0}$.
\end{itemize}
Accordingly, if we hold $k - 1$ variables $J = (\x_2,\ldots,\x_k)$ fixed, at values different from $\frac{c}{2}$: for $r > 0$ small enough, for all $t \in \mathcal{D}\left(0\,; \frac{r}{\rho_0}\right)$, $\mathrm{W}_k^{(g)}(\x,J)$ is holomorphic for $\x \in \C\setminus\mathcal{D}(\frac{c}{2}\,; r)$, $\mathcal{C}^{\infty}$ in $t$, and \mbox{$\mathcal{D}(\frac{c}{2}\,; r)\cap\overline{\mathcal{D}}(-\frac{c}{2}\,; r) = \emptyset$}.

More precisely, there exists $t_0 > 0$ (depending only on $d_{\max}, \hat{t}_3, \dots, \hat{t}_{d_{\max}}, \nn$ and $c$) and two formal series $a(t)$, $b(t)$ in $\sqrt{t}$, whose radius of convergence in $\sqrt{t}$ is greater than  $\sqrt{t_0}$ (non zero), such that (we write $\gamma(t)$ the segment $[a(t),b(t)]$):
\begin{itemize}
\item[$\bullet$] $a(t) = \frac{c}{2} - c_1\sqrt{t} + O(\sqrt{|t|})$ and $b(t) = \frac{c}{2} + c_1\sqrt{t} + O(\sqrt{|t|})$ for some $c_1 \in \mathbb{C}$.
\item[$\bullet$] $\mathrm{W}_1^{(0)}(\x)$ is absolutely convergent for $t \in \mathcal{D}(0\,; t_0)$, holomorphic on $\C\setminus \gamma(t)$, and has a discontinuity on $\gamma(t)$. On a neighborhood of $\gamma(t)$, it takes the form $\mathrm{h}(\x)\times\sqrt{(\x - a)(\x - b)}$, where $\emph{h}$ is meromorphic in a neighborhood of $\gamma(t)$, has no pole except maybe in $a(t)$ and $b(t)$, and has no zeroes on $\gamma(t)$. We call this behavior a square root discontinuity.
\item[$\bullet$] For all the other $k,g$'s, $\mathrm{W}_k^{(g)}(\x_1,\dots,\x_k)$ has a square root discontinuity in each variable on $\gamma(t)$, and is holomorphic on $\mathbb{C}\setminus \gamma(t)$.
\end{itemize}
\end{lemma}

This lemma is quite technical, and we give its proof in Appendix~\ref{app:1cut}. As a brief sketch of the proof, let us say that, by a very rough bound on $\#{\mathbb M}_1^{(0)}(v)$, we first prove that $\mathrm{W}_1^{(0)}$ is convergent in the domain $ \mathbb{C}\setminus\mathcal{D}\left(\frac{c}{2}\,;\,\frac{|c|}{2}\right)$ when $|t|< t^*$ for some $t^*$, and thus $\mathrm{W}_1^{(0)}(\x)$ can have singularities only in a small disc centered in $c$, of radius smaller than $\frac{|c|}{2}$. This implies that $\mathrm{W}_1^{(0)}(-\x)$ is analytical for x in this disc. Then, from the master loop equation (Thm.~\ref{MLP}), $\mathrm{W}_1^{(0)}$ can only have (and must have) square-root discontinuity in the disc, at points $\x = a(t)$ and $\x = b(t)$. Eventually, the series $a(t)$ and $b(t)$ are determined by the master loop equation. This property is called the 1-cut assumption in physics (although it is not an assumption here), and is closely related to Brown's lemma in combinatorics. The loop equations themselves do not have a unique solution, but we shall see that there is a unique one satisfying this lemma.

\subsection{Remark: convergent matrix integrals}
\label{sec:cvm}
The 1-cut assumption holds for formal matrix integrals, i.e  generating functions of the $O(\nn)$ model configurations.

However, one could be interested in studying the matrix integral of Eqn.~\ref{eq:mint}, not as a formal matrix integral, but as a genuine convergent integral. In this case, a 1-cut lemma can hold or not, depending on the choice of $\mathrm{V}(M)$, and in fact on the choice of the integration domain for the eigenvalues of M. In some sense, it holds if the integration domain is a "steepest descent" integration path for the potential V, but those considerations are beyond the scope of our article. When this is the case, it is known that, for $|t|$ small enough according to the bound of the 1-cut lemma, $\mathrm{W}_k$ does have an asymptotic expansion of the form:
 \beq
 \mathrm{W}_k^{(g)} = \sum_{g = 0}^{\infty} \left(\frac{N}{t}\right)^{2 - 2g - k}\mathrm{W}_k^{(g)}(t)
 \eeq
Of course, these $\mathrm{W}_k^{(g)}$ satisfy then the loop equations (Thm~\ref{MLPgk}). In this article, we shall consider only the situation (realized in combinatorics, because of Lemma~\ref{Lemma1cut}) where the 1-cut property holds, with the cut $[a,b] \subseteq \mathbb{R}_+$. We leave the "multi-cut solution" to $\mathcal{O}(\mathfrak{n})$ model loop equations for a future work.

\section{The linear equation}
\label{sec:lineq}
\label{sec:spe}

We write $\nn = -2\cos(\pi\mu)$ and we assume $\mu \in \left.]0,1[\right.\setminus\{0\}$. $\mu\,\in \left.]0,1[\right.$ is in bijection with $\nn \in ]-2,2[$. With few technical modifications, the loop equations can also be solved for $\mathfrak{n}$ outside of that range\footnote{Notice the $\mathcal{O}(-2)$ model is dual to the $\mathcal{O}(0)$ model by the change of variable $M \rightarrow \sqrt{M}$, i.e. to the 1-hermitian matrix model with a non analytic potential $\widetilde{V}(M) = V(\sqrt{M})$}, but the nature of their solution is quite different. The range $\mathfrak{n} \in \mathbb{R}$ was analyzed in \cite{EK2}. We now review the techniques developed in \cite{E1} to solve the master loop equation (Thm.~\ref{MLP}), and introduce enough algebraic geometry to present the solution of the higher loop equations.

\subsection{Saddle point equation}
\label{sec:dessin1}
Due to the analytical structure of $\mathrm{W}_1^{(0)}$ (square root discontinuity with end points $a(t)$,$b(t)$), we can transform the master loop equation, which is quadratic, into a linear one. The latter was originally called "saddle point equation" because it coincides with the saddle point approximation for the density of eigenvalues \cite{EZJ}.

\begin{proposition}

\beq\label{eq:linW10}
\encadremath{
\forall \x \in [a(t),b(t)] \quad\quad  \mathrm{W}_1^{(0)}(\x + i\epsilon) + \nn \mathrm{W}_1^{(0)}(-\x) + \mathrm{W}_1^{(0)}(\x - i\epsilon) \mathop{=}_{\epsilon \rightarrow 0} \mathrm{V}'(\x)
}\eeq
\end{proposition}

\proof{
We start from Thm.~\ref{MLP}
\bea
&& \mathrm{W}^{(0)}_1(\x)^2 + \nn \mathrm{W}^{(0)}_1(\x)\mathrm{W}^{(0)}_1(-\x) + \mathrm{W}^{(0)}_1(-\x)^2 \cr
&=& \mathrm{V}'(\x)\mathrm{W}^{(0)}_1(\x) + \mathrm{V}'(-\x)\mathrm{W}^{(0)}_1(-\x) - \mathrm{P}^{(0)}_1(\x) - \mathrm{P}^{(0)}_1(-\x) \nonumber
\eea
where $\mathrm{P}_1^{(0)}(\x)$ is a polynomial in x of degree ($\mathrm{deg}\,\mathrm{V} - 2$). Because of the 1-cut lemma:
\beq
\forall \x \in [a(t),b(t)] \quad\quad \mathrm{W}_1^{(0)}(-\x + i\epsilon) - \mathrm{W}_1^{(0)}(-\x - i\epsilon) \mathop{=}_{\epsilon \rightarrow 0} 0\nonumber
\eeq
Besides, let us define:
\bea
&& 2\mathrm{P}_e(\x) \nonumber \\ & = & \mathrm{P}_1^{(0)}(\x) + \mathrm{P}_1^{(0)}(-\x) \nonumber \\
& = & \mathrm{W}^{(0)}_1(\x)^2 + \nn \mathrm{W}^{(0)}_1(\x)\mathrm{W}^{(0)}_1(-\x) + \mathrm{W}^{(0)}_1(-\x)^2  - \mathrm{V}'(\x)\mathrm{W}^{(0)}_1(\x) - \mathrm{V}'(-\x)\mathrm{W}^{(0)}_1(-\x) \nonumber
\eea
$\mathrm{P}_e(\x)$ is a polynomial, so:
\beq
\forall \x \in [a(t),b(t)] \quad\quad  \mathrm{P}_e(\x+i\epsilon) - \mathrm{P}_e(\x-i\epsilon)\mathop{=}_{\epsilon \rightarrow 0} 0\nonumber
\eeq
This equation factorizes into:
\beq
 (\mathrm{W}_1^{(0)}(\x+i\epsilon) - \mathrm{W}_1^{(0)}(\x-i\epsilon))\,\left[\mathrm{W}_1^{(0)}(\x+i\epsilon) + \nn \mathrm{W}_1^{(0)}(-\x) + \mathrm{W}_1^{(0)}(\x-i\epsilon) - \mathrm{V}'(\x) \right] \mathop{=}_{\epsilon \rightarrow 0} 0
\nonumber \eeq
Since on $\x\in[a(t),b(t)]$ we have $\mathrm{W}_1^{(0)}(\x+i\epsilon)\neq \mathrm{W}_1^{(0)}(\x - i\epsilon)$ in the limit $\epsilon \rightarrow 0$, we find Eqn.~\ref{eq:linW10}.}

\medskip

Eqn.~\ref{eq:linW10} is linear, provided that $a(t)$ and $b(t)$ are known. The non-linearity is hidden in the determination of $a(t)$ and $b(t)$. Given a segment $[a,b]$ of the positive real line, we shall study the general 1-cut solutions of the homogeneous linear equation:
\beq
\label{eq:00}\forall \x \in [a,b]\quad\quad \mathrm{W}(\x + i\epsilon) + \nn \mathrm{W}(-\x) + \mathrm{W}(\x - i\epsilon) \mathop{=}_{\epsilon \rightarrow 0} 0
\eeq
The extension to an arbitrary path $[a,b]$ in the complex plane presents no difficulty, but is not needed for combinatorics.

\subsection{Algebraic geometry construction}
\label{sec:dessin2}

We look for a solution of Eqn.~\ref{eq:linW10}, with only one cut $[a,b]$, and in particular which is analytical on $[-b,-a]$. Since our equation involves both $\x\in[a,b]$ and $-\x\in [-b,-a]$, it is convenient to introduce a conformal mapping between the complex plane with two cuts $[a,b]\cup [-b,-a]$ and the hyperelliptical curve \mbox{$\mathcal{S}: \sigma^2 = (\x^2-a^2)(\x^2-b^2)$}.

We assume $0 < a < b < \infty$. The cases $a = 0$ of $b = \infty$, treated directly in Section~\ref{sec:dege}, are easier and somewhat more explicit, and they correspond to critical points \cite{KS}, as we review in Section~\ref{sec:princip}.

\newpage

\begin{figure}[h!]
\includegraphics[width=\textwidth]{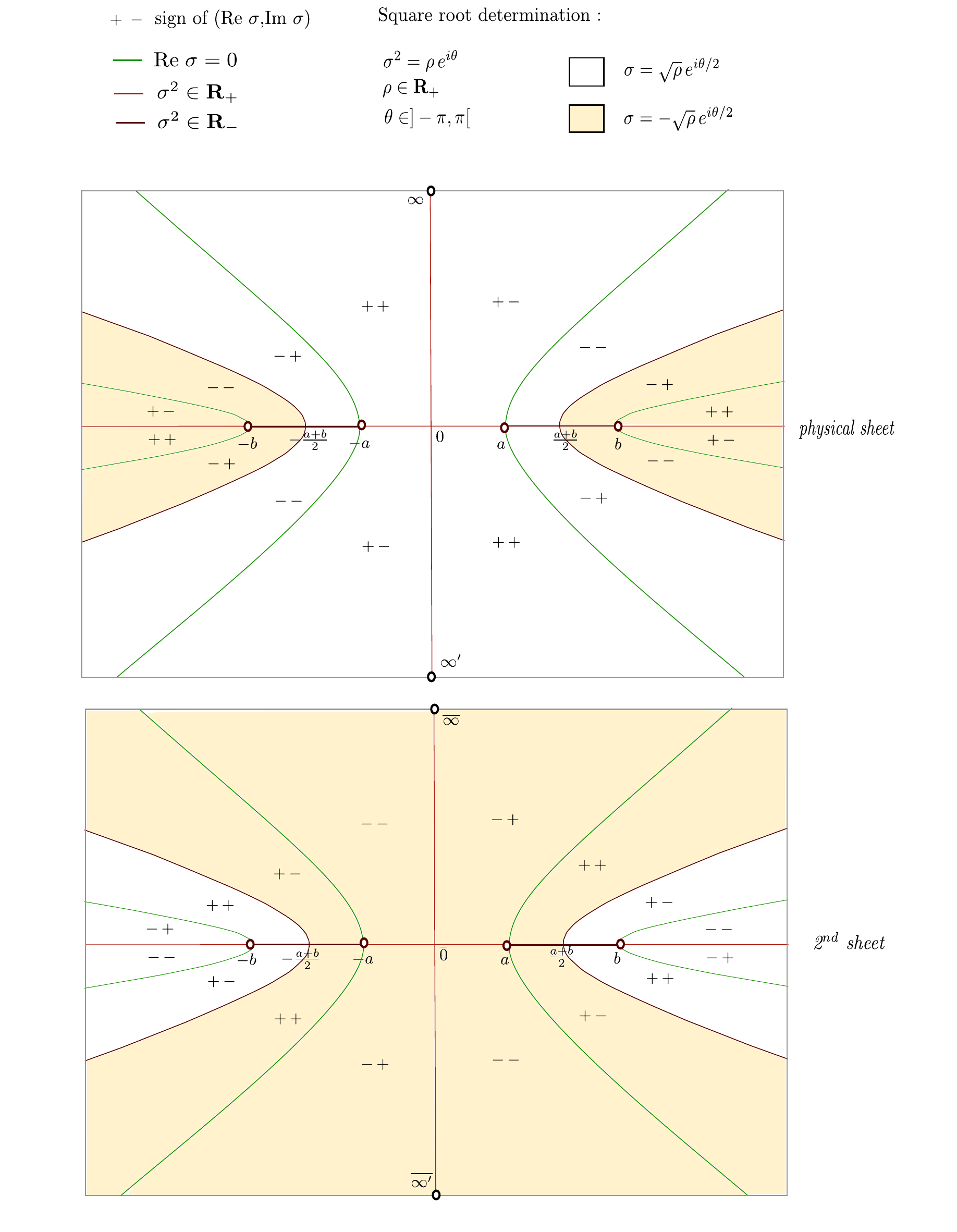}
\caption{\label{fig:2sheets}
$\mathcal{S}: \sigma^2 = (\x^2 - a^2)(\x^2 - b^2)$ has two sheets, corresponding to opposite square roots, and we call one of them the \emph{physical sheet}. The determination (white or yellow) of the square root changes when one crosses the lines where $\sigma^2 \in \mathbb{R}_-$, so that $\sigma$ is an analytical function of x on each sheet, outside $[-b,-a]\cup[a,b]$. We have indicated the corresponding signs for $\mathrm{Re}(\sigma)$ and $\mathrm{Im}(\sigma)$ on the two sheets.}
\end{figure}

\newpage

\begin{figure}[h!]
\includegraphics[width=\textwidth]{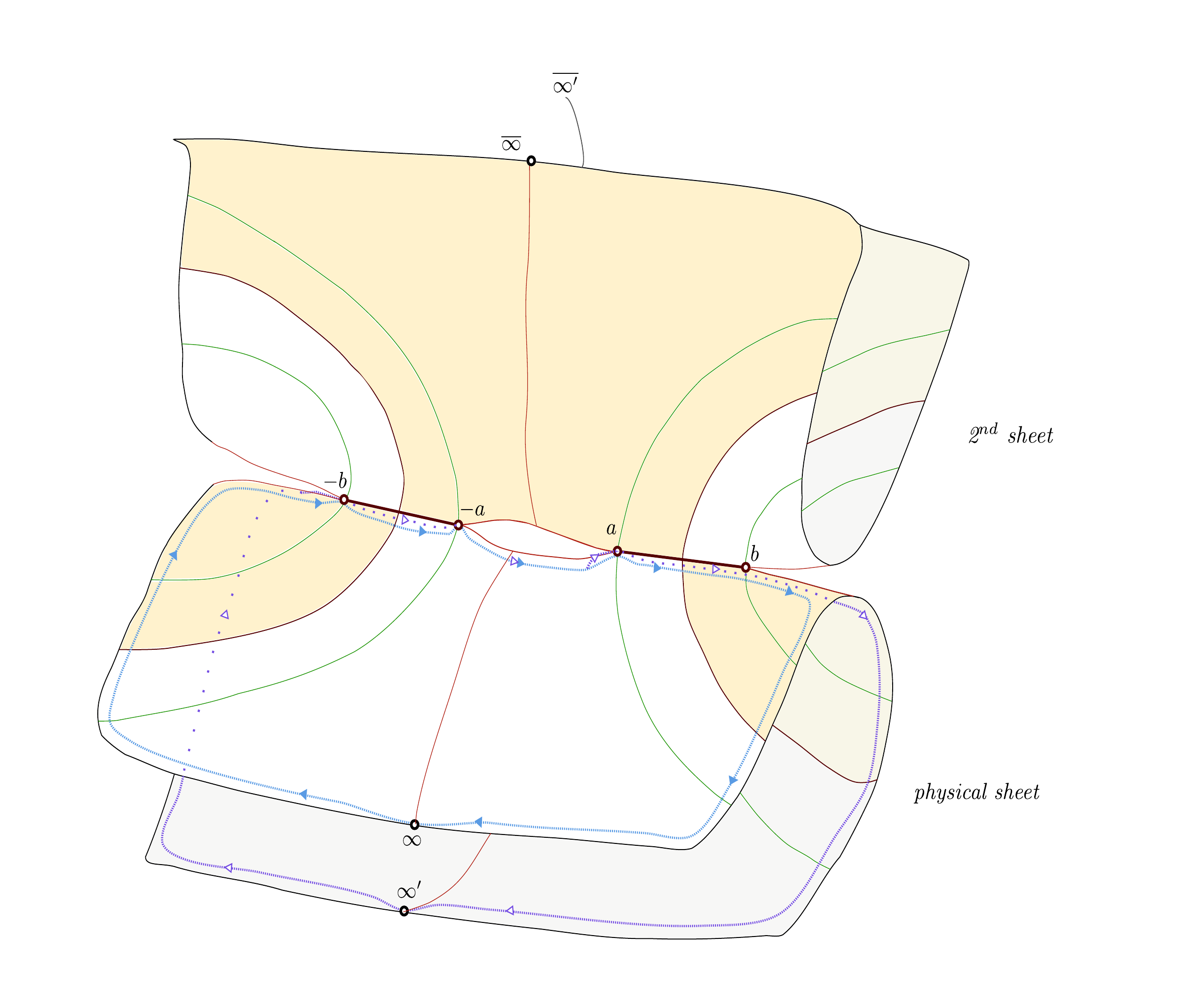}
\caption{\label{fig:sigma2}}
\end{figure}
$\mathcal{S}$ is topologically a torus. So, the space of holomorphic differential form is of complex dimension $1$ and is generated by $\mathrm{d}x/\sigma$. To parametrize the surface, we define
\beq
\mathrm{u(x)} = \frac{ib}{2K'}\int_{-a}^{\x} \frac{\mathrm{dx}'}{\sigma(\x')} \nonumber
\eeq
which is path dependant. We fix $K' \in \mathbb{R}_+^*$ such that $\mathrm{u}(-b) = \frac{1}{2}$ along the blue path. Then, $\mathrm{u}(a) = \tau$, shown in the next paragraph to be half of the modulus of the torus, lies in $i\mathbb{R}_+$ for $a,b \in \mathbb{R}_+$, $a < b$. We have drawn the paths followed on the first sheet only: they follow the real line, the blue one on the side $\mathrm{Im}\, \x > 0$, the purple one on the side $\mathrm{Im}\, \x < 0$. Because the square root on the second sheet is opposite to its determination on the physical sheet, the analogue integration paths on the second sheet lead to opposite $u$ for the same value of x.

\newpage

\begin{figure}[h!]
\includegraphics[width=\textwidth]{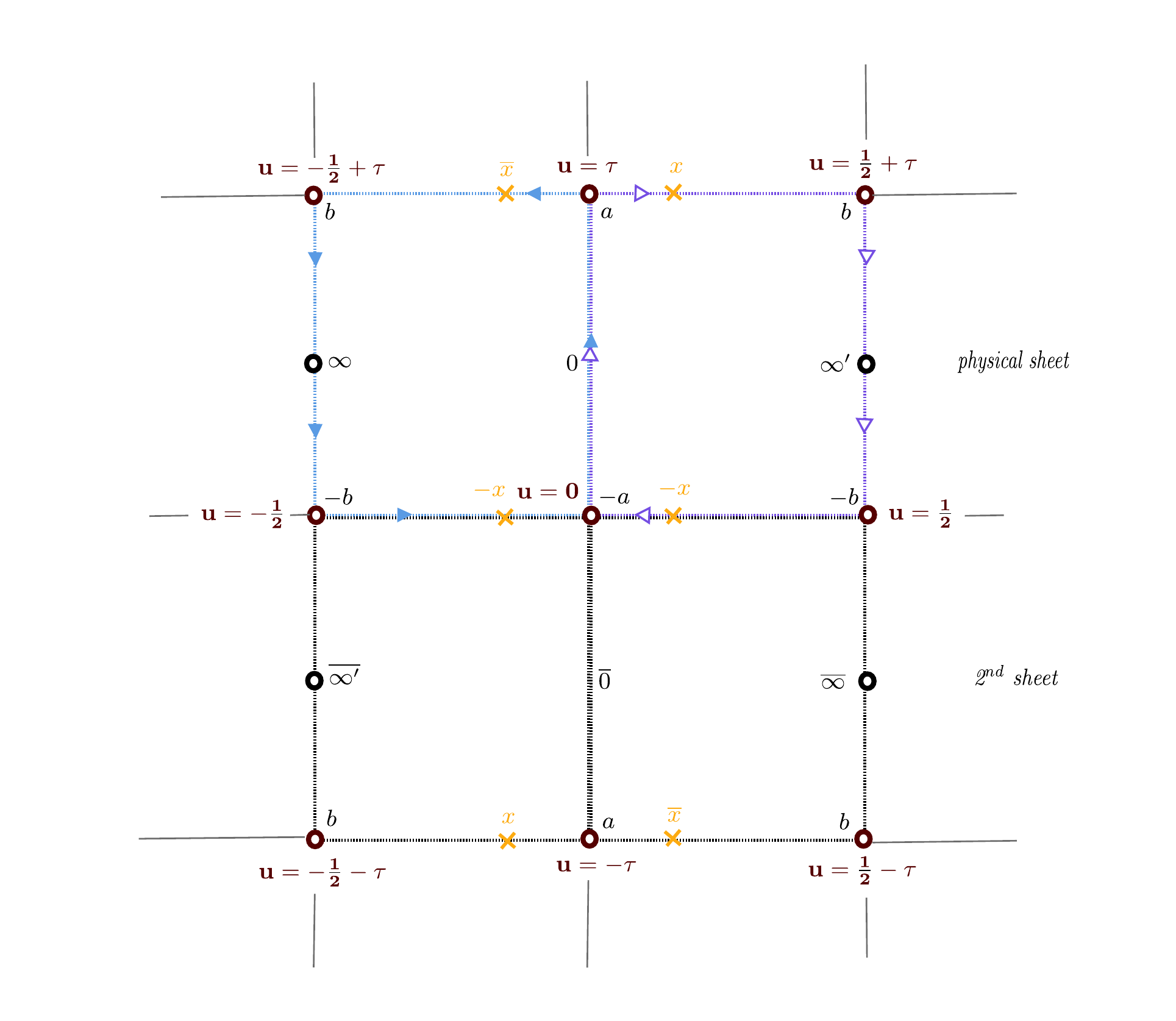}
\caption{\label{fig:torus}}
\end{figure}
We present the $u$-plane at the end of the construction. Circling only once along the paths in the two sheets gives the rectangular region $[-\frac{1}{2},\frac{1}{2}]\times[-\tau,\tau]$. The path corresponding to $u \in [-\frac{1}{2},\frac{1}{2}]$ or $u \in [-\tau,\tau]$ are non contractible loops in $\mathcal{S}$. Following these paths adds 1 (or $2\tau$) to $u$ and leads to the same starting point. Then, it is Abel's theorem that $u$ induces an isomorphism between $\mathcal{S}$ and the Riemann surface $\mathbb{C}/(\mathbb{Z} \oplus 2\tau\mathbb{Z})$. Moreover, we have marked the points corresponding to $\x + i\epsilon$, $\x - i\epsilon$ ($\epsilon > 0$), and $-\x$ for some $\x \in [a,b]$. We see that, in the variable $u$, they are merely $\tau$-translations of each other:
\vspace{0.2cm}
\begin{center}
\begin{tabular}{rcl}
\textbf{generic point} & & \textbf{images in the torus} \\
$\x$ & & $u$, $ -u$ \\
$-\x$ & & $\tau - u$, $-(\tau - u)$ \\
$\overline{\x}$ & & $2\tau - u$,$u - 2\tau$ \end{tabular}
\end{center}

\newpage

In a nutshell, our parametrization in terms of elliptic functions \cite{GW}
reads:\beq \x = a\,\mathrm{sn}_k(\varphi),\qquad \varphi = \int_0^{\x/a}
\frac{\mathrm{dx}}{\sqrt{(1 - \x'^2)(1 - k^2\x'^2)}},\qquad u =
\frac{i\varphi}{2K'} + \frac{\tau}{2} \nonumber \eeq where sn$_k$ is the
odd solution of $y' = \sqrt{(1 - y^2)(1 - k^2y^2)}$, and:
\bea
&& \textrm{cn}_k = \sqrt{1 - \textrm{sn}_k^2},\qquad \textrm{dn}_k = \sqrt{1 - k^2\textrm{sn}_k^2},\qquad k = a/b,\qquad \tau = \frac{iK}{K'} \nonumber\\
&& K = K(k) = \int_0^1 \frac{\mathrm{dx}'}{\sqrt{(1 - \x'^2)(1 - k^2\x'^2)}}\qquad\textrm{is the complete elliptic integral} \nonumber \\
&& K' = K(\sqrt{1 - k^2}) = \int_0^{\infty}\frac{\mathrm{dx}'}{\sqrt{(1 + \x'^2)(1 + k^2\x'^2)}} \nonumber
\eea
With help of the properties \cite{GW} under translation and rotation to imaginary argument of the elliptic functions $\textrm{sn},\textrm{cn},\textrm{dn}$, we obtain:
\beq
\x = a\,\mathrm{dn}_{\sqrt{1 - a^2/b^2}}\left(\frac{2u}{K'}\right) \nonumber
\eeq
We mention it to be explicit, though it is not at all necessary to know this formula in what follows.

\subsection{Change of variable}
\label{sec:dessin3}

This construction provides a nice change of variable to solve Eqn.~\ref{eq:00}. Let us note the inverse function of $\x \mapsto \uu(x)$:
\beq
\begin{array}{rccl}x\,: & [-\frac{1}{2},\frac{1}{2}]\times[0,\tau] & \rightarrow & \mathbb{C}, \textrm{physical sheet} \\
 & u & \mapsto & x(u)\end{array} \nonumber
\eeq
Then, any function:
\beq
\begin{array}{rccl}
\mathrm{W}\,: & \mathbb{C} & \rightarrow & \mathbb{C} \\ & \x & \mapsto & \mathrm{W}(\x)\end{array}\nonumber
\eeq
which is at least analytic in the $\x$-plane with two cuts $[-b,-a]$ and $[a,b]$, defines without ambiguity an analytic function:
\beq
\begin{array}{rccl}
W\,: & [-\frac{1}{2},\frac{1}{2}]\times[0,\tau] & \rightarrow & \mathbb{C} \\
& u & \mapsto & \mathrm{W}(x(u)) \end{array}\nonumber
\eeq
In some cases, $W$ may be extended to the whole $u$-plane because values on the boundary of the initial region match. Eventually, properties of W(x) are translated into properties of $W(u)$ and vice versa.

\begin{figure}[h!]
\begin{center}
\includegraphics[width=0.8\textwidth]{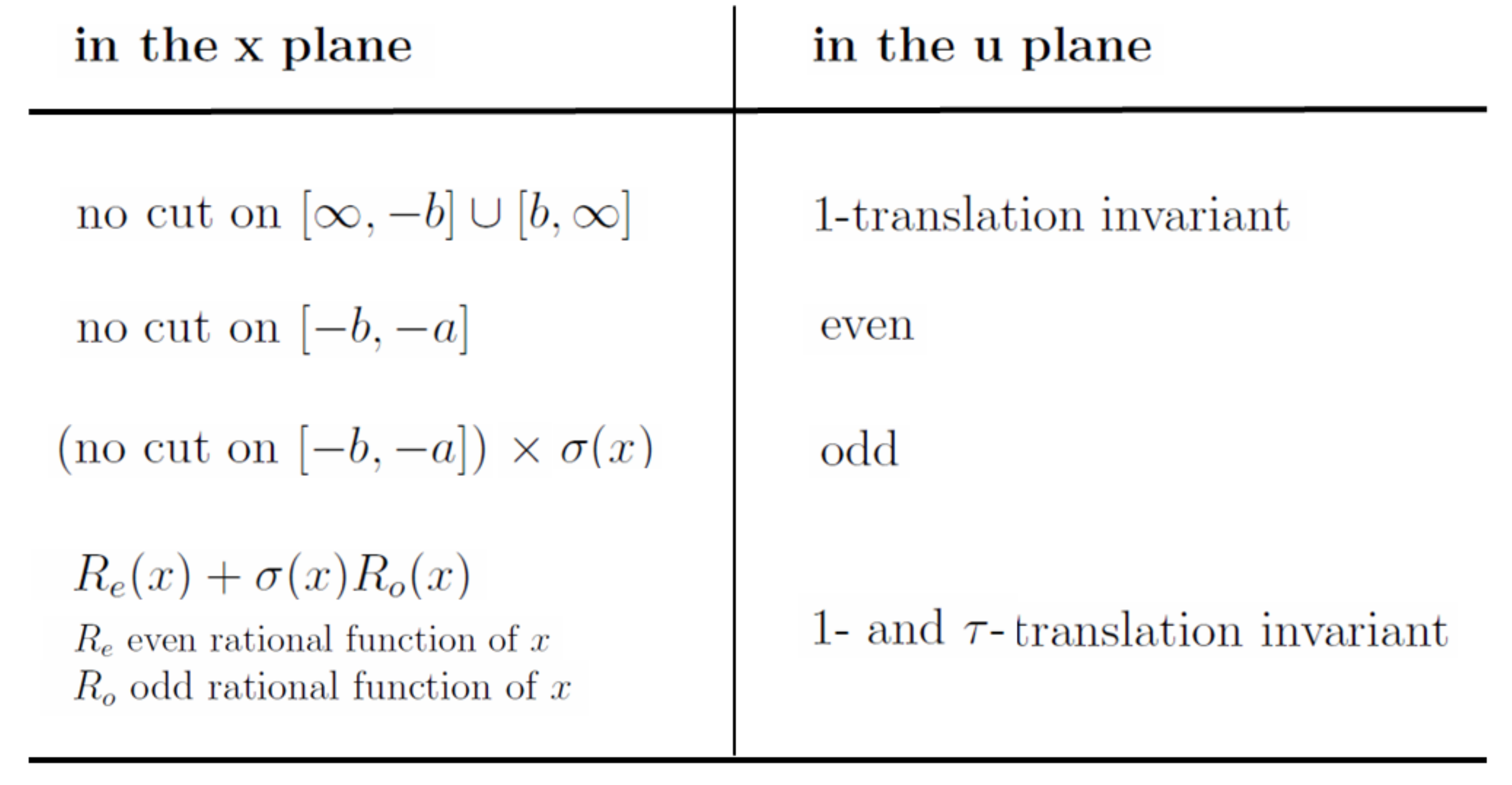}
\caption{\label{fig:dict}Dictionary between invariances in variable $u$ and analyticity properties in variable $x$.}
\end{center}
\end{figure}

\medskip

Now, we apply this to any 1-cut solution W of Eqn.~\ref{eq:00}. We end up with a meromorphic function $W$ defined on the whole $u$-plane, which is $1$-translation invariant, $u$-even, and satisfies:
\beq
\label{eq:02}
\forall u \in \tau + \left[-\frac{1}{2},\frac{1}{2}\right] \quad\quad W(2\tau - u) + \nn{}W(\tau - u) + W(u) = 0
\eeq
Since $W$ is analytic, it must be true on the whole $u$-plane. Using the parity property, and defining the operators of $1$-translation $\mathbf{T}_1$, and of $\tau$-translation \textbf{T}, we rewrite Eqn.~\ref{eq:02} as:
\beq
\label{eq:01}\forall u \in \mathbb{C} \quad\quad \left\{\begin{array}{l}\left(\mathbf{T}^2 + \nn\mathbf{T} + \mathrm{id}\right)(W)(u) = 0 \\ \left(\mathbf{T}_1 - \mathrm{id}\right)(W)(u) = 0 \end{array}\right.
\eeq

\subsection{Special 1-cut solutions}
\label{sec:ssol}
$\mathbf{T}$ is a linear operator on the space of meromorphic functions in the $u$-plane, over the field $\mathbf{k}$ of $1$ and $\tau$-translation invariant functions (the general form in the $\x$ variable of these biperiodic functions is described Fig.~\ref{fig:dict}). With the notation $\nn = -2\cos(\pi\mu)$, the space of solutions of Eqn.~\ref{eq:01} is the intersection of $\mathrm{Ker}\left(\mathbf{T}_1 - \mathrm{id}\right)$ and:
\beq
\mathrm{Ker}\left(\mathbf{T}^2 + \nn\mathbf{T} + \mathrm{id}\right) = \mathrm{Ker}\left(\mathbf{T} - e^{i\pi\mu}\mathrm{id}\right) \oplus \mathrm{Ker}\left(\mathbf{T} - e^{-i\pi\mu}\mathrm{id}\right)\nonumber
\eeq
Thus, it has dimension two. Let us pick up \cite{E1,EK1} in $\mathrm{Ker}(\mathbf{T}_1 - \mathrm{id})\cap\mathrm{Ker}(\mathbf{T} - e^{i\pi\mu}\mathrm{id})$ a special function $D_{\mu}$ (i.e satisfying $D_{\mu}(u  + 1) = D_{\mu}(u)$ and $D_{\mu}(u + \tau) = e^{-i\pi\mu}D_{\mu}(u)$), having the asymptotic properties:
\begin{itemize}
\item[$\bullet$] $D_{\mu}(\mathrm{u(x)}) \sim 1/\x$ when $\x \rightarrow \infty$ in the physical sheet.
\item[$\bullet$] $D_{\mu}(\mathrm{u(x)}) \propto 1/\sigma(\x)$ when $\x \rightarrow a,b,-a,-b$.
\end{itemize}
This determines $D_{\mu}$ uniquely, and we can actually construct it in terms of theta functions of modulus $\tau$ (see Appendix~\ref{app:ell} for a reminder on theta functions):
\beq
\label{eq:Dmudef} D_{\mu}(u) = \frac{x(u)}{\sigma(x(u))}\,\frac{\vartheta_1\left(u - \frac{\tau}{2} + \frac{\mu}{2}\Big|\tau\right)}{\vartheta_1\left(u - \frac{\tau}{2}\Big|\tau\right)}\,\frac{\vartheta_1\left(-\frac{1}{2}\Big|\tau\right)}{\vartheta_1\left(-\frac{1}{2} + \frac{\mu}{2}\Big|\tau\right)}
\eeq
By definition, it has two simple poles (mod $\mathbb{Z}\oplus\tau\mathbb{Z}$) at $0$ and $\frac{1}{2}$, and a simple zero at $\mathrm{u}(\infty) = -\frac{1}{2} + \frac{\tau}{2}$. It also admits a second simple zero at $\mathrm{u}(e_{\mu}) = \frac{\tau - \mu}{2}$. This special function is our elementary brick to define a suitable basis of 1-cut solutions of Eqn.~\ref{eq:01}. Let us notice before that $u\mapsto D_{\mu}(-u)$ generates \mbox{$\mathrm{Ker}(\mathbf{T}_1 - \mathrm{id})\cap\mathrm{Ker}(\mathbf{T} - e^{-i\pi\mu}\mathrm{id})$}.

To describe the space of 1-cut solutions of Eqn.~\ref{eq:01}, we rather work on the subfield $\mathbf{k}'$ of $\mathbf{k}$, consisting of $u$-even functions belonging to $\mathbf{k}$ (in the $\x$ variable, $\mathbf{k}'$ consists of rational functions of $\x^2$). Then, this space of 1-cut solutions has again dimension two on $\mathbf{k}'$. We now describe a basis which we have found convenient for our purposes:
\begin{theorem}
There exists a unique couple of meromorphic functions ($f_{\mu}$,$\widehat{f}_{\mu}$), 1-cut solutions of Eqn.~\ref{eq:01}, such that:
\begin{itemize}
\item[$\bullet$] $f_{\mu}(\mathrm{u(x)})$ is holomorphic for $\x \in \mathbb{C}\setminus[a,b]$ in the physical sheet (it has only one cut).
\item[$\bullet$] $f_{\mu}(\mathrm{u(x)}) \propto 1/\sigma(\x)$ when $u \rightarrow \uu(a) = \tau$, or $\uu(b) = \frac{1}{2} + \tau$.
\item[$\bullet$] $f_{\mu}(\mathrm{u(x)}) \sim 1/\x$ when $\x \rightarrow \infty$ in the physical sheet.
\item[$\bullet$] $\widehat{f}_{\mu}(\mathrm{u(x)})$ is holomorphic for $\x \in \mathbb{C}\setminus([a,b]\cup\{0\})$ in the physical sheet (it has only one cut).
\item[$\bullet$] $\widehat{f}_{\mu}(u)$ is finite when $u \rightarrow \uu(a)$ or $\uu(b)$.
\item[$\bullet$] $\widehat{f}_{\mu}(\mathrm{u(x)}) \propto 1/\x$ when $\x \rightarrow 0$ (in the two sheets).
\item[$\bullet$] $\widehat{f}_{\mu}(\mathrm{u(x)}) \sim 1$ when $\x \rightarrow \infty$ in the physical sheet.
\end{itemize}
\end{theorem}

\subsection{Bilinear form}

The properties of these functions are listed in Appendix~\ref{app:Afmu}. They were constructed such that $f_{\mu}\,\bot\,\widehat{f}_{\mu} = 0$ for the bilinear form (on the field $\textbf{k}$) which appear in the loop equations (Thm.~\ref{MLPgk}):
\beq
(\mathrm{g}\bot \mathrm{h})(\x) \equiv \mathrm{g}(\x)\mathrm{h}(\x) + \mathrm{g}(-\x)\mathrm{h}(-\x) + \frac{\nn}{2}\left(\mathrm{g}(\x)\mathrm{h}(-\x) + \mathrm{g}(-\x)\mathrm{h}(\x)\right) \nonumber
\eeq
Equivalently in the $u$ variable:
\beq
\label{eq:bot}(g\bot h)(u) \equiv g(u)h(u) + g(u-\tau)h(u - \tau) + \frac{\nn}{2}\left(g(u)h(u - \tau) + g(u - \tau)h(u)\right)
\eeq

\begin{lemma}
\label{lemma:fond}$\phantom{blabla}$
\begin{itemize}
\item[$\bullet$] If $f$ and $g$ are solutions of the homogeneous linear equation \ref{eq:00}, then $(f\bot g)(u)$ is $1$- and $2\tau$-translation invariant.
\item[$\bullet$] Besides, if f and g have only one cut, then $\mathrm{f}\bot \mathrm{g}$ is an even rational function of $\x$.
\item[$\bullet$] If $f$ is a 1-cut solution of \ref{eq:00}, $g$ has one cut, and $f\bot g$ has no cut on $[a,b]$, then $g$ is a 1-cut solution of \ref{eq:00} as well.
\end{itemize}
\end{lemma}

\proof{
For the first point, we rather work with the $u$ variable than with $\x$, and use the dictionary of Section~\ref{fig:dict}: recall for example that "having one cut" translates into "being $u$-even".  If $\mathrm{f}$ and $\mathrm{g}$ are solution of the loop equation:
\bea
\mathbf{T}(f\bot g) & = & \mathbf{T}f\cdot\mathbf{T}g + \mathbf{T}^2f\cdot\mathbf{T}^2g + \frac{\nn}{2}\left(\mathbf{T^2}f\cdot\mathbf{T}g + \mathbf{T}f\cdot\mathbf{T}^2g\right)\nonumber \\
& = & (f\bot g)\nonumber\eea
If on the top of that $f$ and $g$ have only one cut, so does $f \bot g$ for it is obviously invariant under $u \mapsto \tau - u$. Hence the second point. Now, assume f is a  1-cut solution of Eqn.~\ref{eq:00} and $\mathrm{f}\bot\mathrm{g}$ has no cut on $[a,b]$ (for example, $\mathrm{f}\bot\mathrm{g}$ is an even rational function of $\x$). Then, for $\x \in [a,b]$ and $\epsilon \rightarrow 0^+$:
\bea
0 & = & (\mathrm{f}\bot \mathrm{g})(\x + i\epsilon) - (\mathrm{f}\bot \mathrm{g})(\x - i\epsilon) \nonumber\\
& = & \left(f(\x + i\epsilon) + \frac{\nn}{2}\mathrm{f}(-\x)\right)\mathrm{g}(\x + i\epsilon) - \left(\mathrm{f}(\x - i\epsilon) + \frac{\nn}{2}\mathrm{f}(-\x)\right)\mathrm{g}(\x - i\epsilon) \nonumber\\
& & + \frac{\nn}{2}\left(\mathrm{f}(\x + i\epsilon) - \mathrm{f}(\x - i\epsilon)\right)\mathrm{g}(-\x)\nonumber \\
& = & \left(\mathrm{f}(\x + i\epsilon) - \mathrm{f}(\x - i\epsilon)\right)\,\left[\mathrm{g}(\x + i\epsilon) + \mathrm{g}(\x - i\epsilon) + \frac{\nn}{2}\mathrm{g}(-\x)\right]\nonumber
\eea
Since f has a discontinuity on $\left.]a,b[\right.$, g must satisfy the homogeneous linear equation.}

Thus, the norms $\mathrm{R}_{\mu}(\x^2) = (\f_{\mu}\bot \f_{\mu})(\x)$ and $\widehat{\mathrm{R}}_{\mu}(\x^2) = (\widehat{\f}_{\mu}\bot\widehat{\f}_{\mu})(\x)$ are rational functions of $\x^2$. They can be identified \cite{E1,EK1} using the analytical properties of $(\mathrm{f}_{\mu},\widehat{\mathrm{f}}_{\mu})$ (see Appendix~\ref{app:Afmu}).

\subsection{General 1-cut solutions}
\label{sec:gensol}
We arrive at the description found in \cite{E1,EK1}.
\begin{theorem}
\label{sol}The 1-cut solutions of Eqn.~\ref{eq:00} are of the form
\beq
\mathrm{W}(\x) = \mathrm{A}(\x^2)\,\frac{\f_{\mu}(\x)}{\mathrm{R}_{\mu}(\x^2)} + \mathrm{B}(\x^2)\,\frac{\widehat{\f}_{\mu}(\x)}{\widehat{\mathrm{R}}_{\mu}(\x^2)} \nonumber
\eeq
where $\mathrm{A}(\x^2)$ and $\mathrm{B}(\x^2)$ are rational functions of $\x^2$.
\end{theorem}
As rational functions of $\x^2$, $\mathrm{A}$ and $\mathrm{B}$ can be determined from the required analytic properties (behavior at poles and zeroes) of the solution of Eqn.~\ref{eq:00} we are looking for.

\section{Unstable generating functions ($2 - 2g - k \geq 0$)}
\label{sec:W10W20}

We review the results of \cite{E1,EK1,EK2}, concerning $\mathrm{W}_1^{(0)}$ and $\mathrm{W}_2^{(0)}$. They can be proved in a straightforward way from Thm.~\ref{sol}. Though the solution for $\mathrm{W}_2^{(0)}$ requires lengthy computations. We have developed an other approach to compute $\mathrm{W}_2^{(0)}$ (Section~\ref{sec:series}) which is closer to the spirit of algebraic geometry, that is, finding a good description of the ring of meromorphic functions on the spectral curve.

\subsection{$\mathrm{W}_1^{(0)}$ (rooted disks)}
\label{sec:W10}
 $\mathrm{W}_1^{(0)}(\x)$ is a 1-cut solution of the complete linear equation with \textsc{rhs} $\mathrm{V}'(\x)$. It is easy to find a particular solution of this equation:
 \beq
 \mathrm{V}_s'(\x) = \frac{2\mathrm{V}'(\x) - \mathfrak{n}\mathrm{V}'(-\x)}{4 - \mathfrak{n}^2} \nonumber
 \eeq
 So, $\overline{\mathrm{W}}_{1}^{(0)}(\x) = \mathrm{W}_1^{(0)}(\x) - \mathrm{V}_s'(\x)$ is a $1$-cut solution of Eqn~\ref{eq:00}. It must satisfy:
 \begin{itemize}
\item[$\bullet$] $\mathrm{W}_1^{(0)}$ has no pole in the physical sheet (i.e  in $\mathbb{C}\setminus[a,b]$).
 \item[$\bullet$] $\mathrm{W}_1^{(0)}(\x) \sim t/\x$ when $\x \rightarrow \infty$ in the
 physical sheet.
 \item[$\bullet$] $\mathrm{W}_1^{(0)}(\x)$ is finite when $\x = a_i$ ($a$ or $b$), and $\mathrm{W}_1^{(0)}(\x) - \mathrm{W}_1^{(0)}(a_i) \propto \sqrt{(\x - a_i)}$ when $\x \rightarrow a_i$.
 \end{itemize}

\begin{theorem}\cite{E1,EK1}
\bea
\label{eq:W10}\mathrm{W}_1^{(0)}(\x) & = & \frac{2\mathrm{V}'(\x) - \mathfrak{n}\mathrm{V}(-\x)}{4 - \mathfrak{n}^2} + \mathrm{A}(\x^2)\frac{\f_{\mu}(\x)}{\mathrm{R}_{\mu}(\x^2)} + \mathrm{B}(\x^2)\frac{\widehat{\f}_{\mu}(\x)}{\widehat{\mathrm{R}}_{\mu}(\x^2)}
\eea
where $\mathrm{A}(\x^2)$ and $\mathrm{B}(\x^2)$ are even polynomials:
\bea
\mathrm{A}(\x^2) & = & - \frac{1}{2}\big(\mathrm{V}'(\x)\f_{\mu}(\x) + \mathrm{V}'(-\x)\f_{\mu}(-\x)\big)_+ \quad \nonumber \\
\mathrm{B}(\x^2) & = &  - \frac{1}{2}\big(\mathrm{V}'(\x)\widehat{\f}_{\mu}(\x) + \mathrm{V}'(-\x)\widehat{\f}_{\mu}(-\x)\big) \nonumber
\eea
$a$ and $b$ must be solution to the compatibility equations:
\bea
\Res_{\x \rightarrow \infty} \mathrm{dx}\,\left(\x{}\mathrm{V}'(\x)\f_{\mu}(\x)\right) & = & (- 2 + \nn)t \nonumber \\
\Res_{\x \rightarrow \infty} \mathrm{dx}\,\left(\x{}\mathrm{V}'(\x)\widehat{\f}_{\mu}(\x)\right) & = & (2 + \nn)\Res_{\x \rightarrow \infty}\mathrm{dx}\,\left(\x{}\mathrm{W}_1^{(0)}(\x)\right) \nonumber
\eea
Generically, they admit a unique solution $\{a(t),b(t)\}$ which is a power series in $\sqrt{t}$ for small $t$ (this $\sqrt{t}$ behavior is required by the 1-cut lemma).
\end{theorem}
Though $\mathrm{R}_{\mu}(\x)$ and $\widehat{R}_{\mu}(\x)$ have a zero when $\x^2 = e_{\mu}^2$, $\mathrm{W}_1^{(0)}$ must be regular at $\x = \pm e_{\mu}$, and this is expressed by the second compatibility equation.

The polynomial parts can be rewritten by taking residues at $\infty$. By moving the contour, we can represent the solution as a contour integral around $[a,b]$ for the solution.
\begin{theorem}\cite{E1,EK1}
\label{thW10}\beq
\encadremath{\mathrm{W}_1^{(0)}(\x_0) = \frac{1}{2i\pi}\oint_{[a,b]}
 \mathrm{dx}\,\frac{\x{}\mathrm{V}'(\x)}{\x_0^2 - \x^2}\left(\frac{\f_{\mu}(\x_0)}{\mathrm{R}_{\mu}(\x_0^2)}\f_{\mu}(\x) + \frac{\widehat{\f}_{\mu}(\x_0)}{\widehat{\mathrm{R}}_\mu(\x_0^2)}\widehat{\f}_{\mu}(\x)\right)} \nonumber
\eeq
\end{theorem}
This formula is the analog for $\mathfrak{n} \neq 0$ of Tricomi's solution \cite{Tri}.

\subsection{y(x) (the spectral curve)}
\label{sec:scurve}
The object which plays the role of a spectral curve is the discontinuity of the resolvent $\mathrm{W}_1^{(0)}(\x)$.
We have:
\beq
\mathrm{y}(\x) \equiv \lim_{\epsilon \rightarrow 0} \left(\mathrm{W}_1^{(0)}(\x+i\epsilon)-\mathrm{W}_1^{(0)}(x-i\epsilon)\right) = 2\mathrm{W}_1^{(0)}(\x) +\nn \mathrm{W}_1^{(0)}(-\x) - \mathrm{V}'(\x) \nonumber
\eeq
A short computation with help of Appendix~\ref{app:Afmu} gives:
\beq
\mathrm{y}(\x) = \mathrm{A}(\x^2)\mathrm{y}_{\mu}(\x) + \mathrm{B}(\x^2)\hat{\mathrm{y}}_{\mu}(\x) \nonumber
\eeq
where the two basic blocks are given by:
\beq
\mathrm{y}_{\mu}(\x) = \frac{\x\sigma(\x)}{\x^2 - e_{\mu}^2}\hat{\f}_{\mu}(-\x),\qquad \hat{\mathrm{y}}_{\mu}(\x) = - \frac{\x\sigma(\x)}{\x^2 - e_{\mu}^2}\f_{\mu}(-\x) \nonumber
\eeq

\subsection{$\mathrm{W}_2^{(0)}$ (rooted cylinders), first method}
\label{sec:W20}
$\mathrm{W}_2^{(0)}(\x_0,\x)$ is obtained by application of the loop insertion operator $\partial/\partial \mathrm{V}(\x)$ to $\mathrm{W}_1^{(0)}(\x_0)$. Accordingly, it is a 1-cut solution (in both $\x_0$ and $\x$) of the complete linear equation with \textsc{rhs} $-1/(\x_0 - \x)^2$. Again, we can decompose into a particular solution, and a solution $\overline{\mathrm{W}}_2^{(0)}$ of Eqn.~\ref{eq:00}:
\beq
\mathrm{W}_2^{(0)}(\x_0,\x) = \frac{-2}{(\x + \x_0)^2} + \frac{\mathfrak{n}}{(\x + \x_0)^2} + \overline{\mathrm{W}}_2^{(0)}(\x_0,\x) \nonumber
\eeq
We must find the solution characterized by:
\begin{itemize}
\item[$(i)$] $\mathrm{W}_2^{(0)}(\x_0,\x)$ is symmetric in $\x_0$ and $\x$.
\item[$(ii)$] $\mathrm{W}_2^{(0)}(\x_0,\x)$ has no pole when $\x_0$ and $\x$ are in the physical sheet.
\item[$(iii)$] $\mathrm{W}_2^{(0)}(\x_0,\x) \propto 1/\x_0^2$ when $\x_0 \rightarrow \infty$ in the physical sheet ($\x$ fixed).
\item[$(iv)$] $\mathrm{W}_2^{(0)}(\x_0,\x) \propto 1/\sqrt{(\x_0 - a)(\x_0 - b)}$ when $\x_0 \rightarrow a,b$ ($\x$ fixed).
\end{itemize}
Indeed: $(i)$ expresses the invariance of the generating function of cylinder maps by exchange of the marked faces ; $(ii)$ is implied by the 1-cut lemma ; $(iii)$ comes from the definition of $\mathrm{W}_2^{(0)}$ as a connected correlation function ; $(iv)$ follows from application of $\partial/\partial \mathrm{V}(\x)$ to $\mathrm{W}_1^{(0)}(\x_0)$, given that $a$ and $b$ are $V$ dependent. Actually, it is easier to determine first the primitive of $\overline{\mathrm{W}}_2^{(0)}$
\beq
\label{eqH} \mathrm{H}(\x_0,\x) = \int_{\infty}^{\x} \mathrm{d}\xi\,\overline{W}_2^{(0)}(\x_0,\xi)
\eeq
which satisfy assertions deduced from $(i)-(iv)$. Although it is convenient to use the 1-cut basis $(\f_{\mu}(\x),\widehat{\f}_{\mu}(\x))$ to derive the result, it can be expressed afterwards in another basis. Let us introduce:
\beq
\mathrm{I(x)} = \frac{\x\sigma(\x) + e_{\mu}\sigma(e_{\mu})}{\x^2 - e_{\mu}^2}
\eeq
which is related to the derivative of $\mathrm{D}_{\mu}$ (see Appendix~\ref{app:Afmu}).
\begin{theorem}\cite{E1,EK1} Primitive of $\mathrm{W}_2^{(0)}$.
\label{H}
\bea
\mathrm{H}(\x_0,\x) & = & \frac{1}{4 - \nn^2}\left(e^{-i\pi\mu}\mathrm{H}_{+}(\x_0,\x) + \mathrm{H}_{+}(-\x_0,\x) - \mathrm{H}_{+}(\x_0,-\x) - e^{i\pi\mu}\mathrm{H}_{+}(-\x_0,-\x)\right) \nonumber\\
\mathrm{H}_+(\x_0,\x) & \equiv & \mathrm{D}_{\mu}(\x_0)\,\sigma(\x)\mathrm{D}_{\mu}(\x)\,\frac{\mathrm{I}(\x) - \mathrm{I}(\x_0)}{\x^2 - \x_0^2} \nonumber
\eea
\end{theorem}

\begin{corollary}
\label{coco}\bea
\mathrm{W}_2^{(0)}(\x_0,\x) & = & \frac{1}{4 - \nn^2}\left(e^{-i\pi\mu}\mathrm{W}_{2|+}^{(0)}(\x_0,\x) + \mathrm{W}_{2|+}^{(0)}(-\x_0,\x) \right. \nonumber \\
& & \phantom{\frac{1}{4 - \nn^2}}\; \left. + \mathrm{W}_{2|+}^{(0)}(\x_0,-\x) + e^{i\pi\mu}\mathrm{W}_{2|+}^{(0)}(-\x_0,-\x)\right) \nonumber\\
\mathrm{W}_{2|+}^{(0)}(\x_0,\x) & \equiv & \mathrm{D}_{\mu}(\x_0)\,\mathrm{D}_{\mu}(\x)\left[1 - \left(\alpha_1 + \frac{\x\sigma(x) - \x_0\sigma(\x_0)}{\x^2 - \x_0^2}\right)\frac{\mathrm{I}(\x) - \mathrm{I}(\x_0)}{\x^2 - \x_0^2}\right] - \frac{1}{(\x + \x_0)^2} \nonumber
\eea
We also have:
\bea
\mathrm{W}_2^{(0)}(\x,\x) & = & \frac{1}{4 - \nn^2}\left(\frac{\nn}{4\x^2} - \nn(\partial_2 \mathrm{H})(\x,\x) + 2\mathrm{W}_{2|+}^{(0)}(\x,-\x)\right)\nonumber \\
\mathrm{W}_{2|+}^{(0)}(\x,-\x) & = & \frac{1}{2}\frac{e_{\mu}^2 - \alpha_1^2}{\sigma^2(\x)} + \frac{\x^2}{4}\frac{(b^2 -
a^2)^2}{\sigma^4(\x)} \nonumber
\eea
\end{corollary}

\subsection{$\mathrm{W}_2^{(0)}$, infinite series representation}
\label{sec:series}

\subsubsection{Decomposition and characterization}
We present here an alternative construction of $\mathrm{W}_2^{(0)}$, which takes advantage of the change of variable $\x \rightarrow \mathrm{u}(\x)$. The natural object to look at is the differential form $\mathrm{W}_2^{(0)}(\x_0,\x)\mathrm{d}\x_0\mathrm{d}\x$. Let us define $s(u) = \frac{\mathrm{d}x}{\mathrm{d}u} = \frac{ib}{2K'}\sigma(x(u))$. Its properties are:
\beq
s(u) = s(u + 1),\qquad s(-u) = -s(u),\qquad s(\tau - u) = s(u) \nonumber
\eeq
We want to determine:
\beq
\overline{\omega}_2^{(0)}(u,u_0) = s(u)s(u_0)\,\overline{W}_2^{(0)}(u,u_0) \nonumber
\eeq
We shall consider the function $\mathcal{B}$ defined by:
\beq
\mathcal{B}(u_0,u) \stackrel{\textrm{def}}{=} -e^{-i\pi\mu}\overline{\omega}_2^{(0)}(u,u_0) + \overline{\omega}_2^{(0)}(\tau - u,u_0) + \overline{\omega}_2^{(0)}(u,\tau - u_0) - e^{i\pi\mu}\overline{\omega}_2^{(0)}(\tau - u,\tau - u_0) \nonumber
\eeq
Conversely, $\overline{\omega}_{2}^{(0)}$ can be decomposed in terms of $\mathcal{B}$:
\begin{small}\bea
\overline{\omega}_2(u,u_0) & = & \frac{1}{4 - \mathfrak{n}^2}\big(e^{-i\pi\mu}\mathcal{B}(u,u_0) + \mathcal{B}(\tau - u,u_0) + \mathcal{B}(u,\tau - u_0) + e^{i\pi\mu}\mathcal{B}(\tau - u,\tau - u_0)\big) \nonumber \\
\label{eq:decom2}&& \eea
\end{small}
$\mathcal{B}$ has the following properties:
\begin{itemize}
\item[$\bullet$] $\mathcal{B}$ is a meromorphic function of $u, u_0 \in \mathbb{C}$.
\item[$\bullet$] $\mathcal{B}$ has for only pole $u = (\tau - u_0)\;\textrm{mod}\;\mathbb{Z}\oplus\tau\mathbb{Z}$, and:
\beq
\mathcal{B}(u) = \frac{1}{(u + u_0 - \tau)^2}\,+\,O(1)\quad\quad\,\textrm{when}\;u \rightarrow \tau - u_0.\nonumber
\eeq
\item[$\bullet$] $\mathcal{B}(u,u_0) = \mathcal{B}(u_0,u)$.
\item[$\bullet$] $\mathcal{B}(u,u_0) = \mathcal{B}(u + 1,u_0)$ and $\mathcal{B}(u + \tau,u_0) = e^{i\pi(1 - \mu)}\mathcal{B}(u,u_0)$.
\end{itemize}
\subsubsection{Construction of a solution}
We can construct such a function explicitly by a deformation of the Weierstra\ss{} function $\wp$ (see Appendix~\ref{app:Weier} for a reminder of $\wp$). We assume $e^{i\pi\mu} \neq -1$ (i.e $\nn \neq 2$), and define a function $\wp_{\mu}$:
\bea
\label{eq:Bmu} \wp_{\mu}(w) & = & \sum_{m \in \mathbb{Z}} e^{i\pi(\mu - 1)m}\,\frac{\pi^2}{\sin^2\pi(w + m\tau)}
\eea
Notice that the series converges absolutely for $\tau \in i \mathbb{R}_+^*$. $\wp_{\mu}$ is $1$-translation invariant and take a phase $e^{i\pi(1 - \mu)}$ when $w \mapsto w + \tau$. It coincides with $\wp + c_0$ when $\mu = 1\,\textrm{mod}\, 2\mathbb{Z}$ (i.e. $\nn = 2$), where $c_0$ is a constant depending on $\tau$. It has for only pole $w = 0\;\textrm{mod}\,(\mathbb{Z}\,\oplus\,\tau\mathbb{Z})$, which is a double pole without residues. As for $D_{\mu}$, one can represent $\wp_{\mu}$ as a quotient of theta functions, and we find that it has two distinct zeroes :
\beq
w_1 \qquad \textrm{and}\qquad w_2 = - w_1 + \frac{1 - \mu}{2} \nonumber
\eeq
$\wp_{\mu}$ enjoys properties generalizing those of $\wp$ and presented in Appendix~\ref{app:Weier}. Now, $\mathcal{B}(u,u_0) - \wp_{\mu}(u + u_0 - \tau)$ is holomorphic, $1$-translation invariant and takes a phase $e^{i\pi(1 - \mu)}$ when one of the arguments is shifted by $\tau$. Since $\mu \in \mathbb{R}$, this difference is a entire bounded function\footnote{This argument is only valid when $\mu \in \mathbb{R}$, i.e requires at least $|\mathfrak{n}| \leq 2$.}, so is constant, and this constant must vanish for $e^{i\pi\mu} \neq -1$. Therefore, $\mathcal{B}(u,u_0) = \wp_{\mu}(u + u_0 - \tau)$. Collecting the terms of Eqn.~\ref{eq:decom2}:
\begin{theorem}
\label{eq:WWW}
\begin{small}
\beq
\encadremath{\overline{\omega}_2^{(0)}(u,u_0) = \sum_{m \in \mathbb{Z}} \frac{2\pi^2\,\cos(\pi(1 - \mu)m)}{4 - \mathfrak{n}^2}\,\left(\frac{1}{\sin^2\pi(u - u_0 + m\tau)} - \frac{1}{\sin^2\pi(u + u_0 + m\tau)}\right)}  \nonumber
\eeq
\end{small}
\end{theorem}
Compared to Cor.~\ref{coco}, this expression is well suited if one wishes to integrate $\mathrm{W}_2^{(0)}(\x_0,\x)$.

\subsubsection{Comment}

As usual in matrix models, we find that $\overline{\omega}_2^{(0)}(u_0,u)$ is a universal object. Universal means here that the dependence in V of $\mathrm{W}_2^{(0)}$ arise only through $a$ and $b$ (it was already manifest in Cor.~\ref{coco}).

In the Coulomb gas picture (Eqn.~\ref{pol}), it characterizes the correlation of two eigenvalues in the large $N$ limit. If we regard $\tau$ and $\mu \in \mathbb{R}/2\mathbb{Z}$ as independent parameters, Thm.~\ref{eq:WWW} gives the Fourier series in $\mu$ of $\overline{W}_2^{(0)}$. Besides, it depends on $\mathrm{u(x)} - \mathrm{u}(\x_0)$ and $\mathrm{u(x)} - \mathrm{u}(-\x_0)$. This has a simple interpretation. Consider two eigenvalues located at $\x_0$ and $\x$. In the $\mathcal{O}(\nn)$ model, $\x_0$ also feels the mirror charge located at $-\x$. We found that the two point correlation function in the large $N$ limit depends on the 'differences' between the position of the interacting sources. "Position" and "difference" have a natural meaning on the spectral curve, which is equipped with the addition law $(u_0,u) \mapsto u_0 + u$.

\section{Stable generating functions ($2 - 2g - k < 0$)}
\label{sec:Wkg}

We will see that the loop equations \ref{eq:loopeqkg} imply that $\mathrm{W}_k^{(g)}$ for $2g - 2 + k > 0$ satisfies the homogeneous linear equation \ref{eq:00}. In principle, one could try to identify the even rational functions of $\x$, $\mathrm{A}_k^{(g)}(\x^2,J)$ and $\mathrm{B}_k^{(g)}(\x^2,J)$ in a decomposition:
\beq
\mathrm{W}_k^{(g)}(\x,I) = \mathrm{A}_k^{(g)}(\x^2,I)\f_{\mu}(\x) + \mathrm{B}_k^{(g)}(\x^2,I)\widehat{\f}_{\mu}(\x) \nonumber
\eeq
This is basically the method of \cite{Amb} to compute higher genus correlators, adapted to the $\mathcal{O}(\mathfrak{n})$ model in \cite{EK1}. Inspired by \cite{E1MM}, we shall take another route, which consists in finding a Cauchy residue formula for 1-cut solutions of Eqn.~\ref{eq:00}.

\subsection{Cauchy residue formula}

Also, we construct the appropriate Cauchy kernel $\mathrm{G}(\x_0,\x)$ for our spectral curve:
\bea
\mathrm{G}(\x_0,\x) & = & \int_{\infty}^{\x}\mathrm{dx}'\left(2\overline{\mathrm{W}}_2^{(0)}(\x_0,\x') + \nn\overline{\mathrm{W}}_2^{(0)}(\x_0,-\x')\right) \\
G(\x_0,u) & = & \mathrm{G}(\x_0,x(u)) \nonumber
\eea
It is possible to derive a infinite series representation of $s(u_0)s(u)\,\mathrm{G}(x(u_0),x(u))$ from Thm.~\ref{eq:WWW}, but we will not use it. The essential properties of G are:
\begin{itemize}
\item[$\bullet$] $G(\x_0,u)$ is a meromorphic function of $u \in \mathbb{C}$, with \mbox{$u = \mathrm{u(x)}\;\textrm{mod}\;(\mathbb{Z}\oplus 2\tau\mathbb{Z})$} as only pole.
\item[$\bullet$] $\mathrm{G}(\x_0,\x) = \frac{1}{\x_0 - \x} + O(1)$ when $\x_0 \rightarrow \x$.
\item[$\bullet$] $G(\x_0,-u) = -G(\x_0,u)$.
\end{itemize}
\begin{theorem} \label{eq:Cauchy} Cauchy formula. Let $\mathrm{W}$ (in the $\x$-plane) or $W$ (in the $u$-plane) be a 1-cut solution of Eqn.~\ref{eq:00}. If $\mathrm{W}$ is holomorphic on $\mathbb{C}\setminus[a,b]$, and has no residue at $\x = \infty$, we have:
\beq
\mathrm{W}(\x_0) = \frac{1}{2}\Res_{u \rightarrow \tau,\tau + \frac{1}{2}} \mathrm{d}us(u)\,G(\x_0,u)W(u) \nonumber
\eeq
where $s(u) = \frac{\mathrm{d}x}{\mathrm{d}u}$. One can add to $G$ a constant $\gamma_i(\x_0)$, depending on the branch point $\mathrm{u}(a_i)$, $a_i \in \{a,b\}$, without changing the formula.
\end{theorem}

To prove this, we write the usual Cauchy formula and recast it as residues on the branch points $a$ and $b$ only.
\bea
\mathrm{W}(\x_0) & = & -\Res_{\x \rightarrow \x_0} \mathrm{dx}\,{}\mathrm{G}(\x_0,\x)\mathrm{W}(\x) \nonumber\\
& = & -\Res_{\x \rightarrow \x_0}\mathrm{dx}\left(\mathrm{G}(\x_0,\x)\mathrm{W}(\x) - \underbrace{\mathrm{G}(\x_0,-\x)\mathrm{W}(-\x)}_{\textrm{no pole at}\;\x_0}\right) \nonumber\\
& = & -\frac{1}{2}\Res_{\x \rightarrow \x_0,-\x_0}\mathrm{dx}\left(\mathrm{G}(\x_0,\x)\mathrm{W}(\x) - \mathrm{G}(\x_0,-\x)\mathrm{W}(-\x)\right) \quad\quad \textrm{by parity}\nonumber
\eea
We now switch to $u$ variable:  let $u_0 = \mathrm{u}(\x_0)$.
\bea
\mathrm{W}(\x_0) & = & -\frac{1}{4}\Res_{\substack{u \rightarrow u_0,-u_0 \\ \phantom{u \rightarrow}\tau - u_0,-(\tau - u_0)}}\mathrm{d}us(u)\left(G(\x_0,u)W(u) - G(\x_0,u - \tau)W(u - \tau)\right) \nonumber\\
& = & -\frac{1}{4}\Res_{\substack{u \rightarrow u_0,-u_0 \\ \phantom{u \rightarrow }\tau - u_0,-(\tau - u_0)}}\mathrm{d}u\left(s(u)G(\x_0,u)W(u) + s(u - \tau)G(\x_0,u - \tau)W(u - \tau)\right)\nonumber \\
& = & -\frac{1}{4}\Res_{\substack{u \rightarrow u_0,-u_0 \\ \phantom{u \rightarrow}\tau - u_0,-(\tau - u_0)}}\mathrm{d}u\left(sH(\x_0,\cdot)\bot{}W\right)(u)
\nonumber\eea
where $H$ is defined in Eqn.~\ref{eqH}. One knows that $s(u)H(\x_0,u)$ (up to a constant in $u$, irrelevant in the residue) satisfies the linear equation in $u$.  So, Lemma \ref{lemma:fond} tells us that \mbox{$(sH(\x_0,\cdot)\bot{}W)(u)$} is $1$- and $2\tau$-translation invariant, i.e is elliptic. Since the sum of residues of an elliptic function vanishes, we have:
\beq
\mathrm{W}(\x_0) = \frac{1}{4}\Res_{\substack{u \rightarrow 0, \frac{1}{2}\\ \phantom{u \rightarrow} \tau, \tau + \frac{1}{2}}}\mathrm{d}u\left(sH(\x_0,\cdot)\bot{}W\right)(u) \nonumber
\eeq
Eventually, the assumption of $u$-parity/1-cut property for $W$ allows us to rewrite:
\beq
\mathrm{W}(\x_0) = \frac{1}{2}\Res_{u \rightarrow \tau, \tau + \frac{1}{2}}\mathrm{d}us(u)\,G(\x_0,u)W(u) \nonumber
\eeq

\subsection{$\mathrm{W}_k^{(g)}$ (all topologies)}

\subsubsection{Recursive residue formula for $2g - 2 + k > 0$}
In the matrix model, $W_k^{(g)}(\x_1,\ldots,\x_k)$ represents correlation of densities of eigenvalues, hence have to be integrated on some interval of $[a,b]$ to give physical results. It is natural to look at the differential forms:
\beq
\mathrm{d}x_1\cdots\mathrm{d}x_k\,\mathrm{W}_k^{(g)}(\x_1,\ldots,\x_k) \nonumber
\eeq
These forms can be pushed backwards under $u \mapsto x(u)$ to define differential forms $\omega_k^{(g)}$ on the $u$-domain $\left]-\frac{1}{2},\frac{1}{2}\right[\times]-\tau,\tau[$.
\bea
\omega_k^{(g)}(u_1,\ldots,u_k) & = & \mathrm{d}u_1s(u_1)\cdots\mathrm{d}u_ks(u_k)\,W(u_1,\ldots,u_k) \nonumber \\
 & = & \mathrm{d}x(u_1)\cdots\mathrm{d}x(u_k)\,\mathrm{W}(x(u_1),\ldots,x(u_k)) \nonumber
\eea
and we recall that $s(u) = \frac{\mathrm{d}x}{\mathrm{d}u}$. The main result of this article is:
\begin{theorem}
\label{thm:toporec}For $2g - 2 + k > 0$, $\mathrm{W}_k^{(g)}$ is a 1-cut solution of the homogeneous linear equation in each variable:
\beq
\mathrm{W}_k^{(g)}(\x + i\epsilon,I) + \nn \mathrm{W}_k^{(g)}(-\x,I) + \mathrm{W}_k^{(g)}(\x - i\epsilon,I) \mathop{=}_{\epsilon \rightarrow 0} 0 \nonumber
\eeq
Thus, $\omega_k^{(g)}(u_1,\ldots,u_k)$ can be extended for $u_1,\ldots,u_k \in \mathbb{C}$,  as a meromorphic form in each $u_i$, with poles only at points where $x(u_i) = a$ or $b$. We have the recursive formula:
\begin{small}
\beq
\label{eq:toporec}\encadremath{\omega_k(u_0,I) = \Res_{u \rightarrow \tau, \tau + \frac{1}{2}} \mathrm{d}u\,\mathcal{K}(u_0,u)\left["\omega_{k + 1}^{(g - 1)}(u,\overline{u},I)" + \sum^{'}_{J \subseteq I,\, 0 \leq h \leq g} \overline{\omega}_{|J| + 1}^{(h)}(u,J)\,\overline{\omega}_{k - |J|}^{(g - h)}(\overline{u},{}^{c}J)\right]}
\eeq
\end{small}\begin{itemize}
\item[$\bullet$] $\sum'_{(J,h)}$ is a sum over $J \subseteq I, 0 \leq h \leq g$, excluding $(J,h) = (\emptyset,0)$ and $(I,g)$.
\item[$\bullet$] $\overline{u} = 2\tau - 1 - u$ is such that $x(u) = x(\overline{u})$.
\item[$\bullet$] The recursion kernel $\mathcal{K}(u_0,u)$ is a differential form in $u$, and the inverse of a differential form in $u$, given by:
\beq
\label{eq:kernel}
\mathcal{K}(u_0,u) = -\frac{1}{2}\frac{\int_{\overline{u}}^u\,\overline{\omega}_2^{(0)}(u_0,u')}{(y\mathrm{d}x)(u)}
\eeq
\item[$\bullet$] $"\omega_{k'}^{(g')}(u,\overline{u},I)"$ differs from $\omega_{k'}^{(g')}$ only for $(k',g') = (2,0)$, and:
    \beq
    \label{eq:REG}"\omega_2^{(0)}(u,\overline{u})" = -\left(\omega_2^{(0)}(u,u) + \nn\omega_2^{(0)}(u,\tau - u) + \omega_2^{(0)}(\tau - u,\tau - u)\right)
    \eeq
\item[$\bullet$] When $(k,g) \neq (1,1)$, the same formula is also true if one replaces $\omega_{k + 1}^{(g - 1)}(u,\overline{u},I)$ by $\omega_{k + 1}^{(g - 1)}(u,u,I)$. However, one cannot in general replace $\overline{u}$ by $u$ in the terms of $\sum_{J,h}'$.
\end{itemize}
\end{theorem}

\subsubsection{Comment}

We have found that the generating functions for genus $g$ maps, carrying closed loops, with $k$ boundaries, are given by the same recursive structure than the generating function for maps without decorations. Such a structure is called "topological recursion" \cite{EOFg}. It is a general fact that the counting of stable\footnote{A Riemann surface with $k$ marked points and genus $g$ is called stable if it has only a finite number of automorphisms. This happens iff $2g - 2 + k > 0$. We carry this notion to discrete maps and correlation function.} maps is more uniform than the counting of the unstable ones (rooted disks and rooted cylinders). This result provides a recursive algorithm to compute these generating functions. Minus the Euler characteristics of maps $|\chi| = 2g - 2 + k$ decreases by $1$ at each step, and $\mathrm{W}_k^{(g)}$ is reached with a stack $2g - 2 + k$ residues.

This result provides a first example where a topological recursion holds for a model which admits a non-algebraic plane curve as spectral curve:
\beq
\mathcal{L}_{\nn} = \left(\mathcal{T} = \mathbb{C}/(\mathbb{Z} \oplus \tau\mathbb{Z}), x, y\right) \nonumber
\eeq
$\nn = -2\cos(\pi\mu)$ is the deformation parameter. The set of branch points is \mbox{$\{0,\frac{1}{2}\} \subseteq \mathcal{T}$}. But $y(u)$, $\mathcal{K}(u_0,u)$, \ldots$\,$ are multivalued function on $\mathcal{T}$, basically constructed with the function $D_{\mu}$ which takes a phase $e^{i\pi\mu}$ under $\tau$-translation. We shall give a univocal meaning to the notions of "function" or "differential form on the spectral curve" in Section~\ref{fnspc}, where we need it for computations. Though, we keep their polysemy in the next paragraph.

The initial data to run the algorithm of the topological recursion is not only a plane curve ($\mathcal{C},x,y$), but also of a Bergman kernel\footnote{Recall that a Bergman kernel $\mathrm{d}u_0\mathrm{d}u\,\mathcal{B}(u_0,u)$ is a meromorphic form on the spectral curve with a pole only when $x(u) = x(u_0)$, or order $2$ and without residues.} which is closely related to $\overline{\omega}_2^{(0)}$. For example, in the $1$-matrix model with two cuts, the plane curve associated to the model is a torus, and the Weierstrass function $\wp(u - u_0) + \mathrm{cte}$ are the possible Bergman kernel of the curve. Here, the notion of Bergman kernel is slightly deformed, and one may consider the Weierstrass $\wp_{\mu}$ function as the appropriate Bergman kernel. The recursion kernel should always be given by:
\beq
\mathcal{K}(u,u_0) = -\frac{1}{2}\frac{\int_{\overline{u}}^{u} \overline{\omega}_2^{(0)}(\cdot,u_0)}{(y\mathrm{d}x)(u)} \nonumber
\eeq

The introduction of $\overline{u}$ (especially in the bracket term of Eqn.~\ref{eq:toporec}) is essential in the formulation of the topological recursion. If one try to replace $\overline{u}$ by $u$, the expression becomes wrong outside of the $1$ hermitian matrix model. The reason of being of this $\overline{u}$ begins to be understood in the geometrical interpretation of the topological recursion \cite{EOGeo}.

\subsubsection{Proof}
$\mathrm{W}_k^{(g)}$ satisfies the loop equation:
\beq
2\left(\mathrm{W}_k^{(g)}(\cdot,I)\bot\overline{\mathrm{W}}_1^{(0)}\right)(\x) = - \mathrm{E}_k^{(g)}(\x,I) - \mathrm{Q}_k^{(g)}(\x,I) \nonumber
\eeq
where
\begin{small}
\bea
\mathrm{E}_k^{(g)}(\x,I) & = & \mathrm{W}_{k + 1}^{(g - 1)}(\x,\x,I) + \mathrm{W}_{k + 1}^{(g - 1)}(-\x,-\x,I) + \nn \mathrm{W}_{k + 1}^{(g - 1)}(\x,-\x,I) \nonumber \\
& + & \sum_{J \subseteq I,\,0 \leq h \leq g}^{'} \big(\mathrm{W}_{|J| + 1}^{(h)}(\cdot,J)\bot \mathrm{W}_{k - |J|}^{(g - h)}(\cdot,{}^{c}J)\big)(\x) \nonumber \\
\mathrm{Q}_{k}^{(g)}(\x,I) & = & \mathrm{P}_{k}^{(g)}(\x,I) + P_{k}^{(g)}(-\x,I) + \sum_{\x_i \in I} \frac{\partial}{\partial \x_i}\, \left[\mathrm{W}_{k-1}^{(g)}(I) \left(\frac{1}{\x - \x_i} - \frac{1}{\x + \x_i}\right)\right] \nonumber
\eea
\end{small}
It is obvious that:
\begin{itemize}
\item[$\bullet$] $\mathrm{Q}_k^{(g)}(\cdot,I)$ is finite when $\x = a$ or $b$.
 \item[$\bullet$] $\mathrm{Q}_k^{(g)}$ has no cut on $[a,b]$.
\end{itemize}
 One can check easily as well (using symmetry in all variables of generating functions) that the combination of \mbox{$\mathrm{W}_{k + 1}^{(g - 1)}(\cdot,\cdot,I)$} entering in $\mathrm{E}_k^{(g)}(\x_0,I)$ has no cut on $[a,b]$. Since $\mathrm{Q}_k^{(g)}$ has no cut on $[a,b]$ as well, an easy recursion on $2g - 2 + k$ and the second point in Lemma~\ref{lemma:fond} show that $\mathrm{E}_k^{(g)}(\cdot,I)$ has no cut on $[a,b]$. Hence, $\big(\mathrm{W}_k^{(g)}(\cdot,I)\bot\overline{\mathrm{W}}_1^{(0)}\big)$ has no cut on $[a,b]$. Furthermore, according to the 1-cut lemma (Lemma~\ref{Lemma1cut}), $\mathrm{W}_k^{(g)}(\cdot,I)$ has only one cut. Now, the assumptions in the third point of Lemma \ref{lemma:fond} are satisfied, with $\mathrm{f} = \mathrm{W}_k^{(g)}(\cdot,I)$ and $\mathrm{g} = \overline{\mathrm{W}}_1^{(0)}$. So, $\mathrm{W}_k^{(g)}(\cdot,I)$ must be a 1-cut solution of the homogeneous linear equation.

Now, we apply the Cauchy formula (Thm.~\ref{eq:Cauchy}), with arbitrary constants $\gamma_{i}(\x_0)$:
\bea
&& \mathrm{W}_k^{(g)}(\x_0,I) \nonumber \\
& = & \frac{1}{2}\Res_{u \rightarrow \tau, \tau + \frac{1}{2}}\mathrm{d}us(u)\left(G(\x_0,u) + \gamma_i(\x_0)\right)W_k^{(g)}(u,I)\nonumber \\
& = & \frac{1}{2}\Res_{u \rightarrow \tau, \tau + \frac{1}{2}}\mathrm{d}us(u)\frac{G(\x_0,u) + \gamma_i(\x_0)}{y(u)}\left(W_k^{(g)}(u,I)y(u) + \underbrace{W_k^{(g)}(u - \tau,I)y(u - \tau)}_{\textrm{no pole at}\;\tau, \tau + \frac{1}{2}}\right) \nonumber\\
& = & -\frac{1}{2}\Res_{u \rightarrow \tau, \tau + \frac{1}{2}}\mathrm{d}us(u)\frac{G(\x_0,u) + \gamma_i(\x_0)}{y(u)}\left(E_k^{(g)}(u,I) + \underbrace{Q_k^{(g)}(u,I)}_{\textrm{is finite at}\; \tau, \tau + \frac{1}{2}}\right) \nonumber\\
& = & \Res_{u \rightarrow \tau, \tau + \frac{1}{2}}\mathrm{d}u\left[-\frac{1}{2}\frac{G(\x_0,u) + \gamma_i(\x_0)}{s(u)y(u)}\right]E_k^{(g)}(u,I)\left(s(u)\right)^2 \nonumber
\eea
Besides, we have:
\bea
G(\x_0,u) & = & \int_{\mathrm{u}(\infty)}^{u}\mathrm{d}u's(u')\left(2\overline{W}_2^{(0)}(\x_0,u') + \nn\overline{W}_2^{(0)}(\x_0,\tau - u')\right) \nonumber\\
& = & \int_{\mathrm{u}(\infty)}^{u}\mathrm{d}u's(u')\left(\overline{W}_2^{(0)}(\x_0,u') - \overline{W}_2^{(0)}(x_0,u' - 2\tau)\right) \nonumber\\
& = & \left(\int_{\mathrm{u}(\infty)}^{u}\mathrm{d}u's(u')\overline{W}_2^{(0)}(\x_0,u')\right) - \left(\int_{2\tau - \mathrm{u}(\infty)}^{2\tau - u}\mathrm{d}u's(u')\overline{W}_2^{(0)}(\x_0,u')\right)\nonumber \\
& = & \left(\int_{2\tau - u}^{u}\mathrm{d}u's(u')\overline{W}_2^{(0)}(\x_0,u')\right) + \left(\int_{\mathrm{u}(\infty)}^{2\tau - \mathrm{u}(\infty)}\mathrm{d}u's(u')\overline{W}_2^{(0)}(\x_0,u')\right)\nonumber\\
\textrm{or} & = & \left(\int_{2\tau + 1 - u}^{u}\mathrm{d}u's(u')\overline{W}_2^{(0)}(\x_0,u')\right) + \left(\int_{\mathrm{u}(\infty)}^{2\tau - \mathrm{u}(\infty)}\mathrm{d}u's(u')\overline{W}_2^{(0)}(\x_0,u')\right) \nonumber\\
& & - \left(\int_{0}^{1}\mathrm{d}u's(u')\overline{W}_2^{(0)}(\x_0,2\tau - u')\right) \nonumber
\eea
We used the $1$-translation invariance for the last line. We see that, in a neighborhood of $\mathrm{u}(a_i) = \tau$ (resp. $\mathrm{u}(a_i) = \tau + \frac{1}{2}$):
\beq
G(\x_0,u) = \int_{\overline{u}}^u\mathrm{d}u's(u')\overline{W}_2^{(0)}(\x_0,u') + \textrm{cte}_i(\x_0) \nonumber
\eeq
and we can cancel this constant by the choice of $\gamma_i(\x_0)$. Thus, the quantity in bracket (the recursion kernel) is indeed given by Eqn.~\ref{eq:kernel} (written completely in variable $u$).

\medskip

We assume now $(k,g) \neq (1,1)$. Some terms in $E_k^{(g)}$ have $u - \tau$ as arguments, hence are regular at $\mathrm{u}(a) = \tau$ and $\mathrm{u}(b) = \tau + \frac{1}{2}$ and do not contribute to the residue. We make use of the homogeneous linear equation to write them apart. Since $(k + 1,g - 1) \neq (2,0)$, $W_{k + 1}^{(g-1)}(u,\overline{u},I)$ exists, and:
\bea
E_k^{(g)}(u,I) & = & - W_{k + 1}^{(g - 1)}(u,\overline{u},I) - \sum_{J \subseteq I,\,0 \leq h \leq g}^{'} \overline{W}_{|J| + 1}^{(h)}(u,J)\overline{W}_{k - |J|}^{(g - h)}(\overline{u},{}^{c}J)\nonumber \\
 & + & \sum_{J \subseteq I,\,0 \leq h \leq g}^{'} \overline{W}_{|J| + 1}^{(h)}(u - \tau,J)\overline{W}_{k - |J|}^{(g - h)}(u - \tau,{}^{c}) \nonumber \\
 & + & W_{k + 1}^{(g - 1)}(u - \tau,u - \tau,I) \nonumber
\eea
Accordingly, we end up with:
\begin{footnotesize}\beq
W_k^{(g)}(u_0,I) = \Res_{u \rightarrow \tau, \tau + \frac{1}{2}} \mathrm{d}u\,\mathcal{K}(u_0,u)\left[W_{k + 1}^{(g)}(u,\overline{u},I) + \sum_{J \subseteq I,\,0 \leq h \leq g} \overline{W}_{|J| + 1}^{(h)}(u,J)\overline{W}_{k - |J|}^{(g - h)}(\overline{u},{}^{c})\right]s(u)(-s(u)) \nonumber
\eeq\end{footnotesize}
and we know that $s(\overline{u}) = -s(u)$. This can be rewritten with differential forms and we find Eqn.~\ref{eq:toporec}.

Eventually, we have to write separately the case $(k,g) = (1,1)$, because the polar structure of $W_2^{(0)}$ is different. In $E_k^{(g)}(u,I)$, there is only one term, $-"W_2^{(0)}(u,\overline{u})"$, defined by
\beq
"W_2^{(0)}(u,\overline{u})" \equiv -\left(W_2^{(0)}(u,u) + \nn{}W_2^{(0)}(u,\tau - u) + W_2^{(0)}(\tau - u,\tau - u)\right) \nonumber
\eeq
If we set:
\bea
"\omega_2^{(0)}(u,\overline{u})" & \equiv & -(\mathrm{d}x(u))^2\,\left(W_2^{(0)}(u,u) + \nn W_2^{(0)}(u,\tau - u) + W_2^{(0)}(\tau - u,\tau - u)\right) \nonumber \\
& = & -\left(\omega_2^{(0)}(u,u) + \nn\omega_2^{(0)}(u,\tau - u) + \omega_2^{(0)}(\tau - u,\tau - u)\right), \nonumber
\eea
then Eqn.~\ref{eq:toporec} is still correct.

\subsection{Examples}

Finding $\textrm{W}_3^{(0)}$ and $\textrm{W}_1^{(1)}$ involve only one residue computation. Let us call $v_i$ the branch points $\mathrm{u}(a) = \tau$ and $\mathrm{u}(b) = \tau + \frac{1}{2}$. We shall use the following notation:
\bea
\overline{\omega}_2^{(0)}(u_0,v_i) & \equiv &  \left(\frac{\overline{\omega}_2^{(0)}(u_0,u)}{\mathrm{d}u}\right)\Big|_{u = v_i} \nonumber \\
\textrm{and}\; (\partial_2^{m}\overline{\omega}_2^{(0)})(u_0,v_i) & \equiv & \frac{\partial^m}{\partial u^m}\left(\frac{\overline{\omega}_2^{(0)}(u_0,u)}{\mathrm{d}u}\right)\Big|_{u = v_i} \qquad \textrm{for}\; m \geq 1 \nonumber
\eea
I.e., we identify differentials forms and families of functions indexed by an atlas of $\mathbb{C}$ and having the proper transformation under a change of local coordinate, and we read all differential forms in the local coordinate $u$.

\subsubsection{Recursion kernel at the branchpoints}
As a preliminary to the computations, we give the Laurent expansion of $\mathcal{K}(u,u_0)$ when $\delta = u - v_i \rightarrow 0$. In the local coordinate $u$:
\bea
\mathcal{K}(u_0,v_i + \delta) & = & -\frac{\overline{\omega}_2^{(0)}(u_0,v_i)}{y'(v_i)s'(v_i)}\,\frac{1}{\delta} + \nonumber \\ & & \left[\frac{1}{4}\frac{\overline{\omega}_2^{(0)}(u_0,v_i)}{y'(v_i)s'(v_i)}\left(\frac{y'''(v_i)}{y'(v_i)} + \frac{s'''(v_i)}{s'(v_i)}\right) - \frac{1}{6}\frac{\partial_2^2\overline{\omega}_2^{(0)}(u_0,v_i)}{y'(v_i)s'(v_i)}\right]\,\delta +\, o(\delta)\nonumber \\
\label{eq:KL}& &
\eea
We notice $\mathcal{K}(u_0,v_i + \delta)$ is odd in $\delta$ since $\mathcal{K}$ is changed in $-\mathcal{K}$ when $u \leftrightarrow \overline{u}$.
All the same, only the odd order derivative (with respect to $u$) of $s$ and $y$ do not vanish a priori at $u = v_i$.

\subsubsection{$\mathrm{W}_3^{(0)}$ (pairs of pants)}
\begin{theorem}
\label{eq:W30} \beq \encadremath{\omega_3^{(0)}(u_0,u_1,u_2) = \sum_i
\frac{-2}{y'(v_i)s'(v_i)}\,\overline{\omega}_2^{(0)}(u_0,v_i)\,\overline{\omega}_2^{(0)}(u_1,v_i)\,\overline{\omega}_2^{(0)}(u_2,v_i)} \nonumber
\eeq
\end{theorem}

\subsubsection{$\mathrm{W}_1^{(1)}$ (rooted disk with one handle)}

$\mathrm{W}_1^{(1)}$ is the generating function for rooted toroidal maps. According to the topological recursion:
\beq
\omega_1^{(1)}(u_0) = \sum_i\Res_{u \rightarrow v_i} \mathcal{K}(u,u_0)\,"\omega_2^{(0)}(u,\overline{u})" \nonumber
\eeq
Using the Laurent expansion of $\mathcal{K}(u_0,u)$ when $u \rightarrow v_i$ (Eqn.~\ref{eq:KL}), we find:
\bea
\omega_1^{(1)}(u_0) & = & - \frac{\overline{\omega}_2^{(0)}(u_0,v_i)}{y'(v_i)s'(v_i)}\,\Res_{u \rightarrow v_i} \frac{"\omega_2^{(0)}(u,\overline{u})"}{u - v_i} \nonumber\\
& & + \frac{1}{4}\frac{\overline{\omega}_2^{(0)}(u_0,v_i)}{y'(v_i)s'(v_i)}\left(\frac{y'''(v_i)}{y'(v_i)} + \frac{s'''(v_i)}{s'(v_i)}\right)\,\Res_{u \rightarrow v_i} (u - v_i)\,"\omega_2^{(0)}(u,\overline{u})" \nonumber \\
& & - \frac{1}{6}\frac{\partial_2^2 \overline{\omega}_2^{(0)}(u_0,v_i)}{y'(v_i)s'(v_i)}\,\Res_{u \rightarrow v_i} (u - v_i)\,"\omega_2^{(0)}(u,\overline{u})" \nonumber
\eea
Besides, we can make use of the representation of $\overline{\omega}_2^{(0)}$ as an infinite series (Thm.~\ref{eq:WWW}) and compute explicitly:
\bea
\frac{"\omega_2^{(0)}(u,\overline{u})"}{(\mathrm{d}u)^2} & = & - \frac{1}{(\mathrm{d}u)^2}\left(\omega_2^{(0)}(u,u) + \nn \omega_2^{(0)}(u,\tau - u) + \omega_2^{(0)}(\tau - u,\tau - u)\right) \nonumber \\
& = &  \lim_{u \rightarrow u_0}\left(\frac{s(u)s(u_0)}{(x(u) - x(u_0))^2} - \sum_{m \in \mathbb{Z}} \frac{(2r_m - \nn r_{m + 1})\pi^2}{\sin^2\pi(u - u_0 + m\tau)}\right) \nonumber \\
& = & \frac{1}{6}(\mathrm{S}x)(u) - \frac{\pi^2}{3} + \sum_{m \geq 1} \frac{\pi^2\,\cos\pi(1 - \mu)m}{\sin^2\pi m\tau}\nonumber
\eea
We have introduced the schwartzian derivative of $x$ with respect to $u$:
\beq
(\mathrm{S}x)(u) \equiv \frac{x'''(u)}{x'(u)} - \frac{3}{2}\left(\frac{x''(u)}{x'(u)}\right)^2 \nonumber
\eeq
When $u \rightarrow v_i$, the first term is finite whereas the second one has a double pole:
\beq
(\mathrm{S}x)(u) = -\frac{3}{2}\frac{1}{(u - v_i)^2} + O((u - v_i)^2) \qquad \textrm{when} \; u \rightarrow v_i \nonumber
\eeq
Hence, $\Res_{u \rightarrow v_i} (u - v_i)"\omega_2^{(0)}(u,\overline{u})" = -\frac{1}{4}$. Eventually, let us introduce the so-called \emph{connective projection}:
\bea
S_{B,\x}(v_i) & \stackrel{\textrm{def}}{=} & - 6\,\Res_{u \rightarrow v_i} \frac{"\omega_2^{(0)}(u,\overline{u})"}{\mathrm{d}x(u)} \nonumber \\
& = & \frac{1}{s'(v_i)}\left(2\pi^2 + \sum_{m \in \mathbb{Z}\setminus\{0\}} \frac{6q_m\pi^2}{\sin^2\pi{}m\tau}\right) \nonumber
\eea
We have in terms of $S_{B,\x}$:
\beq
\Res_{u \rightarrow v_i} \frac{"\omega_2^{(0)}(u,\overline{u})"}{(u - v_i)} = -\frac{1}{6}S_{B,\x}(v_i) \nonumber
\eeq
The final result is:
\begin{theorem}
\label{W11}\begin{small}\beq
\encadremath{\omega_1^{(1)}(u_0) = \sum_i \frac{1}{24}\frac{(\partial_2^2\overline{\omega}_2^{(0)})(u_0,v_i)}{y'(v_i)s'(v_i)} + \frac{\overline{\omega}_2^{(0)}(u_0,v_i)}{y'(v_i)}\left(\frac{S_{B,\x}(v_i)}{6} - \frac{1}{16}\frac{s'''(v_i)}{s'(v_i)^2} - \frac{1}{16}\frac{y'''(v_i)}{y'(v_i)s'(v_i)}\right)}
\nonumber
\eeq\end{small}
\end{theorem}
This expression is, again, similar to the 1-matrix model. We have given the details of the computation to illustrate that the steps are the same as in the 1-matrix model, provided the definition of $"\omega_2^{(0)}(u,\overline{u})"$ is adapted to the $\mathcal{O}(\nn)$ model according to Eqn.~\ref{eq:REG}.

\section{Properties of the $\mathrm{W}_k^{(g)}$'s}
\label{sec:prop}

In this section, we show that the properties found with the topological recursion in the 1-matrix model can be completely generalized for the deformed topological recursion relevant in the $\mathcal{O}(\nn)$ model. The special geometry structure is present, the stable $\mathrm{F}_g$'s ($g \geq 2$) are computed by the same integration formula, and (almost) the same formula exists for F$_0$ and F$_1$.

Once the properties for unstable quantities are checked, many proofs can be done by induction like in the 1-hermitian matrix model. Those proofs only use the residue recursion formula Eqn.~\ref{eq:toporec} and a few other simple properties which are satisfied here. We choose not to reproduce the derivation of these known results, for sake of brevity, and only stress the particularities in the $\mathcal{O}(\nn)$ model. Also, we refer to \cite{EOFg} for the complete proofs.

\subsection{Symmetry}

Though it is not obvious, the residue formula Eqn.~\ref{eq:toporec} yields symmetric $\omega_k^{(g)}$'s. This statement is based on the symmetry of $\omega_2^{(0)}$, and the loop equations Eqn.~\ref{eq:loopeqkg}. The proof is similar to the case of the one hermitian matrix model.

\subsection{Homogeneity} The partition function of the $\mathcal{O}(\nn)$ model (Eqn.~\ref{eq:mint}), $\mathrm{Z}(t,t_3,t_4,\ldots)$, is invariant under
\beq
(t,t_0,t_1,t_2,t_3,\ldots) \mapsto (\lambda t,\lambda t_0,\lambda t_1,\lambda t_2,\lambda t_3,\ldots)\quad\quad\quad\lambda > 0 \nonumber
\eeq
Therefore:
\beq
\label{eq:hom}\forall k,g \geq 0, \qquad \left(t\frac{\partial}{\partial t} + \sum_{j \geq 0} t_j\frac{\partial}{\partial t_j}\right)\mathrm{W}_k^{(g)} = (2 - 2g - k)\mathrm{W}_k^{(g)}
\eeq

\subsection{Special geometry}
\label{fnspc}
What we call special geometry is the data of a non degenerate pairing, which to a differential form $\widehat{\Omega}$ on the spectral curve, associates:
\begin{itemize}
\item[$\bullet$] a cycle $\Omega^* \subseteq \mathbb{C}$, i.e  in the $u$-plane,
\item[$\bullet$] a germ of holomorphic function $\Lambda_{\Omega}$ on $\Omega^*$,
\end{itemize}
with the following property, for all $k,g \geq 0$:
\bea
\label{eq:sgeom}\delta_{\epsilon}\, \overline{\omega}_{k}^{(g)}(I) & = & \int_{\Omega^*}\mathrm{d}u\Lambda_{\Omega}(u)\overline{\omega}_{k + 1}^{(g)}(u,I) \\
\textrm{when}\; y\mathrm{d}x & \rightarrow & y\mathrm{d}x + \epsilon\widehat{\Omega}\qquad \epsilon \rightarrow 0 \nonumber
\eea
Though, for the $\mathcal{O}(\nn)$ model, we understand the notion of "differential form on the spectral curve" in a slightly deformed way. The manifold for the spectral curve can be considered as $\mathcal{T}$ or $\mathbb{C}$. If $\widehat{\Omega}$ is a differential meromorphic form on $\mathbb{C}$, we define:
\beq
\Omega = \frac{1}{4 - \nn^2}\left(2\widehat{\Omega}(u) + \nn\widehat{\Omega}(\tau - u)\right) \nonumber
\eeq
For us, $\widehat{\Omega}$ is a differential form on the spectral curve if $\Omega/\mathrm{d}u$ is a $u$-even solution of Eqn.~\ref{eq:01} (that is, $\widehat{\Omega}$ has the same properties as $y\mathrm{d}x$).
A consequence of Eqn.~\ref{eq:sgeom} is that the pairing is given by integration of $\overline{\omega}_2^{(0)}$ against $\Omega^*$:
\bea
\Omega(u_0) & = & \int_{\Omega^*}\mathrm{d}u\,\Lambda_{\Omega}(u)\overline{\omega}_2^{(0)}(u_0,u) \nonumber \\
\widehat{\Omega}(u_0) & = & \int_{\Omega^*}\mathrm{d}u\,\Lambda_{\Omega}(u)\left(2\overline{\omega}_2^{(0)}(u_0,u) - \nn\overline{\omega}_2^{(0)}(\tau - u_0,u)\right) \nonumber
\eea

In the next paragraphs, we compute the variations of the
$\omega_k^{(g)}$'s with respect to the $t_j$'s (which is a simple task) and to
$t$ (which gives an interesting result). Roughly speaking, we hold $\x$ fixed,
but the value $\mathrm{u(x)}$ changes since $a$ and $b$ are determined by the
consistency relations and depend on $t$ and on the $t_j$'s. To avoid confusions, we note $\partial_t|_{x}$, $\partial_{t_j}|_{x}$ these variations. By definition:
\beq
\left(\frac{\partial}{\partial t_{\bullet}}\Big|_{\x}\,\omega_{k}^{(g)}\right)(u_1,\ldots,u_k) = \mathrm{d}x(u_1)\cdots\mathrm{d}x(u_k)\,\left(\frac{\partial}{\partial t_{\bullet}}\,\mathrm{W}\right)(x(u_1),\ldots,x(u_k)) \nonumber
\eeq
We show that for each of these parameters, one can find a dual cycle such that special geometry holds. If we had considered $\mathrm{W}_1^{(0)}(\x)$ with several cuts in the $\x$ plane, one would also like to compute variations with respect to filling fractions and identify the corresponding dual cycle.

\subsubsection{Variations of the $t_j$'s}

\begin{theorem}
 \label{eq:tjvar}\beq\forall k,g \geq 0,\;\forall j \geq 3
\qquad\left(\frac{\partial}{\partial t_j}\Big|_{\x}\,
\omega_k^{(g)}\right)(I) = \Res_{u \rightarrow
\mathrm{u}(\infty)}\mathrm{d}u\,\frac{x(u)^j}{j}\,\omega_{k+1}^{(g)}(u,I) \nonumber \eeq
The cycle associated to $t_j$ is $(\Omega_j^*,\Lambda_j) =
(\frac{1}{2i\pi}\mathcal{C}_{\infty},\frac{x(u)^j}{j})$,
where $\mathcal{C}_{\infty}$ is a contour surrounding $\mathrm{u}(\infty)$
and no other special point.
\end{theorem}

We start from the definition of the correlation function before the topological expansion:
\bea
\left(\frac{\partial}{\partial t_1} \mathrm{W}_k\right)(I) & = & -\Big\langle\frac{N}{t}\frac{{\mathrm{Tr} M^j}}{j} \, \prod_{\x_i \in I}\mathrm{Tr}\frac{1}{\x_i - M}\Big\rangle_C \nonumber\\
 & = & \Res_{\x \rightarrow \infty} \mathrm{dx}\,\Big\langle\frac{N}{t}\frac{\x^j}{j} \mathrm{Tr}\frac{1}{\x - M}\,\prod_{\x_i \in I}\mathrm{Tr}\frac{1}{\x_i - M}\Big\rangle_C \nonumber
\eea
After expansion in powers of $\frac{N}{t}$, and translation into the language of differential forms, we obtain Thm.~\ref{eq:tjvar}.

\subsubsection{Variation of $t$}

We begin with the obvious remark that $\partial_t|_{\x}\,\overline{\omega}_k^{(g)} = \partial_t|_{\x}\,\omega_k^{(g)}$. Let us compute first the variations of the stable correlation forms $\omega_1^{(0)}$ and $\omega_2^{(0)}$.
\begin{theorem}\label{W10vart}
\beq
\left(\frac{\partial}{\partial t}\Big|_{\x}\,\overline{\omega}_1^{(0)}\right)(u) = \int_{\mathrm{u(\infty)}}^{\overline{\mathrm{u}(\infty)}} \overline{\omega}_2^{(0)}(\cdot,u) \nonumber
\eeq
We have defined the point $\overline{\mathrm{u}(\infty)} = 2\tau - 1 - \mathrm{u}(\infty) = \frac{-1 + 3\tau}{2}$. The cycle associated to $t$ is \mbox{$(\Omega_t^*,\Lambda_t) = ([\frac{- 1 +
\tau}{2},\frac{- 1 + 3\tau}{2}],1)$}. There is a corresponding form on
the spectral curve: $\Omega_t(u) = \mathrm{d}x(u)\,f_{\mu}(u)$.
\end{theorem}

\begin{theorem}\label{W20vart}
\bea
\left(\frac{\partial}{\partial t}\Big|_{\x} \overline{\omega}_2^{(0)}\right)(u_1,u_2) & = & \int_{\Omega_t^*}\omega_3^{(0)}(\cdot,u_1,u_2) \nonumber \\
& = & \sum_i\Res_{u \rightarrow
v_i}\mathrm{d}u\,\left[\mathcal{K}(u_1,u)\overline{\omega}_2^{(0)}(u,u_2)\left(\frac{\partial}{\partial
t}\Big|_{\x}\,\overline{\omega}_1^{(0)}\right)(u)\right] + (u_1
\leftrightarrow u_2) \nonumber \\ \label{eq:W20vart} & & \eea
\end{theorem}
To prove the formula \ref{eq:sgeom} for $2g - 2 + k > 0$, we need first a lemma for the variation of the recursive kernel $\mathcal{K}$ (Eqn.~\ref{eq:kernel}).
\begin{lemma}\label{Kvart}
\beq
\left[\left(\frac{\partial}{\partial t}\Big|_{\x} + \frac{\partial
\ln y}{\partial t}(u)\right)\mathcal{K}\right](u_0,u) =
\sum_i\Res_{u' \rightarrow
v_i}\mathrm{d}u'\,\left[\mathcal{K}(u_0,u')\mathcal{K}(u',u)\,\left(\frac{\partial}{\partial
t}\Big|_{\x}\,\overline{\omega}_1^{(0)}\right)(u')\right] \nonumber
\eeq
\end{lemma}

\noindent Then, the general result follows:
\begin{theorem}\label{Wkgvart}
For all $k,g \neq (0,0)$: \beq \left(\frac{\partial}{\partial
t}\Big|_{\x}\,\overline{\omega_k^{(g)}}\right)(I) = \int_{\Omega_t^*}\overline{\omega}_{k + 1}^{(g)}(\cdot,I) \nonumber \eeq
\end{theorem}
\medskip
This will be completed in Section~\ref{sec:CF0} with expressions for the derivatives of $\mathrm{F}_0$.

\subsubsection*{Proof of Thm.~\ref{W10vart}}

\label{sec:HH} We derive the properties of $\partial_t|_{\x}\,\omega_1^{(0)}$
from those of $\mathrm{W}_1^{(0)}$ listed in Section~\ref{sec:W10}.
They coincide with those of $\f_{\mu}$ listed in Section~\ref{sec:ssol},
and both of these functions are 1-cut solution of the homogeneous
linear equation. Hence
$\left(\frac{\partial}{\partial t}\Big|_{\x}\,\mathrm{W}_1^{(0)}\right)(\x) =
\f_{\mu}(\x)$. Now, we would like to represent $\f_{\mu}(\x_0)\mathrm{dx_0}$ as an
integral of $\overline{\omega}_2^{(0)}$ over some path in the
$u$-plane. Having a look at $H(x_0,x)$ in Thm.~\ref{H}, it is
possible to write: \beq f_{\mu}(u_0)\mathrm{d}x(u_0) = \mathrm{cte}\times
\int_{\frac{- 1 +
\tau}{2}}^{u^*} \overline{\omega}_2^{(0)}(\cdot,u_0)
\eeq if we take $u^* \neq \mathrm{u}(\infty)$ such that $f_{\mu}(u^*) = 0$.
An accurate value is:
\beq
u^* = \overline{\mathrm{u}(\infty)} = 2\tau - 1 - \mathrm{u}(\infty) = \frac{-1 + 3\tau}{2} \nonumber
\eeq
Using the homogeneous linear equation, one computes:
\beq f_{\mu}(u_0) \sim \frac{\nn - 1}{x(u_0)} \qquad\textrm{when}\; u_0 \rightarrow \overline{\mathrm{u}(\infty)} \nonumber
\eeq
and find that $f_{\mu}(u_0)\mathrm{d}x(u_0) = H(u_0,\overline{\mathrm{u}(\infty)})\mathrm{d}x(u_0)$.

\subsubsection*{Proof of Thm.~\ref{W20vart}-Lemma~\ref{Kvart}}

We compute with help successively of Eqn.~\ref{eq:hom}, Thm.~\ref{eq:tjvar}, Thm.~\ref{eq:W30}, Thm.~\ref{W10vart}, and comparison to Eqn.~\ref{eq:KL}:
\bea
\left(t\frac{\partial}{\partial t}\Big|_{\x}\,\overline{\omega}_2^{(0)}\right)(u_0,u_1) & = & - \sum_j \left(t_j\frac{\partial}{\partial t_j}\Big|_{\x}\,\overline{\omega}_2^{(0)}\right)(u_0,u_1) \nonumber \\
& = & - \Res_{u \rightarrow \mathrm{u}(\infty)} \mathrm{V}(x(u))\omega_3^{(0)}(u,u_0,u_1)\nonumber \\
& = & 2\sum_i\frac{\overline{\omega}_2^{(0)}(v_i,u_0)\,\overline{\omega}_2^{(0)}(v_i,u_1)}{y'(v_i)s'(v_i)}\,\Res_{u \rightarrow \mathrm{u}(\infty)} \left(\mathrm{V}(x(u))\overline{\omega}_2^{(0)}(v_i,u)\right)\nonumber \\
& = & 2\sum_i\frac{\overline{\omega}_2^{(0)}(v_i,u_0)\,\overline{\omega}_2^{(0)}(v_i,u_1)}{y'(v_i)s'(v_i)}\,\left[- \left(t\frac{\partial}{\partial t}\Big|_{\x}\,\overline{\omega}_1^{(0)}\right)(v_i) + \overline{\omega}_1^{(0)}(v_i)\right]\nonumber \\
& = &-2\sum_i\frac{\overline{\omega}_2^{(0)}(v_i,u_0)\,\overline{\omega}_2^{(0)}(v_i,u_1)}{y'(v_i)s'(v_i)}\,\left(t\int_{\frac{- 1 + \tau}{2}}^{\frac{- 1 + 3\tau}{2}} \overline{\omega}_2^{(0)}(v_i,u)\right)\nonumber \\
& = & t\int_{\frac{ - 1 + \tau}{2}}^{\frac{- 1 + 3\tau}{2}} \omega_3^{(0)}(u,u_0,u_1)\nonumber \\
\mathrm{or} & = & 2\sum_i\Res_{u \rightarrow v_i}
\left[\mathcal{K}(u_0,u)\,\overline{\omega}_2^{(0)}(u,u_1)\,{}\left(t\frac{\partial}{\partial
t}\Big|_{\x}\,\overline{\omega}_1^{(0)}\right)(u)\right] \nonumber
\eea The last line is an application of the lemma above. By symmetry
of $(\partial_t|_{\x}\, \overline{\omega}_2^{(0)})(u_0,u_1)$, we
can also distribute $u_0$ and $u_1$ in two ways, as in
Thm.~\ref{W20vart}.

Then, it is easy to prove Thm.~\ref{Kvart} from the expression of
$\partial_t|_{\x}\,\overline{\omega}_2^{(0)}$ in the last line.

\subsubsection*{Proof of Thm.~\ref{Wkgvart}}
For $(k,g) \neq (0,0)$, the proof is similar to the 1-matrix model. We shall consider the case of $\partial_t \mathrm{F}_0$ later, in Section~\ref{sec:CF0}.

\subsection{Integration formula}
\label{sec:intform}
The inverse operation, consisting in recovering $\omega_{k}^{(g)}$ from $\omega_{k + 1}^{(g)}$, can also be performed by a residue calculation at the branch points.
\begin{theorem}
For all $k,g$ such that $2g - 2 + k \geq 0$: \beq \label{eq:dilaton}
\encadremath{\sum_i \Res_{u \rightarrow v_i} \phi(u)\,\omega_{k +
1}^{(g)}(u,I) = (2 - 2g - k)\omega_k^{(g)}(I)} \eeq where
$\mathrm{d}\phi = y\mathrm{d}x$. In particular, for $g \geq 2$: \beq
\mathrm{F}_g = \frac{1}{2 - 2g}\sum_i\Res_{u \rightarrow v_i}
\phi(u)\,\omega_1^{(g)}(u) \eeq
\end{theorem}

\proof{Similar to the 1-matrix model.}
\medskip
\noindent Though, $\mathrm{F}_0$ and $\mathrm{F}_1$ cannot be found by this method.

\subsection{$\mathrm{F}_0$ (spherical maps)}
\label{sec:CF0}
\subsubsection{Earlier results}
\label{sec:F0F1}
In the matrix model, $\mathrm{F}_0$ is minus the free energy of the model in the thermodynamic limit. In other words, it gives the leading order of $\ln \mathrm{Z}$ modulo the remark about convergent matrix integrals (Section~\ref{sec:cvm}). The saddle point technique applied to the partition function Eqn.~\ref{eq:mint} gives a heuristic formula:
\begin{proposition}
\beq
\mathrm{F}_0 = -\frac{t}{2}\int_{a}^b \mathrm{d}x\,\varrho(\x)"\mathrm{V}(\x)" \nonumber
\eeq
where $\varrho(\x) = -\frac{1}{2i\pi t} \mathrm{y}(\x)$ is the density of eigenvalues (which has $[a,b]$ as support on the real line) in the thermodynamic limit. $"\mathrm{V}(\x)"$ is the primitive of $\mathrm{V}'(\x)$ which satisfies:
\beq
\forall \x_0 \in [a,b],\qquad \int_{a}^b \mathrm{d}x\,\rho(x)\big(2\ln|\x - \x_0| - \frac{\mathfrak{n}}{2}\ln|\x + \x_0|\big) = "\mathrm{V}(\x_0)" \nonumber
\eeq
(It differs from $\mathrm{V}(\x)$ only by a constant term).
\end{proposition}
In fact, it is easier to compute derivatives of $\mathrm{F}_0$ with respect to $t$. The most explicit expressions obtained in \cite{EK2} were:
\begin{theorem}
\beq \label{eq:F0}\frac{\partial^3 \mathrm{F}_0}{\partial t^3} = \frac{(2 - \nn)}{a^2 - b^2}\left[\frac{a^2 -
e_{\mu}^2}{a}\frac{\partial a}{\partial t} - \frac{b^2 - e_{\mu}^2}{b}\frac{\partial b}{\partial t}\right]
\eeq
It can be integrated once:
\beq
\frac{\partial^2 \mathrm{F}_0}{\partial t^2} = \left(1 - \frac{\nn}{2}\right)\ln\left[a^2\,g\left(1 -\frac{a^2}{b^2}\right)\right] + \hat{\mathrm{C}_2} \nonumber
\eeq
where $\hat{C}_2$ is a constant independent of $t$, and $g$ a function defined by:
\beq
\frac{\mathrm{d} \ln g}{\mathrm{d} m} = \frac{1}{m(1-m)\,\mathrm{cd}^2_{\sqrt{m}}(\mu K(\sqrt{m}))} \nonumber
\eeq
and $\mathrm{cd}_k(w) = \frac{\mathrm{dn}_k(w)}{\mathrm{cn}_k(w)}$.
\end{theorem}

\subsubsection{New results}

With special geometry, we can compute $\partial_t^k \mathrm{F}_0$ by integrating $\omega_k^{(0)}$ on $(\Omega_t^*)^k$. This is true for $k \geq 3$, but some adaptations are required for $k = 2$ and $k = 1$ because those integrals diverge. For these cases, we only give the result, which is very similar to the 1-hermitian matrix model. We recall that $s(u) = x'(u)$.
\begin{theorem}
\begin{small}
\bea
\frac{\partial^2 \mathrm{F}_0}{\partial t^2} & = & \lim_{\epsilon \rightarrow 0}\left((2 - \mathfrak{n})\ln x(\mathrm{u}(\infty) + \epsilon) + \int_{\mathrm{u}(\infty) + \epsilon}^{\mathrm{u}(b)} \mathrm{d}u\,\big((s\,f_{\mu})(u) + (s\,f_{\mu})(\overline{u})\big)\right) + \mathrm{C}_2 \nonumber \\
\label{ehq}\frac{\partial \mathrm{F}_0}{\partial t} & = & \lim_{\epsilon \rightarrow 0}\left((2 - \mathfrak{n})t\ln x(\mathrm{u}(\infty) + \epsilon) - \mathrm{V}\big(x(\mathrm{u}(\infty) + \epsilon)\big) + \int_{\mathrm{u}(\infty) + \epsilon}^{\mathrm{u}(b)} \mathrm{d}u\,y(u)\right)
\eea
\end{small}
\end{theorem}
By homogeneity (Eqn.~\ref{eq:hom}), it is straightforward to obtain $\mathrm{F}_0$ from Eqn.~\ref{ehq}
\begin{theorem}
\beq
\mathrm{F}_0 =  \frac{1}{2} \Res_{\mathrm{x} \rightarrow \infty} \mathrm{dx}\,\mathrm{V}(\x)\,\mathrm{W}_1^{(0)}(\x) + \frac{t}{2}\,\frac{\partial \mathrm{F}_0}{\partial t} \nonumber
\eeq
\end{theorem}
With the infinite series representation, we can compute further. As for $\mathrm{W}_2^{(0)}$, $\partial_t^2 \mathrm{F}_0$ depends on the potential $V$ only through the parameters $a$ and $b$, or equivalently through $\tau$ and $K$ introduced in Section~\ref{sec:dessin2}.
\begin{theorem}
\begin{small}\beq
\encadremath{\frac{\partial^2 \mathrm{F}_0}{\partial t^2} = (2 - \mathfrak{n})\ln\left(\frac{\pi{}b|\tau|}{2K}\right) + 2\sqrt{4 - \mathfrak{n}^2} \sum_{m \geq 1}  T_m(\mathfrak{n}/2)\,\ln\big(\sin\pi(1 - \mu)m\big)} \nonumber
\eeq
\end{small}
$T_{m}(\mathfrak{n}/2) = \cos\pi(1 - \mu)m$ is the $m$-th Chebyshev polynomial.
\end{theorem}
($\partial_t^2 \mathrm{F}_0$ counts spherical maps with $v$ vertices with a weight $v(v - 1)t^{v - 2}$. In a rough way, we can say that this weight is proportional to the area of the dual map). The situation is very different for $\partial_t \mathrm{F}_0$ and $\mathrm{F}_0$, which depends fully on the potential $V$. We have obtained an explicit formula for $\mathrm{F}_0$ for any polynomial $V$, involving an infinite series. We do not reproduce it here for lisibility, but it is available on request.

\subsection{$\mathrm{F}_1$ (toroidal maps)}
\label{sec:F1F1}
We can use special geometry to compute $\partial_t \mathrm{F}_1$.
\bea
\frac{\partial \mathrm{F}_1}{\partial t} & = & \int_{\Omega_t^*} \omega_1^{(1)} \nonumber \\
& = & \sum_i \frac{1}{24}\frac{\int_{u_0 \in \Omega_t^*}(\partial_2^2\overline{\omega}_2^{(0)})(u_0,v_i)}{y'(v_i)s'(v_i)} + \nonumber \\
& & + \frac{\int_{u_0 \in \Omega_t^*}\overline{\omega}_2^{(0)}(u_0,v_i)}{y'(v_i)}\left(\frac{S_{B}(v_i)}{6} - \frac{1}{16}\frac{s'''(v_i)}{s'(v_i)^2} - \frac{1}{16}\frac{y'''(v_i)}{y'(v_i)s'(v_i)}\right) \nonumber \\
\label{eq:F111} & = & \sum_i \frac{1}{24}\frac{(sf_{\mu})''(v_i)}{y'(v_i)s'(v_i)} + \frac{(sf_{\mu})(v_i)}{y'(v_i)}\left(\frac{S_{B}(v_i)}{6} - \frac{1}{16}\frac{s'''(v_i)}{s'(v_i)^2} - \frac{1}{16}\frac{y'''(v_i)}{y'(v_i)s'(v_i)}\right) \nonumber
\eea
The integration of the right hand side with respect to $t$ has been studied in the literature \cite{EKK}, and relies on the computation of $\partial_t|_{\x}(y'(v_i))$ and the comparison with Eqn.~\ref{eq:F111}. This equation is essentially the same as the one encountered in the one matrix model. So, we give directly the result:
\begin{theorem}
\beq
\label{eq:F110}\encadremath{\mathrm{F}_1 = \frac{1}{12}\ln [\tau_B((a_i)_i)] + \frac{1}{24}\ln\left( \prod_{i} y'(a_i)\right)} \nonumber
\eeq
$\tau_B$ is the Bergman tau function of our problem, which is a function of the position of the branch points in the $\x$-plane, $(a_i)_i$, and which is defined by:
\beq
\forall i, \qquad \frac{\partial \ln \tau_B}{\partial a_i} = S_B(\mathrm{u}(a_i)) \nonumber
\eeq
The very special expression of $\omega_3^{(0)}$ (Thm.~\ref{eq:W30}, which is a sort of WDVV formula) ensures that the form $\sum_i \mathrm{d}a_i\,S_B(\mathrm{u}(a_i))$ is closed.
\end{theorem}

\section{Large maps}
\label{sec:lmaps}
As an application of the topological recursion, we can derive the critical exponents for the $\mathcal{O}(\nn)$ model without loop at boundaries, thus extending the proofs existing for planar maps (genus $0$). In this section, we recall known facts about the critical points, outline the method to take the critical limit in the formalism we used, and give the main results. We recall that:
\beq
\nn = -2\cos(\pi\mu),\qquad \textrm{with}\; \mu \in \left.]0,1[\right. \nonumber
\eeq

\subsection{Principles}
\label{sec:princip}
\subsubsection{Asymptotic of maps}
Let $\hat{\mathrm{V}}$ be a fixed potential. The analysis of singularities (also called critical points) in $t$ of $\mathrm{W}_k^{(g)}(t)$ allows to find the asymptotic of the (weighted) number of $\mathcal{O}(\nn)$ maps with genus $g$ and $k$ boundaries. We illustrate this on $\mathrm{F}_0(t)$.
\beq
\mathrm{F}_0(t) = \sum_{v \geq 0} \mathrm{F}_{0|v}\,t^v\nonumber
\eeq
was defined as a generating function of spherical $\mathcal{O}(\nn)$ maps with $v$ vertices. It is a formal power series in $t$, and we know (see next paragraph) by a corollary of the 1-cut lemma that its radius of convergence $\rho_0(\hat{\mathrm{V}})$ is strictly positive. Hence, there exists $t^* \in \mathcal{C}(0\,;\,\rho_0(\hat{\mathrm{V}}))$ such that $\mathrm{F}_0(t)$ has a singularity when $t \rightarrow t^*$. We can decompose $\mathrm{F}_0(t) = f_r(t) + f_s(t)$ with $f_r$ analytic in a neighborhood of $t^*$ and $f_s$ nonanalytic (maybe divergent) at $t = t^*$. Then, we have the well-known relation:
\bea
\textrm{if}\;\; f_s(t) \sim A\left(1 - \frac{t}{t^*}\right)^{\alpha} && \qquad \textrm{when}\; t \rightarrow t^*  \nonumber \\
\textrm{then}\;\; \mathrm{F}_{0|v} \sim \,\frac{A}{\Gamma(-\alpha)}v^{-(\alpha + 1)}(t^*)^{-v} &&\qquad \textrm{when}\; v \rightarrow \infty
\eea

\subsubsection{Critical points of the $\mathcal{O}(\nn)$ model}

We have to find the singularities of the $\mathrm{W}_k^{(g)}$'s which are the closest to $0$ in order to study observables on large maps. We claim that this singularity is common to $k,g$. Indeed, according to the 1-cut lemma, $\mathrm{W}_1^{(0)}(t)$ has a strictly positive radius $\rho_0(\hat{\mathrm{V}})$ and a singularity at some $t^* \in \mathcal{C}(0\,;\,\rho_0(\hat{\mathrm{V}}))$. Successive applications of the loop insertion operator $\frac{\partial}{\partial \hat{\mathrm{V}}(x)}$ preserve the radius and the singularity, and yield $\mathrm{W}_k^{(0)}$ ($k \geq 1$). We have to argue that the residue formula do not change the radius either, but this granted, $t^*$ is common to all stable $\mathrm{W}_k^{(g)}$'s. The unstable functions, $\mathrm{F}_0$ and $\mathrm{F}_1$ are obtained by an integration formula, which preserves the radius and the singularity at $t^*$.

 In general, when one solves the saddle point equation, one finds $\mathrm{y(x)} \sim 2\eta_i\sqrt{\x - a_i}$ for some $\eta_i \in \mathbb{C}$ when $x \rightarrow a_i$. This means that, in the local parameter on the spectral curve $u \propto \sqrt{\x - a_i}$, $y$ has a simple zero at $u = \mathrm{u}(a_i)$. This property falls down for some exceptional potentials at some value $t = t^*$, and we rather have $\mathrm{y(\x)} \propto (\x - a_i)^{\lambda}$ for $\lambda \geq \frac{1}{2}$. Then, the stable $\mathrm{W}_k^{(g)}$'s expressed by the residue formula diverge when $t \rightarrow t^*$. The intuition can be supported by the Coulomb gas picture and one finds \cite{KS,E1} different kinds of critical points $t^*$.

\begin{itemize}
\item[$\bullet$] \underline{Pure gravity}. $a_i^2(t^*)$ is a zero of order $p$ of $\mathrm{A}(\x^2)$ and $\mathrm{B}(\x^2)$ in Eqn.~\ref{eq:W10}. Then, $\mathrm{y(x)} \propto (\x - a_i)^{p + \frac{1}{2}}$ when $\x \rightarrow a_i$. It happens when the liquid of eigenvalues (in the thermodynamic limit) crosses a potential barrier, i.e. at a point $a_i$ where $\mathrm{y}(a_i) = 0$ and where the effective potential for an eigenvalues, $\mathrm{V}_{\textrm{eff}}(\x) = \frac{2\mathrm{V}'(\x) - \nn \mathrm{V}'(-\x)}{4 - \nn^2}$, safisfies $\mathrm{V}_{\textrm{eff}}^{(l)}(a_i) = 0$ for $1 \leq l \leq p +1$. Then, the model has a well-defined limit for $\x$ close to $a_i$, which does not feel the presence of the interface at $\x = 0$. Back to combinatorics, such a singularity is associated with large maps without macroscopic loops. This limit exists already in the limit of the 1-matrix model and describes pure gravity. In the CFT classification, it is described by the $(p,2)$ minimal model.
\item[$\bullet$] \underline{$a(t^*) = 0$}. The cuts $[-b,-a]$, $[a,b]$ merge. It happens when the eigenvalues can be as close as they want to their mirrors. Then, the method of Section~\ref{sec:spe} is not valid stricto sensu, because the torus is pinched at the point $\x = 0$. The singularity lie in $\mathrm{D}_{\mu}$, and not in the polynomials $A$ and $B$. These critical points are specific to the $\mathcal{O}(\nn)$ model and associated to large maps where macroscopic loops are densely drawn.
\item[$\bullet$] \underline{Combinations of the two situations}. Merging cuts ($a = 0$) and tuning of $V$ such that $\mathrm{A}(\x^2)$ and $\mathrm{B}(\x^2)$ have zeroes of order $m$. These multicritical points are also associated to large maps where loops are macroscopic. Although, they are not dense : we rather have cohabitation of regions dominated by gravity, and regions dominated by macroscopic loops \cite{KS}.
\end{itemize}

We let aside pure gravity, and concentrate ourselves on the critical point $t \rightarrow t^*$ corresponding to $a \rightarrow 0$.

\subsubsection{Taking the limit $a \rightarrow 0$}
\label{sec:bjbj}
When $a \rightarrow 0$:
\beq
K \sim \frac{\pi}{2},\qquad\qquad K' \simeq \ln\left(\frac{4b}{a}\right) \rightarrow \infty,\qquad \tau = \frac{iK}{K'}\nonumber
\eeq
The limit of the functions involved depends on the position of x in the complex plane (of $u$ in the torus). We are first interested in the vicinity of $a$, i.e. $x/a = x^*$ finite. We may use the parametrization:
\beq
\label{eq:paramb}
x^* = \sin\varphi = \mathrm{ch}\,\chi
\eeq
When x is in the first quadrant of the physical sheet, $\sigma(\x) = a^2\cos\varphi$, and the cut is located at $\chi \in [0,\infty]$. To take the limit $\tau \rightarrow 0$ in theta functions, it is convenient to perform the modular transformation $\tau \mapsto \tau' = -1/\tau$. Then, in the sum:
\beq
\vartheta_1(w|\tau') = i\sum_{m \in \mathbb{Z}}(-1)^{m}e^{i\pi\left(m - \frac{1}{2}\right)^2\tau' + i\pi{}(2m - 1)w},\nonumber
\eeq
only one or two terms (depending on the position of $w$ we are interested in) are dominant since $i\pi\tau' \rightarrow -\infty$. For instance, the special solution of Eqn.~\ref{eq:01} we introduced in Eqn.~\ref{eq:Dmudef} becomes:
\beq
\mathrm{D}_{\mu}(u) = \frac{\mathrm{x}(u)}{\sigma(\mathrm{x}(u))}\,\frac{\vartheta_1\left(\frac{\varphi}{2K} - \frac{\mu\tau'}{2}\Big|\tau'\right)}{\vartheta_1\left(\frac{\varphi}{2K}\Big|\tau'\right)}\,\frac{\vartheta_1\left(\frac{\tau'}{2}\Big|\tau'\right)}{\vartheta_1\left((1 - \mu)\frac{\tau'}{2}\Big|\tau'\right)}\,{}\exp\left[i\pi\mu\left(-\frac{\varphi}{2K} + \frac{\tau'}{2}\right)\right] \nonumber
\eeq
where $u = \frac{i\varphi}{2K'} + \frac{\tau}{2}$. Taking the limit, we obtain
\beq
\label{eq:fmul}\mathrm{D}_{\mu}(\x) \sim \left(\frac{a}{4b}\right)^{-\mu}\,\frac{e^{-i\mu\varphi}}{2ib\,\cos\varphi}
\eeq
The function for x arbitrary is given by analytic continuation of the expression above. With similar methods, we can derive (see also \cite{E1,EK1}:
\begin{itemize}
\item[$\bullet$] (see \cite{E1,EK1} or Appendix~\ref{app:Afmu}):
\beq
\label{eq:alpha1}\alpha_1 \sim -i\mu b,\qquad e_{\mu}^2 \sim -\frac{1}{4}\,a^{2(1-\mu)}(4b)^{2\mu} \nonumber
\eeq
The behavior of $e_{\mu}$ can be derived by computing $\mathrm{D}_{\mu}(\x)\mathrm{D}_{\mu}(-\x)$ computed directly with Eqn.~\ref{eq:fmul}, and Eqn.~\ref{eq:relD} in Appendix~\ref{app:Afmu}.
\item[$\bullet$] Basis of $1$-cut solution of Eqn.~\ref{eq:01}.
\bea
\label{eq:fmua0}\frac{\f_{\mu}}{\mathrm{R}_{\mu}} & \sim & \textrm{cte}\, \cos\varphi\cos\mu\left(\varphi - \frac{\pi}{2}\right) = -i\textrm{cte}\,\ch\chi\sh\mu\chi  \\
\label{eq:fmua1}\frac{\widehat{\f}_{\mu}}{\widehat{\mathrm{R}}_{\mu}} & \sim & \widehat{\textrm{cte}}\,\sin\varphi\sin\mu\left(\varphi - \frac{\pi}{2}\right) = i\widehat{\textrm{cte}}\,\ch\chi\sh\mu\chi
\eea
\item[$\bullet$] Basic blocks of the spectral curve (Section~\ref{sec:scurve})
\bea
\mathrm{y}_{\mu} & \sim & \mathrm{cte}\,\sin\varphi\cos\mu\left(\varphi + \frac{\pi}{2}\right) = \mathrm{cte'}\,\ch\chi\ch\mu(\chi + i\pi) \nonumber\\
\hat{\mathrm{y}}_{\mu} & \sim & \hat{\mathrm{cte}}'\,\cos\varphi\sin\mu\left(\varphi + \frac{\pi}{2}\right) = -i\hat{\mathrm{cte}}'\,\sh\chi\sh\mu(\chi + i\pi) \nonumber
\eea
\item[$\bullet$] Brick of the Cauchy kernel (Thm.~\ref{H}).
\beq
\label{eq:Hl}H_+(\chi_0,\chi) \sim e^{-i\pi\mu}\frac{\mathrm{d}\chi_0}{\mathrm{dx}^*(\chi_0)}\,e^{\mu(\chi + \chi_0)}( - 1 - i\mathrm{coth}\,(\chi + \chi_0))
\eeq
\item[$\bullet$] Bergman kernel (Section~\ref{sec:series}).
\beq
\label{eq:Bl}\mathcal{B}(\chi_0,\chi) \sim \mathrm{d}\chi_0\mathrm{d}\chi{}e^{\mu(\chi + \chi_0)}\left[\mu + i\mu\mathrm{coth}(\chi + \chi_0) - \frac{1}{\mathrm{sh}^2(\chi + \chi_0)}\right]
\eeq
\end{itemize}

\subsection{Spectral curve in the scaling limit}

\subsubsection*{Description of the critical point}
\begin{figure}[h!]
\label{fig:exp}
\begin{center}
\includegraphics[width=0.95\textwidth]{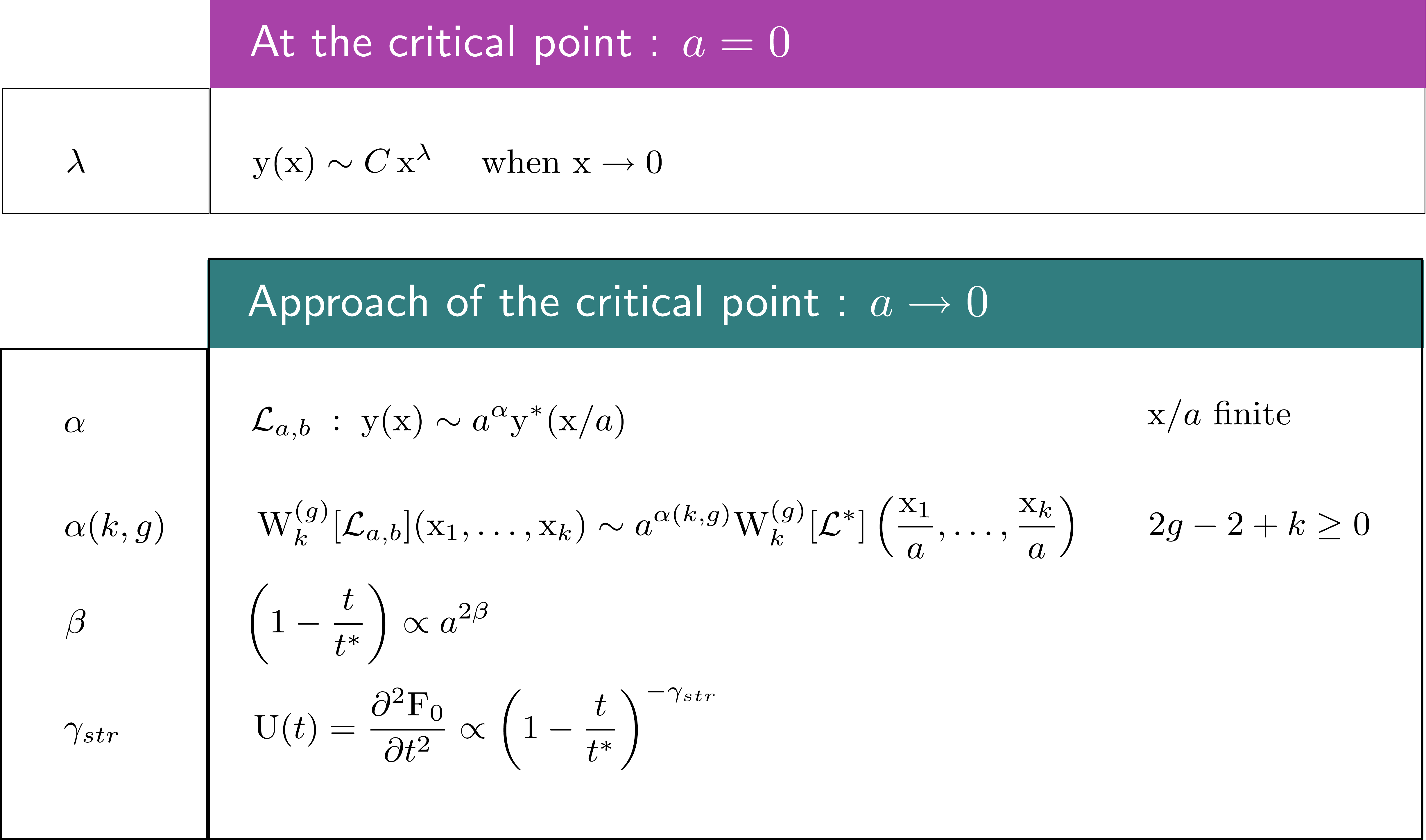}
\end{center}
\end{figure}
We gathered above the definitions of the critical exponents of the bulk $\mathcal{O}(\nn)$ model. $U(t)$ is a fundamental quantity to many aspects. In the correspondence of the critical model with a CFT, the "string susceptibility" $\gamma_{str}$ gives the necessary central charge:
\beq
\textrm{c} = 1 - 6\frac{\gamma_{str}^2}{1 - \gamma_{str}} \nonumber
\eeq
Furthermore, the conformal dimensions (from which the scaling exponents can be extracted) of an operator in a CFT coupled to gravitation ($\widetilde{\Delta}$), or not ($\Delta$), are related by the $\textsc{kpz}$ equation \cite{Liou}:
\beq
\widetilde{\Delta} = \frac{\Delta(\Delta - \gamma_{str})}{1 - \gamma_{str}} \nonumber
\eeq
In other words, this equation relates exponents of a statistical model of a regular lattice to those on the random lattice.

\subsubsection*{Limit curve}
\begin{figure}[h!]
\begin{center}
\includegraphics[width=0.95\textwidth]{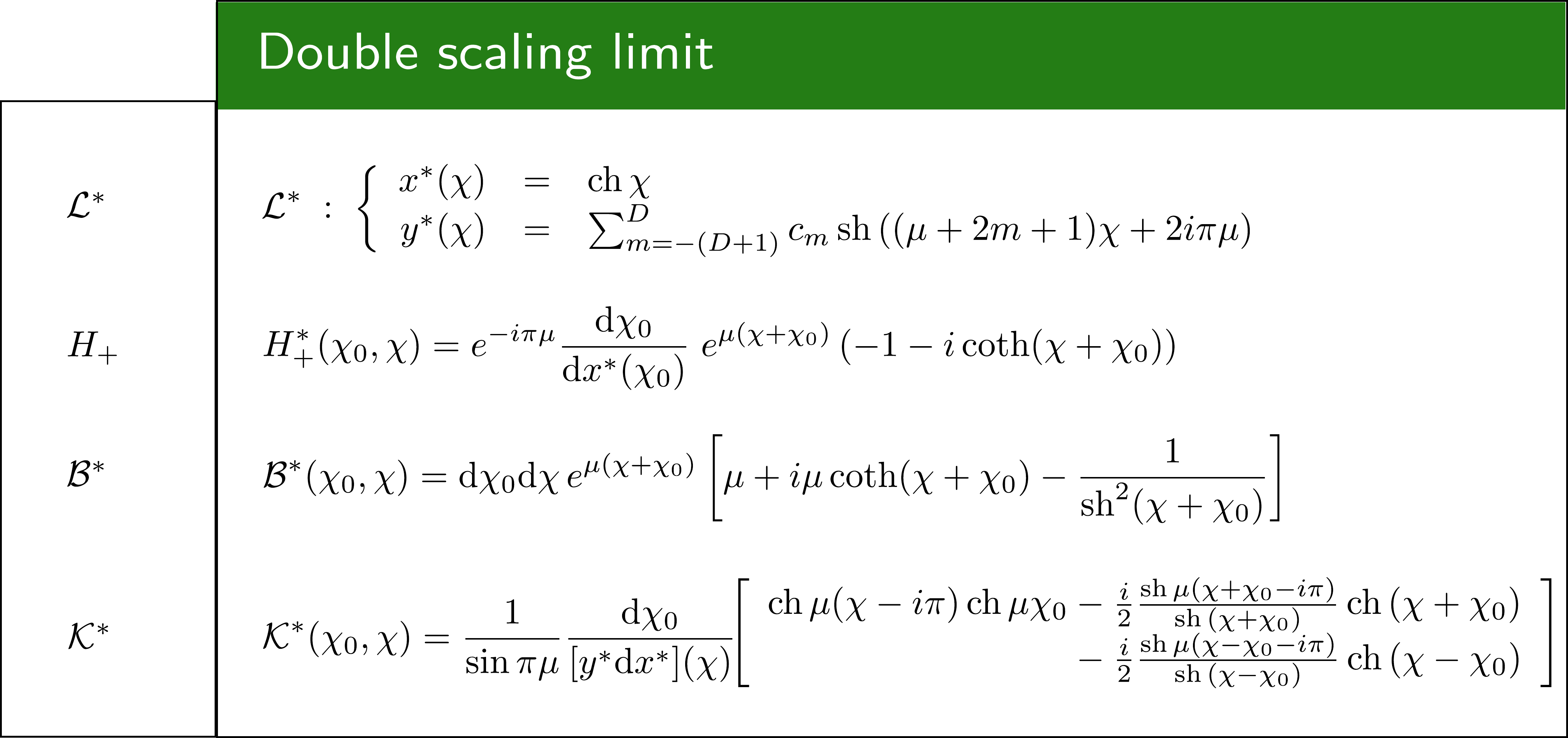}
\caption{\label{fig:limit} Summary on the critical $\mathcal{O}(\nn)$ model.}
\end{center}
\end{figure}

An important feature of the residue formula is its compatibility with limits of curves. So, we ask ourselves what is the spectral curve $\mathcal{L}_{a,b}$ in the limit $a \rightarrow 0$, and we call it $\mathcal{L}^*$. We find, by linear combination of Eqn.~\ref{eq:fmua0} and \ref{eq:fmua1}, that $w_{0}(\chi) = \ch(\mu + 1)\chi$ and $w_{-1}(\chi) = \ch(\mu - 1)\chi$ are solutions. By choosing appropriate polynomials $\mathrm{A}(\ch^2\chi)$ and $\mathrm{B}(\ch^2\chi)$ of maximal degree $D$, which are also linear combinations of $\ch(2m\chi)$ for $0 \leq l \leq D$ integer, we find recursively that the general solution for $w$ is a linear combination of:
\beq
w_m(\chi) = \ch(\mu + 2m + 1)\chi,\qquad\qquad \textrm{for}\; -(D + 1) \leq m \leq D \nonumber
\eeq
The spectral curve associated to a homogeneous part of the resolvent which would be $w_m$ is:
\beq
y_m(\chi) = 2w_m(\chi) + \nn w_m(-(\chi + i\pi)) = -2\sh\left[(\mu + 2m + 1)\chi + 2i\pi\mu\right] \nonumber
\eeq
We obtain:
\begin{theorem}
\label{yl}When $a \rightarrow 0$ and $\x^* = \x/a$ is kept finite, the rescaled limit of the spectral curve is of the form:
\beq
\encadremath{\left\{\begin{array}{l}
x^*(\chi) = \mathrm{ch}\chi \\
y^*(\chi) = \sum_{m = - (D + 1)}^{D} c_m\,\sh\left((\mu + 2m + 1)\chi + 2i\pi\mu\right)\end{array}\right.} \nonumber
\eeq
If the potential V has maximal degree $d_{\textrm{max}}$, then $D = \lfloor(d_{\textrm{max}} - 1)/2\rfloor$ and $c_m \in \mathbb{C}$. The coefficients are subjected to the extra condition $y(\chi = 0) = 0$.
\end{theorem}
Eventually, we gather in Fig.~\ref{fig:limit} the remaining data which initialize the topological recursion in the scaling limit $\x/a$ finite, $a \rightarrow 0$.

\subsection{Topological recursion in the scaling limit}

We see on Eqn.~\ref{eq:Bl} that $\mathcal{B}$ has a well-defined limit, without scaling factor, when $a \rightarrow 0$:
\beq
\mathrm{dx}_0\mathrm{dx}\,\overline{\mathrm{W}}_2^{(0)}(x_0,x) \sim \mathrm{dx}_0^*\mathrm{dx}^*\,(\overline{\mathrm{W}}_2^{(0)})^*(x_0^*,x^*) \nonumber
\eeq
Subsequently, for the recursion kernel:
\beq
\mathrm{K}(\x_0,\x) \sim a^{-(\alpha + 1)}\,\mathrm{K}^*(\x_0^*,\x^*) \nonumber
\eeq
For $g,k$ such that $2g - 2 + k > 0$, $\mathrm{W}_k^{(g)}$ is obtained by a stack of $2g - 2 + k$ residues against a recursion kernel, of a product of $\overline{\mathrm{W}}_2^{(0)}(\cdot,\cdot)$ blocks.
Hence:
\beq
\mathrm{dx}_1\cdots\mathrm{dx}_k\,\mathrm{W}_k^{(g)}(x_1,\ldots,x_k) \sim a^{(\alpha + 1)(2 - 2g - k)}\,\mathrm{dx}_1^*\cdots\mathrm{dx}_k^*\,(\mathrm{W}_k^{(g)})^*(x_1^*,\ldots,x_k^*) \nonumber
\eeq
This yields:
\beq
\encadremath{\alpha(k,g) = (2 - 2g - k)(\alpha + 1) - k}
\eeq
Thus, we have proved the KPZ scaling \cite{KPZ,D,DK,Liou}.

\subsection{Critical exponents}

\begin{figure}
\label{fig:expr}
\begin{center}
\includegraphics[width=0.95\textwidth]{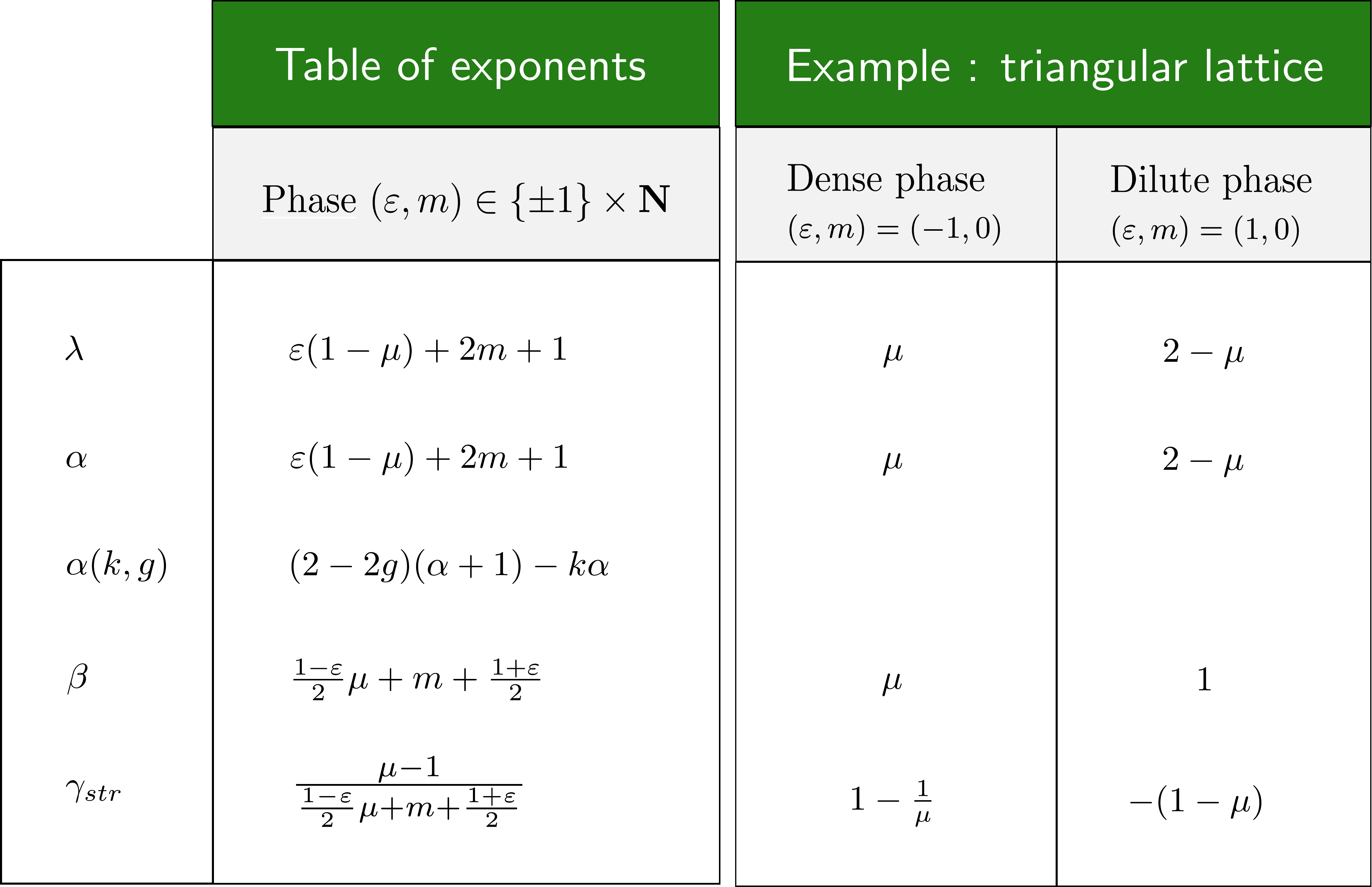}
\caption{Summary on the critical $\mathcal{O}(\nn)$ model.}
\end{center}
\end{figure}

\subsubsection*{Determination of $\lambda$}
\label{sec:lambdas}We have to solve the saddle point equation for $a = 0$ and $\x$ finite. A first method reviewed in Appendix~\ref{sec:dege} consists in a direct guess of a good parametrization or/and of the general solution \cite{E1,EZJ,K}. A second method is to compute directly the limits of theta functions, of $\mathrm{D}_{\mu}$ and then of the basis $(\f_{\mu}/\mathrm{R}_{\mu},\widehat{\f}_{\mu}/\widehat{\mathrm{R}}_{\mu})$, as we did when $\x$ was of order $a$. Of course, the two methods lead to the same results.

\begin{theorem}
\label{ylf}We introduce the parametrization $\x = \frac{b}{\mathrm{ch}\,\zeta}$. Then, the general spectral curve when $a = 0$ and $\x$ is kept finite, is of the form:
\beq
\encadremath{\mathrm{y(\x)} = \mathrm{A}\left(\x^2\right)\mathrm{th}\zeta\,\mathrm{ch}(\mu\zeta) + \mathrm{B}\left(\x^2\right)\mathrm{sh}(\mu\zeta)}
\eeq
where $A$ and $B$ are polynomials. If the potential V has maximal degree $d_{\textrm{max}}$, we have:
\beq
\mathrm{deg}\,(\mathrm{A}) \leq \Big\lfloor\frac{d_{\textrm{max}}}{2}\Big\rfloor - 1,\qquad\qquad \mathrm{deg}\,(\mathrm{B}) \leq \Big\lfloor{\frac{d_{\textrm{max}} - 1}{2}}\Big\rfloor \nonumber
\eeq
\flushright{$\Box$}
\end{theorem}
When $\x \rightarrow 0$, $\x \sim \frac{b}{2}e^{-\zeta}$. Since $\mu\in\,\left.]0,1[\right.$, the two terms ($\mathrm{y}_{\mu}$ and $\hat{\mathrm{y}}_{\mu}$) admit $\x^{-\mu}$ as leading order, and:
\beq
\x^{\mu},\;\x^{- \mu + 2},\;\ldots,\;\x^{(\mu - 1) + 2m + 1},\;\x^{(1 - \mu) + 2m + 1},\;\ldots \nonumber
\eeq
as subleading orders. If we demand\footnote{It is true for all $a \neq 0$ that $\mathrm{y}_a(a) = 0$. Though, we did not find a convincing argument to rule out from the case $\mathrm{y}_{a = 0}(0) = \infty$ for combinatorics. In this case, $\mathrm{y}$ could be divergent as $x^{-\mu}$ but integrable on $[0,b]$ since $\mu \in \left.]0,1[\right.$.} $\lim_{\x \rightarrow  0} \mathrm{y}(\x) = 0$, $A$ and $B$ are such that this leading order disappears. Hence, the first possible term is $x^{\mu}$, and other subleading orders can be canceled by a special choice of $A$ and $B$. Thus, in general, the leading term is $\x^{\lambda}$, with:
\beq
\lambda = \varepsilon(1 - \mu) + 2m + 1,\qquad \varepsilon \in \{\pm 1\},\;m \in \mathbb{N} \nonumber
\eeq
The admissible values for $\varepsilon$ and $m$ depends on $d_{\textrm{max}}$. They determine the "phases" of the $\mathcal{O}(\nn)$ model.
For example :
\begin{itemize}
\item[$\bullet$] On a triangular lattice ($d_{\textrm{max}} = 3$), there exists only two phases, in agreement with \cite{K}:
\bea
\textrm{Dense phase} & \; (\varepsilon,m) = (-1,0) \; & \lambda = \mu \nonumber \\
\textrm{Dilute phase} & \; (\varepsilon,m) = (1,0) \; & \lambda = 2 - \mu \nonumber
\eea
\item[$\bullet$] In the fully packed case ($d_{\textrm{max}} = 2$), only the dense phase $(\varepsilon,m) = (-1,0)$ is present, and the limit spectral curve is (see Appendix~\ref{sec:dege}):
\beq
\mathrm{y}_{\textrm{FPL}}(\x) = \textrm{cte}\,\left(\mathrm{ch}(\mu\zeta) - \mu\,\mathrm{coth}\zeta\,\mathrm{sh}(\mu\zeta)\right)
\eeq
\item[$\bullet$] With the general potential of degree $d_{\max} = 2d' + 1$ ($d' > 1$), there are $2d'$ phases described by $(\varepsilon,m) \in \{\pm 1\}\times[1,d']$.
\item[$\bullet$] With the general potential of degree $d_{\max} = 2d'$, $(\varepsilon,m) = (1,d')$ cannot be reached, so there are $2d' - 1$ phases.
\end{itemize}

\subsubsection*{Determination of $\alpha$}

$\alpha$ is defined such that $\mathrm{y}_a(\x) \sim a^{\alpha}\mathrm{y}^*(\x/a)$ for finite $\x/a$ and $a \rightarrow 0$. On the other hand, $\mathrm{y}_{a = 0}(\x)$ behaves as $\x^{\lambda}$ when $\x \rightarrow 0$. The two solutions match if $\mathrm{y}^*(\x^*) \propto \x^{\alpha'}$ when $\x^* \rightarrow \infty$ with $\alpha = \alpha'$, and $\alpha' = \lambda$. Thms.~\ref{yl} and~\ref{ylf} are compatible in the sense that they produce the same possible values for $\alpha'$ and $\lambda$.

\subsubsection*{Determination of $\beta$}

We have proved that $\left(\partial_t \mathrm{W}_1^{(0)}\right)(\x) = \f_{\mu}(\x)$. Independently, we have in the limit $a \rightarrow 0$:
\bea
\partial_t(\mathrm{W}_1^{(0)})(\x) & \propto & a^{\alpha - 2\beta}\,\times(\textrm{function of}\;\x/a) \qquad \textrm{almost by definition}\nonumber\\
\phantom{\partial_tt}\f_{\mu}(\x) & \sim & a^{-\mu}\,\f^*_{\mu}(\x/a)\phantom{unction of \x/a}\qquad \textrm{from Eqn.~\ref{eq:fmul}} \nonumber
\eea
Hence:
\beq
\encadremath{2\beta = \alpha + \mu} \nonumber
\eeq

\subsubsection*{Determination of $\gamma_{str}$}

We take the limit $a \rightarrow 0$ in Eqn.~\ref{eq:F0} giving the third derivative of $\mathrm{F}_0$ and keep the nonanalytic part at $t = t^*$:
\beq
\left(\frac{\partial^3 \mathrm{F}_0}{\partial t^3}\right)_{\textrm{singular}} \sim \textrm{cte}\,a^{2(1 - \mu) - 2\beta}
\eeq
Hence:
\beq
\encadremath{\gamma_{str} = \frac{\mu - 1}{\beta} = \frac{\mu - 1}{\frac{1 + \varepsilon}{2}\mu + m}}
\eeq
We notice that $\gamma_{str}$ is always negative. Furthermore, the limit of Eqn.~\ref{eq:F0} yields a regular part for $\partial_t^3 \mathrm{F}_0$, i.e which is analytic at $t = t^*$.  This regular part is dominant in $\partial_t^3 \mathrm{F}_0$ when $t \rightarrow t^*$. Also, the same phenomenon occurs for $U(t)$ and $\mathrm{F}_0(t)$.

\subsubsection*{Critical behavior of $\mathrm{F}_1$}

With expression Eqn.~\ref{eq:F110}, it is easy to see that $\mathrm{F}_1 \propto \ln(a)$ when $a \rightarrow 0$.

\subsection{Remark: double scaling limit}
\label{sec:dbllimit}

We have reviewed the fact that, for a given value of $\nn$, the model has many possible continuum limits, described by $(\varepsilon,m)\,\in \{\pm 1\}\times\mathbb{N}$.
We refer to \cite{E1} or \cite{DiFGZJ} for a discussion of the case "$\mu = p/q$ rational" in relation with the $(p,q)$ minimal models of CFT.

For any of these limits, we have seen that the stable $\mathrm{W}_k^{(g)}(t)$ diverge (at least for $g \geq 2$) as $a^{\alpha(k,1) - (2g - 2)(\alpha + 1)}$ when $t \rightarrow t^*$. Take $k \neq 0$ to avoid unstable maps, and consider the formal power series $\mathrm{W}_k(t)$, depending on $N$:
\bea
\mathrm{W}_k(t|N) & = & \sum_{g \geq 0} \left(\frac{N}{t}\right)^{2 - 2g}\,\mathrm{W}_k^{(g)}(t) \nonumber \\
& = & a^{-\alpha(k,1)}\sum_{g \geq 0} \left(\frac{Na^{-(\alpha + 1)}}{t}\right)^{2 - 2g}\,a^{\alpha(k,g)}\mathrm{W}_k^{(g)}(t) \nonumber
\eea
We can also define a function:
\beq
\mathrm{W}_k^*(N^*) = \sum_{g \neq 0} (N^*)^{2 - 2g}\mathrm{W}_k^{(g)}{}^* \nonumber
\eeq
If we send $a \rightarrow 0$ and $N \rightarrow \infty$ while keeping $Na^{-(\alpha + 1)} = N^*$ finite, then:
\beq
\mathrm{W}_k(t|N) \sim \mathrm{W}_k^*(N^*) \nonumber
\eeq
For this reason, $\mathrm{W}_k^*$, and by extension $\mathrm{W}_k^{(g)}{}^*$, are called in matrix model context the "double scaling limit" of $\mathrm{W}_k$, resp. $\mathrm{W}_k^{(g)}$. Within the conjecture relating limits of matrix models to CFT (see Section~\ref{sec:CFTr}), $\mathrm{W}_k^*$ should be solution of PDE's coming from the conformal field theory of central charge:
\beq
\mathrm{c} =  1 - 6\left(\sqrt{\lambda} - \frac{1}{\sqrt{\lambda}}\right) \nonumber
\eeq
Still, it demands a mathematical proof.

\section{Conclusion}

We have defined for the first time the topological recursion for a spectral curve which has monodromies around branchpoints, and extended to that case all its usual properties. The bulk correlation functions $\mathrm{W}_k^{(g)}$ in the $\mathcal{O}(\mathfrak{n})$ matrix model fit in the formalism. So, we have an algorithm to compute number of maps with self avoiding loops of $\nn$ possible colors in any topology and with an arbitrary number of boundaries. In particular, the generating function of cylinders $\mathrm{W}_2^{(0)}$ is universal (as expected) and closely related to a deformation $\wp_{\mu}$ of the Weierstra\ss{} $\wp$ function. We also derived new formulas for $\mathrm{F}_0$ and $\mathrm{F}_1$. Our method consisted in finding a good description of the ring of meromorphic functions on the spectral curve. The theory of elliptic functions can generalized to functions periodic in $1$ direction, and taking a phase in the other direction. $\wp_{\mu}$ give rise to a Bergman kernel (fundamental bi-differential with a double pole at coinciding points), which can be used to express all meromorphic forms on the spectral curve.

The generalization to the multi-cut solutions of the $\mathcal{O}(\mathfrak{n})$ model loop equations seems straightforward: in the residue formula expressing $\mathrm{W}_k^{(g)}$, one just has to pick up a residue at every endpoint of a cut. Some work remains however. We must first express the general multicut solution to the master loop equation (Thm.~\ref{eq:MLP}), and find the appropriate Bergman kernel. We will return to this point in the future. We stress that for combinatorics of maps, the $1$-cut solution that we presented is enough.

The next step is to compute all correlations functions with residue formulas, in the spirit of what was done for the $2$-matrix model \cite{EO2MM}. It would be very interesting to explore the integrability of the model before the continuum limit, and then, the relations with boundary conformal field theory. Since we can compute a lot of observables in the discrete model, we think that many exact results for the continuum limit could be obtained by this method. For instance, we have been able to give a rigourous proof of the KPZ scaling for all topologies.

On the flat lattice, there exists a relation between the $\mathcal{O}(\nn)$ model, and the Potts model with $\mathfrak{q} = \mathfrak{n}^2$ colors. Namely, the critical point of the $\mathfrak{n}^2$ Potts model on a square lattice is equivalent to the critical point of the $\mathcal{O}(\nn)$ model on a square-triangular lattice in the dense phase. On the random lattice, there is again a relation between the two models. They both admit a matrix model representation, and the planar resolvents $\mathrm{W}_1^{(0)}(\x)$ of these two matrix models are (up to some simple change of variable) reciprocal functions \cite{BEP}. It would be interesting to see if a topological recursion could solve the $\mathfrak{q}$-Potts matrix model for all topologies, and then, if a relation with the $\mathcal{O}(\nn)$ model can be extended to all topologies. The Potts model has been up to now more popular in combinatorics and statistical physics than the $\mathcal{O}(\nn)$ model. Nevertheless, supported by the present work, we think that the $\mathcal{O}(\nn)$ on the random lattice is easier to solve, and that this solution might shed light on the Potts model on random lattice (outside and at the critical point).

We also expect that, for many other models for which the leading order resolvent $W_1^{(0)}$ is known, like the $\widehat{A},\widehat{D},\widehat{E}$ models of \cite{KADE}, or the $ABAB$ model considered in \cite{PZJ}, a topological recursion (maybe slightly deformed) can be found to find all genus correlation functions. We are currently working on this problem.

\medskip

\subsection*{Acknowledgments}

We thank I.~Kostov and J.E.~Bourgine for valuable discussions and indication of references, and also J.~Bouttier and N.~Curien.

\newpage

\setcounter{section}{0}

\appendix

\section{Proof of the 1-cut lemma}
\label{app:1cut}

\begin{lemma}
For all $\nn$, $k$, $g$, there exists $\rho > 0$ such that \mbox{$\#
\mathbb{M}_k^{(g)}(v) \in O(\rho^v)$} when $v \rightarrow \infty$.
\end{lemma}

\begin{figure}[h!]
\label{fig:Techlemma}
\begin{center}
\includegraphics[width=\textwidth]{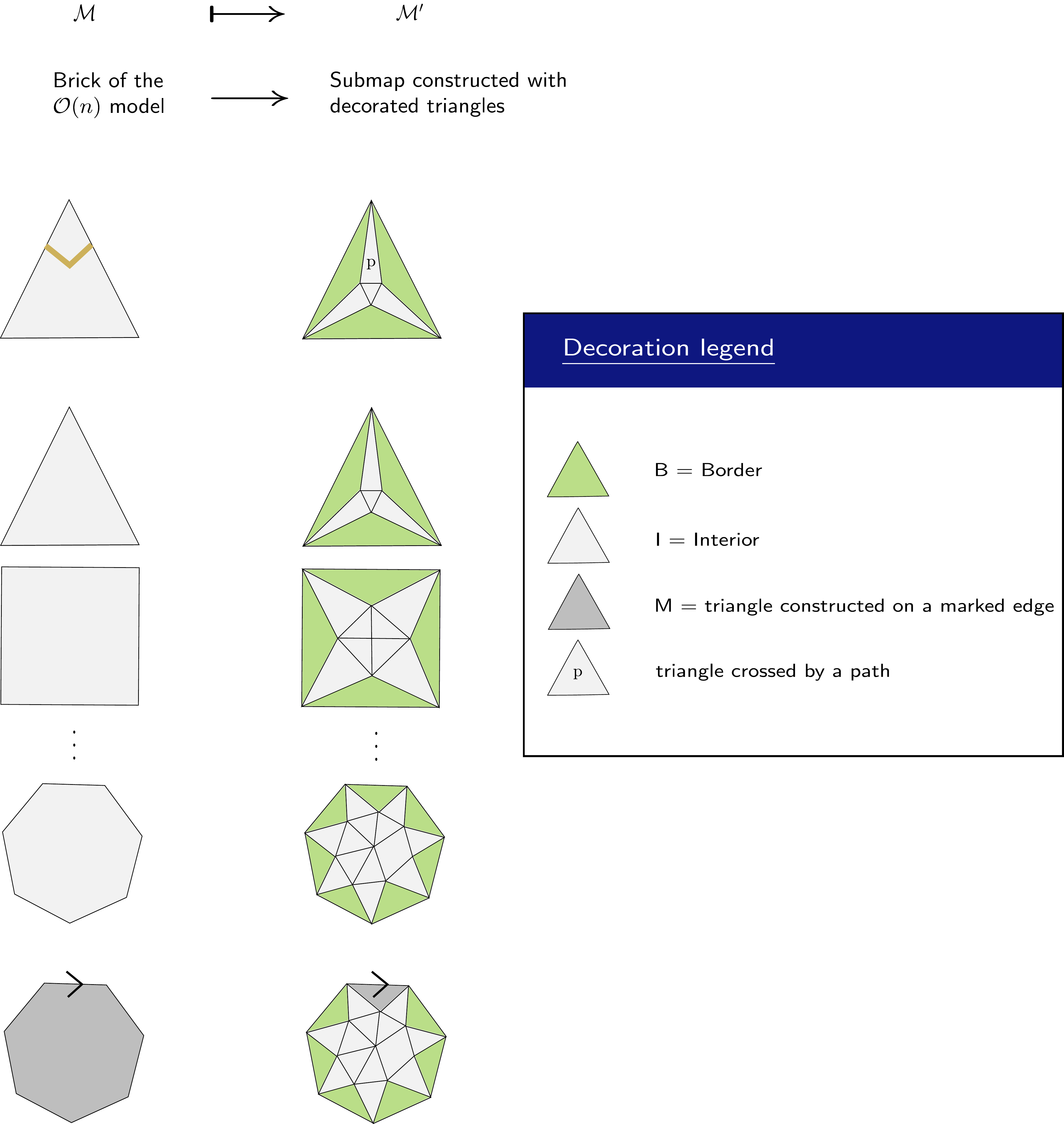}
\caption{Every $j$-gon (with or without loops, marked or unmarked) is itself triangulated with $3j$
decorated triangles: construct a triangle on each edge, and a triangle at each corner. In the inside, there
remains a $j$-gon which you cut in $j$ triangular
sectors from its center, that you label "I". The other labels are distributed as follows: for a triangle carrying a loop, the corner triangle where the loop is drawn is decorated by the color of the loop \mbox{$j \in \{1,\ldots,\nn\}$} ; for a marked face, the triangle constructed on the marked edge is decorated by "M" ; in any other case, the triangles constructed on the edges are decorated by "B", while the other are labelled "I".}
\end{center}
\end{figure}

\proof{This is based on a rough upper bound on the cardinality of
${\mathbb M}_k^{(g)}$. If $\nn = 0$ and $t_j = 0$ for $j \geq 4$,
\mbox{$\mathbb{M}_k^{(g)}(v)$} are sets of usual triangulations, and
it is known \cite{Tutte1} that the lemma is true with
\mbox{$\overline{\rho} = \frac{1}{6\sqrt{3}}$}. We assume $\nn$ an
integer, for it is enough to prove the lemma for $\lceil\nn\rceil$.
Now, we describe in Fig.~\ref{fig:Techlemma} a map from
\mbox{$\mathbb{M}_k^{(g)}(v)$} to the set of triangulations with $v$
vertices where every face carries a label among
\mbox{$\{\textrm{B}, \textrm{I}, \textrm{M},1,\ldots,\nn\}$}.

This association is injective:
\begin{itemize}
\item[$\bullet$] If a decorated triangulation comes from \mbox{$\mathcal{M}
\in \mathbb{M}_k^{(g)}(v)$}, the unmarked edges in $\mathcal{M}$ are those
which glue two "B"-triangles, while the marked edges glue a "B" and a "M" triangle. So, we have the
skeleton of $\mathcal{M}$.
\item[$\bullet$] The polygons of this skeleton admitting an inner triangle labeled \mbox{$j
\in \{1,\ldots,\nn\}$} instead of "I" are necessarily triangles.
They must carry a loop, and the $j$-label indicates where it should
be drawn.
\end{itemize}

By construction:
\beq v(\mathcal{M}') \leq (d_{\textrm{max}} +
1)v(\mathcal{M}),\qquad g(\mathcal{M}) = g(\mathcal{M'}),\qquad k(\mathcal{M}) =
k(\mathcal{M}')\nonumber
\eeq

Hence: \beq \# \mathbb{M}_k^{(g)}(v) \in
O\left(\overline{\rho}^{(\nn + 3)(d_{\textrm{max}}+ 1)v}\right)
\nonumber \eeq}

 We proceed in detail with the proof of the 1-cut lemma, which is now the same as for usual maps and the 1-matrix model.
 We recall that the notion of the color of a loop does not intervene in ${\mathbb M}_k^{(g)}$, but only in the weight of a map: the previous upper bound on \mbox{$\# {\mathbb M}_k^{(g)}$} is valid uniformly for $\nn$ bounded. Moreover, we notice that the association of Fig.~\ref{fig:Techlemma} preserves the number of automorphism. We conclude that the coefficient of $t^v$ is a $O(\rho_0/r)$ where
 \bea
 \rho_0 & = & \max(t_j^{\frac{2}{j - 2}},\nn c^2) \nonumber\\
 r & = & \min_{i}\sqrt{\left|\x_i - \frac{c}{2}\right|} \nonumber
 \eea
which proves the first part of the 1-cut lemma.

\newpage

\medskip

\newpage

Next, we solve the master loop equation in $\mathrm{W}_1^{(0)}(\x)$ which is quadratic:
\bea
\mathrm{W}_1^{(0)}(\x) & = & \frac{1}{2}\left(\mathrm{V}'(\x) - \nn{}\mathrm{W}_1^{(0)}(-\x) + \sqrt{\Delta(\x)}\right) \nonumber \\
\label{eq:199} \Delta(\x) & = & \left(\mathrm{V}'(\x) - \nn{}\mathrm{W}_1^{(0)}(-\x)\right)^2 \nonumber \\
 & & - 4\left(\mathrm{P}_1^{(0)}(\x) + \mathrm{P}_1^{(0)}(-\x) - \mathrm{V}'(-\x)\mathrm{W}_1^{(0)}(-\x) + \mathrm{W}_1^{(0)}(-\x)^2\right) \nonumber
\eea
We have to describe the zeroes of $\Delta$ for $t$ small enough. There exists $r$ small enough such that the discs of centers $\pm \frac{c}{2}$ and radius $r$ do not intersect and such that $\mathrm{V}'$ has no other zero than $\frac{c}{2}$ in $\mathcal{D}(\frac{c}{2}\,;\,r)$. Then, for $t \in \mathcal{D}(0\,;\,\frac{\rho_0}{r})$, we see that $\Delta(\x)$ in holomorphic on $\mathcal{D}(\frac{c}{2}\,;\,r)$. For $t = 0$, $\Delta(\x) = (\mathrm{V}'(\x))^2$ is not the zero function and has a double zero in $c$. By continuity in $t$ and analyticity for $\x \in \mathcal{D}(\frac{c}{2}\,;\,r)$, $\Delta$ has still two zeroes (with multiplicity), which limit is $\frac{c}{2}$ when $t \rightarrow 0$. Thus, there exists an holomorphic function $h_t$ in $\mathcal{D}(\frac{c}{2}\,;\,r)$, without zeroes, and a continuous pair $\{a(t),b(t)\}$ such that $\Delta(\x) = \mathrm{h}_t^2(\x)(\x - a(t))(\x - b(t))$.

Computing $\Delta(a_i(t))$ for a zero $a_i(t)$ of $\Delta$ from Eqn.~\ref{eq:199} allows us to find $a_i(t)$ as a formal series in $\sqrt{t}$, and yields two solutions $a(t)$ and $b(t)$, such that $a(t) + b(t) \in O(t)$. The property of having a finite radius of convergence in $t$ for $\mathrm{W}_1^{(0)}$ implies a finite radius of convergence for $a$ and $b$.

Eventually, the analytical properties of $\mathrm{W}_k^{(g)}$ are derived by a straightforward recursion using the loop equations and Lemma~\ref{lemma:fond}.

\section{Elliptic functions}\label{app:ell}

\subsubsection*{Basic facts} Let us say a few words about biperiodic functions, i.e. satisfying:
\beq
f(w)=f(w + 1)=f(w + \tau)
\eeq
where $(1,\tau)$ is $\mathbb{R}$-free. Such functions are also called "elliptic functions" \cite{GW}. They satisfy a few properties:
\begin{itemize}
\item[$\bullet$] If $f$ has no pole, then $f$ is constant.
\item[$\bullet$] The number of zeroes of $f$ in the rectangle $(0,1,1+\tau,\tau)$ equals the number of its poles.
\item[$\bullet$] The sum of residues of $f$ at its poles in the rectangle, vanishes. In particular, this shows that $f$ cannot have only one simple pole, it must have at least two simple poles, or one multiple pole.
\end{itemize}

\subsubsection*{$\vartheta_1$ function}

Let $\tau \in \mathbb{C}$ such that $\textrm{Im} \tau > 0$. The first Jacobi
theta function is defined by:
\beq
\vartheta_1(w|\tau) = i\sum_{m \in \mathbb{Z}}(-1)^{m}e^{i\pi\left(m - \frac{1}{2}\right)^2\tau + i\pi{}(2m - 1)w}
\eeq

This series is absolutely convergent for all $w\in\mathbb C$, so
$\vartheta_1\left(\cdot|\tau\right)$ is an entire function. It satisfies:
\beq \vartheta_1(w|\tau) = -\vartheta_1(w + 1|\tau),\qquad\qquad \vartheta_1(w +
\tau|\tau) = \e^{-2i\pi\left(w+\frac{\tau}{2}\right)} \eeq
and:
\beq
\vartheta_1(w|\tau) = -\vartheta_1(-w|\tau)
\eeq
It has a unique zero $\textrm{mod}\,(\mathbb{Z}\oplus\tau\mathbb{Z})$, located at 0. Besides:
\beq
(\vartheta_1)'(1/2\left|\tau\right.) = 0
\eeq

\subsubsection*{Description of the meromorphic functions on the torus}

An elliptical function $f$ with poles $w_1,\ldots,w_L$
and poles $w'_1,\ldots{},w'_{L'}$, can always be
written\footnote{Indeed, the ratio of $f$ and this expression would
be an elliptical function with no pole, i.e  a constant.} as:
\beq
f(w) = A\frac{\prod_{l' = 1}^{L'} \vartheta_1(w - w_l|\tau)}{
\prod_{l = 1}^L \vartheta_1(w - w'_{l'}|\tau)}
\eeq
for some constant $A$, and the poles and zeroes are such that:
\beq
\sum_{l'} w'_{l'} - \sum_l w_l = 0 \quad {\rm mod}\,\, \mathbb{Z}\oplus\tau\mathbb{Z}
\eeq

\subsubsection*{Inversion of modulus}
There is a relation between $\vartheta_1(\cdot|\tau)$ and
$\vartheta_1(\cdot|\tau')$, where $\tau' = -1/\tau$, which is an
application of Poisson's summation formula:
\beq
\vartheta_1(w|\tau) =
\frac{-1}{\sqrt{i\tau}}e^{\frac{-i\pi{}w^2}{\tau}}\vartheta_1\left(\frac{w}{\tau}\Big|-\frac{1}{\tau}\right)
\eeq

\subsubsection*{Weierstra\ss{} function}
Let us introduce the Weierstra\ss{} $\wp$ function:
\bea
\wp(w|\tau) & = & \frac{1}{w^2} + \sum_{(l,m) \in \mathbb{Z}^2\setminus\{(0,0)\}}\left( \frac{1}{(w + l + m\tau)^2} - \frac{1}{(l + m\tau)^2}\right) \nonumber \\
& = & \sum_{m \in \mathbb{Z}} \frac{\pi^2}{\sin^2\pi(w + m\tau)} - \sum_{m \in \mathbb{Z}\setminus\{0\}} \frac{\pi^2}{\sin^2\pi{}m\tau} - \frac{\pi^2}{3}
\eea
This function is elliptic, and related to the second derivative of $\ln\vartheta_1$:
\beq
\label{eq:wpt}\wp(w|\tau) = - (\ln\vartheta_1)''(w|\tau) + \tilde{c}_0
\eeq
for some constant $\tilde{c}_0$ depending on $\tau$. It satisfies the differential equations:
\bea
\label{eq:fhi1}\wp'^2 & = & 4\wp^3 - g_2\wp - g_3 \\
\label{eq:fhi2}\wp'' & = & 6\wp^2 - \frac{g_2}{2}
\eea
where $g_2$ and $g_3$ are two constants depending on $\tau$.

\section{Properties of the $\mu$-deformed Weierstra\ss{}  function}
\label{app:Weier}
\subsection*{Definition and basic facts}
We have defined:
\beq
\wp_{\mu}(w) = \sum_{m \in \mathbb{Z}}e^{-i\pi(1 - \mu)m}\,\frac{\pi^2}{\sin^2\pi(w + m\tau)}
\eeq
We shall derive the differential equations it satisfies for \mbox{$\mu \neq 1 \mod 2\mathbb{Z}$}. Let us recall two essential facts about functions which, like $\wp_{\mu}$, belong to \mbox{$\mathrm{Ker}(\mathbf{T}_1 - \mathrm{id})\cap\mathrm{Ker}(\mathbf{T} - e^{i\pi\nu}\mathrm{id})$}.
\begin{itemize}
\item[$(a)$] Any quotient of them is elliptic, and any elliptic function is a rational function of $\wp$ and $\wp'$, modulo \mbox{$\wp'^2 = (4\wp^3 - g_2\wp - g_3)$}.
\item[$(b)$] Any such function has at least a pole and a zero (just represent it by the accurate ratio and products of theta functions).
\end{itemize}
$(a)$ could be used directly to find $\wp_{\mu}'/\wp_{\mu}$ and $\wp_{\mu}''/\wp_{\mu}$ by studying their zeroes and poles. We found easier to take another route.

\subsection*{First order equation}
For $\mu = 1$, $\wp_{\mu}$ reduces to the Weierstrass function up to a constant depending on $\tau$:
\beq
\wp_{1}(w) = \wp(w) + c_0,\qquad c_0 = \frac{\pi^2}{3} + \sum_{m \in \mathbb{Z}\setminus\{0\}} \frac{\pi^2}{\sin^2\pi m\tau}
\eeq
From Eqns.~\ref{eq:fhi1}-\ref{eq:fhi2}:
\bea
\label{eq:111}(\wp_1')^2 & = & 4\wp_1^3 - 12c_0\wp_1^2 + (12c_0^2- g_2)\wp_1 - (g_3 + 4c_0^3) \\
\label{eq:110}\wp''_1 & = & 6\wp_1^2 -12c_0\wp_1 + 6c_0^2 - \frac{g_2}{2}
\eea
where $g_2$, $g_3$ are constants depending on $\tau$.  In the limit $\mu \rightarrow 1$, we ought to recover these equations.
Yet, the differential equation involving only $\wp_{\mu}$ that we look for must be linear. We guess that the correct generalization of $(\wp_1'(w))^2$ when $\mu \neq 0$ is $\wp'_1(w)\wp'_{\mu}(w)$, and we try to match its polar part when $w \rightarrow 0$ with a generalization of the \textsc{rhs} of Eqn.~\ref{eq:111}. $\wp_1$ is even and behaves as:
\beq
\wp_1(w) = \frac{1}{w^2} + c_0 + \frac{1}{2}c_2w^2 + O(w^4)
\eeq
where it turns out (from Eqn.~\ref{eq:111}) that $c_2 = g_2/10$. On the other hand, $\wp_{\mu}$ is not even. So, there exists constants $c_j^{(\mu)}$ ($0 \leq j \leq 3$) such that:
\beq
\wp_{\mu}(w) = \frac{1}{w^2} + c_0^{(\mu)} + c_1^{(\mu)} w + \frac{1}{2}c_2^{(\mu)} w^2 + \frac{1}{6}c_3^{(\mu)}w^3 + O(w^4)
\eeq
We obtain:
\beq
\wp'(w)\wp_{\mu}'(w) = \frac{4}{w^6} - \frac{2c_1^{(\mu)}}{w^3} - \frac{2c_2^{(\mu)} + 2c_2}{w^2} + \frac{c_3^{(\mu)}}{w} + O(1)
\eeq
The simplest candidate in \mbox{$\mathrm{Ker}(\mathbf{T}_1 - \mathrm{id})\cap\mathrm{Ker}(\mathbf{T} - e^{i\pi\nu}\mathrm{id})$} to match this polar part is a linear combination of
 \beq
 \wp_1^2\wp_{\mu},\quad\wp_1\wp_{\mu},\quad\wp_{\mu},\quad\wp_1'\wp_{\mu},\quad\wp_1\wp'_{\mu}\quad\textrm{and}\quad\wp'_{\mu}.
 \eeq
The polar part of these six functions are independent, so they are enough to fix the terms up to $w^{-6}$. As an application of $(b)$, this linear combination must be equal to $\wp'_1\wp'_{\mu}$. We give directly the result:
\beq
\label{eq:112}\wp_1{}'\wp_{\mu}' = 4\wp_1^2\wp_{\mu} - 4(2c_0 + c_0^{(\mu)})\wp_1\wp_{\mu} + \left[4c_0^2 + c_0c_0^{(\mu)} + (c_0^{(\mu)})^2 -  g_2^{(\mu)}\right]\wp_{\mu} + \lambda q_{\mu}
\eeq
where $q_{\mu}$ is the function defined by:
\beq
q_{\mu}(w) = \left[\wp_1(w)\wp_{\mu}'(w) - \wp_1'(w)\wp_{\mu}(w) + (c_0^{(\mu)} - c_0)\wp_{\mu}'(w) - 3c_1^{(\mu)}\wp_{\mu}(w)\right]
\eeq
and:
\bea
g_2^{(\mu)} & \stackrel{{\rm def}}{=} & 6c_2 + 4c_2^{(\mu)} \\
\lambda & \stackrel{{\rm def}}{=} & \frac{-\frac{5}{6}c_3^{(\mu)} + (2c_0 + c_0^{(\mu)})c_1^{(\mu)}}{c_2^{(\mu)} - c_2}
\eea

\subsection*{Second order equation}
Similarly, we can match the polar part when $w \rightarrow 0$ of $\wp''_{\mu}$ by a linear combination of
\beq
\wp_1\wp_{\mu},\quad\wp_{\mu},\quad\wp_1'\wp_{\mu},\quad\wp_1\wp'_{\mu}\quad\textrm{and}\quad\wp'_{\mu}.
\eeq
The result is:
\beq
\label{eq:115}\wp''_{\mu} = 6\wp_1\wp_{\mu} - 6(c_0 + c_0^{(\mu)})\wp_{\mu} + \lambda' q_{\mu}(w)
\eeq
where:
\beq
\lambda' = \frac{-3c_1^{(\mu)}}{c_2^{(\mu)} - c_2}
\eeq

\subsection*{Consistency of the limit $\mu \rightarrow 1$}
Let us consider the limit $\mu \rightarrow 1$ of these differential equations.
$c_{j}^{(\mu)}$ (for $0 \leq j \leq 3$) is equal to $c_j$ (which is zero when $j$ is odd) in the limit $\mu \rightarrow 1$ since we can commute the limit and the residue in:
\beq
c_j^{(\mu)} = \Res_{w \rightarrow 0} \frac{\mathrm{d}w}{w^{l + 1}}\,\wp_{\mu}(w)
\eeq
Comparing to Eqn.~\ref{eq:112} to \ref{eq:111}, and  Eqn.~\ref{eq:115} to \ref{eq:110}, we find:
\bea
\forall w \in \mathbb{C} && \lim_{\mu \rightarrow 1} \lambda q_{\mu}(w) = -(4c_0^3 + g_3) \nonumber \\
&& \lim_{\mu \rightarrow 1} \lambda' q_{\mu}(w) = 6c_0^2 - \frac{g_2}{2}
\eea
But one can compute independently:
\beq
\lambda'q_{\mu}(w) = -\frac{6c_1^{(\mu)}}{w} + \lambda'\left(\frac{5}{6}c_3^{(\mu)} - 2c_1^{(\mu)}c_0^{(\mu)}\right) + O(w)
\eeq
Hence $\lambda$ and $\lambda'$ diverge when $\mu \rightarrow 1$ in a way such that:
\beq
\lim_{\mu \rightarrow 1} \frac{\lambda}{\lambda'} = \frac{g_3 + 4c_0^2}{\frac{g_2}{2} - 6c_0^3}
\eeq

\subsection*{Primitive of $\wp_{\mu}$}

$\wp_{\mu}$ can be integrated in terms of theta functions. We only state the formula:
\beq
\int_{w_1}^{w_2} \mathrm{d}w\,\wp_{\mu}(w) = \pi\frac{\vartheta_1'(0\,|\,\tau)}{\vartheta_1(\nu/2\,|\,\tau)}\left(\frac{\vartheta_1(\nu/2 - w_1\,|\,\tau)}{\vartheta_1(w_1\,|\,\tau)} - \frac{\vartheta_1(\nu/2 - w_2\,|\,\tau)}{\vartheta_1(w_2\,|\,\tau)}\right)
\eeq
From this point, one can also obtain theta function representations for $c_j^{(\mu)}$. For example:
\bea
c_0^{(\mu)} & = & \frac{1}{6}\frac{\vartheta_1'''(0\,|\,\tau)}{\vartheta_1'(0\,|\,\tau)} - \frac{1}{2}\frac{\vartheta_1''(\nu/2\,|\,\tau)}{\vartheta_1(\nu/2\,|\,\tau)} \\
c_1^{(\mu)} & = & \frac{1}{6}\left(\frac{\vartheta_1'''(0\,|\,\tau)}{\vartheta_1'(0\,|\,\tau)}\frac{\vartheta_1'(\nu/2\,|\,\tau)}{\vartheta_1(\nu/2\,|\,\tau)} - \frac{\vartheta_1'''(\nu/2\,|\,\tau)}{\vartheta_1(\nu/2\,|\,\tau)}\right)
\eea

\section{Properties of the special solutions of Eqn.~\ref{eq:01}}
\label{app:Afmu}
\begin{itemize}
\item[$\bullet$] \underline{Theta expression for $D_{\mu}$.}
\beq
\label{eq:Dtau2}D_{\mu}(u) = \frac{x(u)}{\sigma(x(u))}\,\frac{\vartheta_1\left(u - \frac{\tau}{2} + \frac{\mu}{2}\Big|\tau\right)}{\vartheta_1\left(u - \frac{\tau}{2}\left|\tau\right.\right)}\,\frac{\vartheta_1\left(-\frac{1}{2}\left|\tau\right.\right)}{\vartheta_1\left(-\frac{1}{2} + \frac{\mu}{2}\Big|\tau\right)}
\eeq
$u_e = \frac{\tau - \mu}{2}$ is the second zero (mod
$\mathbb{Z}\oplus\tau\mathbb{Z}$) of $D_{\mu}$. We call $e_{\mu} = x(u_e)$,
the corresponding point in the physical x-sheet:
\beq
\label{eq:emu}e_{\mu} =
a\,\textrm{sn}_k(i\mu K')
\eeq
We use the notation $\mathrm{D}_{\mu}(\x) = D_{\mu}(\mathrm{u(x)})$. \\

\item[$\bullet$] \underline{Relation between $D_{\mu}$ and $D_{-\mu}$}. $D^{(2)}: u \mapsto D_{\mu}(u)D_{\mu}(\tau - u)$ is an $u$-even and biperiodic function. Hence $\mathrm{D}^{(2)}_{\mu}(\x) = D^{(2)}_{\mu}(\mathrm{u(x)})$ is a rational fraction of x$^2$. In this variable, it has simple poles at $\x^2 = a^2$ and $\x^2 = b^2$, simple zeroes at $\x^2 = \infty$, $\x^2 = e_{\mu}^2$, and is equivalent to $- 1/\x^2$ when $\x^2 \rightarrow \infty$. Thus:
    \beq
    \label{eq:relD}\mathrm{D}^{(2)}_{\mu}(\x) = - \frac{\x^2 - e_{\mu}^2}{(\x^2 - a^2)(\x^2 - b^2)}
    \eeq
\item[$\bullet$] \underline{Asymptotic of $\mathrm{D}_{\mu}(\x)$}. We define the constants $\alpha_1$ and $\alpha_2$ by:
\beq \mathrm{D}_{\mu}(\x) = \frac{1}{\x} + \frac{\alpha_1}{\x^2} +
\frac{\alpha_2}{\x^3} + O\left(\frac{1}{\x^4}\right) \quad\quad
\textrm{when}\;\x \rightarrow \infty\;\textrm{in the physical sheet}
\eeq Studying Eqn.~\ref{eq:Dtau2} when \mbox{$u \rightarrow \frac{- 1 + \tau}{2}$} yields the expression: \beq \alpha_1 =
\frac{-ib}{2K'}(\ln\vartheta_1)'\left(\frac{1- \mu}{2}\Big|\tau\right). \eeq
Studying Eqn.~\ref{eq:relD} when $x \rightarrow \infty$ yields the expression of $\alpha_2$:
\beq
\label{eq:alpha2}\alpha_2 = \frac{1}{2}(\alpha_1^2 + a^2 + b^2 - e_{\mu}^2)
\eeq
The terms up to $\alpha_2$ are involved in the fully packed case.
\\
\item[$\bullet$] \underline{Derivative of $\mathrm{D}_{\mu}$}. $D_{\mu}$ and $(\partial_u D_{\mu})$ are both in $\mathrm{Ker}\left(\mathbf{T} - e^{i\pi\mu}\right)$, so $\partial_u \ln D_{\mu}$ is 1- and $\tau$- translation invariant. By studying its analytical properties, we obtain:
    \beq\label{eq:derGmu}
    \frac{\mathrm{d}\mathrm{D}_{\mu}}{\mathrm{dx}} = \mathrm{L}(\x)\mathrm{D}_{\mu}(\x)
    \eeq
    where:
    \bea
    \mathrm{L}(\x) & = & \frac{-\alpha_1 + \mathrm{I}(\x)}{\sigma(\x)} - \frac{\mathrm{d}\ln\sigma}{\mathrm{dx}} \\
    \textrm{and}\; \mathrm{I}(\x) & = & \frac{\x\sigma(\x) + e_{\mu}\sigma(e_{\mu})}{\x^2 - e_{\mu}^2}
    \eea
    \\
\item[$\bullet$] \underline{Definition of $f_{\mu}$ and $\widehat{f}_{\mu}$}.
\bea
f_{\mu}(u) & = & \frac{D_{\mu}(u) + D_{\mu}(- u)}{1 - e^{i\pi\mu}} \\
\label{eq:fmu2}\widehat{f}_{\mu}(u) & = & \frac{\sigma(x(u))}{x(u)}\,\frac{D_{\mu}(u) - D_{\mu}(- u)}{1 + e^{i\pi\mu}}
\eea
We recall that $\mathrm{u}(-\x) = \tau - \mathrm{u}(\x)$. So, in the x variable:
\bea
\f_{\mu}(\x) & = & \frac{\mathrm{D}_{\mu}(\x) + e^{i\pi\mu}\mathrm{D}_{\mu}(-\x)}{1 - e^{i\pi\mu}} \\
\label{eq:fmu1}\widehat{\f}_{\mu}(\x) & = & \frac{1}{\nn - 2}\frac{\sigma(\x)}{\x}\left(\nn\f_{\mu}(\x) + 2\f_{\mu}(-\x)\right)
\eea \\

\item[$\bullet$] \underline{Wronskien}. The wronskien function is defined by $\Xi_{\mu} = (\mathbf{T}f_{\mu})\,\widehat{f}_{\mu} - f_{\mu}\,(\mathbf{T}\widehat{f}_{\mu})$. After computation:
\bea
\label{eq:Smu}\Xi_{\mu}(\mathrm{u(x)}) & = & \f_{\mu}(\x)\widehat{\f}_{\mu}(-\x) - \f_{\mu}(-\x)\widehat{\f}_{\mu}(\x) \nonumber\\
& = & - \frac{2\sigma(\x)}{\x}\mathrm{D}^{(2)}_{\mu}(\x) \nonumber \\
& = & \frac{2(\x^2 - e_{\mu}^2)}{\x\sigma(\x)}
\eea \\

\item[$\bullet$] \underline{$\bot$-Norms} of $\f_{\mu}$ and $\widehat{\f}_{\mu}$. From the definitions:
\bea
\mathrm{R}_{\mu}(\x^2) = \left(\f_{\mu}\bot \f_{\mu}\right)(\x) & = & -(2 - \nn)\,{}\mathrm{D}^{(2)}_{\mu}(\x) \nonumber \\
\label{eq:Rmu}& = & (2 - \nn)\frac{\x^2 - e_{\mu}^2}{(\x^2 - a^2)(\x^2 - b^2)} \\
\widehat{\mathrm{R}}_{\mu}(\x^2) = \left(\widehat{\f}_{\mu}\bot\widehat{\f}_{\mu}\right)(\x) & = & -(2 + \nn)\frac{\sigma(\x)^2}{\x^2}\,{}\mathrm{D}^{(2)}_{\mu}(\x) \nonumber \\
& = & (2 + \nn)\frac{\x^2 - e_{\mu}^2}{\x^2}
\eea \\

\item[$\bullet$] \underline{Asymptotic of $\f_{\mu}$ and $\widehat{\f}_{\mu}$}.
\bea
\f_{\mu}(\x) & = & \frac{1}{\x} + \frac{1 + e^{i\pi\mu}}{1 - e^{i\pi\mu}}\frac{\alpha_1}{\x^2} + \frac{\alpha_2}{\x^3} + O\left(\frac{1}{\x^4}\right) \\
\label{eq:asym}\widehat{\f}_{\mu}(\x) & = & 1 + \frac{1 - e^{i\pi\mu}}{1 + e^{i\pi\mu}}\frac{\alpha_1}{\x} + \frac{\alpha_2}{\x^2} +
O\left(\frac{1}{\x^3}\right) \eea when $\x \rightarrow \infty$ in the
physical sheet. \\
\item[$\bullet$] \underline{Values at $e_{\mu}$}. Since $\mathrm{D}_{\mu}(e_{\mu}) = 0$, $\f_{\mu}(e_{\mu})$ and $\widehat{\f}_{\mu}(e_{\mu})$ are both proportional to $\mathrm{D}_{\mu}(-e_{\mu})$ (everything is read in the variable $x$). More precisely:
\beq
\frac{\f_{\mu}(e_{\mu})}{\widehat{\f}_{\mu}(e_{\mu})} = - \frac{1 + e^{i\pi\mu}}{1 - e^{i\pi\mu}}\frac{e_{\mu}}{\sigma(e_{\mu})}
\eeq \\
\item[$\bullet$] \underline{Behavior at branch points}. Thanks to Eqn.~\ref{eq:Rmu}, we can find the behavior of $\f_{\mu}$ near $a_i \in \{a,b\}$:
\beq
\label{eq:behfmu}\lim_{\x \rightarrow a_i} \sigma(\x)\f_{\mu}(\x) = \sqrt{(2 - \nn)(a_i^2 - e_{\mu}^2)}
\eeq \\
\item[$\bullet$] \underline{First order $2\times 2$ differential system}. The differential equation Eqn.~\ref{eq:derGmu} can be turned into a first order differential system defining $(\f_{\mu}, \widehat{\f}_{\mu})$ intrinsically. The solution is unique if we require the asymptotic to be given by Eqn.~\ref{eq:asym} up to $O(\x^{-3})$ when $\x \rightarrow \infty$ in the physical sheet. The data of $e_{\mu}$ (Eqn.~\ref{eq:emu}) and $\alpha_1$ (Eqn.~\ref{eq:alpha1}) ensures that this functions have the correct monodromy, i.e are one cut solutions of the saddle point equation \ref{eq:00}.
\beq
\frac{\mathrm{d}}{\mathrm{dx}} \left(\begin{array}{c} \,\f_{\mu}(\x) \\ \frac{\x\widehat{\f}_{\mu}(\x)}{\sigma(\x)}\end{array}\right) = \left(\begin{array}{cc} \mathrm{L}_o(\x) & \frac{1 + e^{i\pi\mu}}{1 - e^{i\pi\mu}}\,\mathrm{L}_e(\x) \\ \frac{1 - e^{i\pi\mu}}{1 + e^{i\pi\mu}}\,\mathrm{L}_e(\x) & \mathrm{L}_o(\x) \end{array}\right)\cdot\left(\begin{array}{c} \f_{\mu}(\x) \\ \frac{\x\widehat{\f}_{\mu}(\x)}{\sigma(\x)}\end{array}\right)
\eeq
where $\mathrm{L}_o$ and $\mathrm{L}_e$ and the odd and even part of L:
\bea
\mathrm{L}_o(\x) & = & \frac{1}{2}\frac{\mathrm{d}\ln \mathrm{R}_{\mu}}{\mathrm{dx}} = \frac{\mathrm{d}}{\mathrm{dx}}\left[\frac{1}{2}\ln(\x^2 - e_{\mu}^2) - \ln\sigma(\x)\right] \\
\mathrm{L}_e(\x) & = & \frac{1}{\sigma(\x)}\left[-\alpha_1 + \frac{e_{\mu}\sigma(e_{\mu})}{\x^2 - e_{\mu}^2}\right] = \frac{-\alpha_1}{\sigma(\x)}\frac{\x^2 - \widehat{e}_{\mu}^2}{\x^2 - e_{\mu}^2}
\eea
We have called $\widehat{e}_{\mu} = \sqrt{e_{\mu}^2 + \frac{e_{\mu}\sigma(e_{\mu})}{\alpha_1}}$, the zero of $\mathrm{L}_e(\x)$.

\end{itemize}

\section{Resolvent at $a = 0$ and $b = \infty$}
\label{sec:dege}
The master loop equation when $a = 0$ ($\x$ finite), or $b = \infty$ (equivalently $a \rightarrow 0$ and $\x^* = \x/a$ finite), can be solved directly, and quite explicitly in terms of Chebyshev functions. These computations were first done by Kostov et al. at the time of the introduction of the $\mathcal{O}(\nn)$ model. We presented another method to compute when $a \rightarrow 0$ in Section.~\ref{sec:}, and original computations with many examples (in particular, the cases $\mu$ rational) can be found in \cite{EZJ,E1,EK1,EK2}. Here, we review in a self-contained way the solution for $b = \infty$, for sake of completeness.

We want to solve the following problem. Let $\gamma = [0,b]$. Find $\mathrm{W} = \mathrm{W}_1^{(0)}$ such that:
\begin{itemize}
\item[$(i)$] $\mathrm{W}$ is holomorphic on $\mathbb{C}\setminus\gamma$.
\item[$(ii)$] $\mathrm{W}(\x) \sim t/\x$ when $\x \rightarrow \infty$
\item[$(iii)$] $\mathrm{W}(\x) \propto \sqrt{\x - b}$ when $\x \rightarrow b$ (resp 
\item[$(iii)$] For any $\x \in \gamma$, we have the linear equation:
\beq
\mathrm{W}(\x + i0^+) + \mathrm{W}(\x + i0^-) + \nn \mathrm{W}(-\x) = \mathrm{V}'(\x)
\eeq
\end{itemize}
Then, the density $\rho(\x)$, defined for $\x \in \gamma$ is recovered through:
\beq
\rho(\x) = \frac{1}{2i\pi t}\left(\mathrm{W}(\x - i0^+) - \mathrm{W}(\x + i0^+)\right)
\eeq
We also set $\nn  = - 2\cos\pi\mu$, with $\mu \in [0,1]$.
The relation between $b$ (if $\gamma = [0,b]$) or $a$ (if $\gamma = [a,\infty]$), and $t$, must be such that $\int_0^b \rho = 1$.

At the end, we specialize the results to the fully packed loop model, where the potential is quadratic $\mathrm{V}(\x) = \frac{1}{2}\left(\x - \frac{c}{2}\right)^2$.

\subsection*{Properties of the linear equation}

\begin{itemize}
\item[$\bullet$] The linear equation has an easy particular solution, so we can write:
\beq
\mathrm{W}(\x) = \overline{\mathrm{W}}(\x) + \frac{2\mathrm{V}'(\x) - \nn \mathrm{V}'(-\x)}{4 - \nn^2}
\eeq
with $\overline{\mathrm{W}}(\x)$ satisfying:
\beq
\label{eq:H}\overline{\mathrm{W}}(\x + i0^+) + \overline{\mathrm{W}}(\x - i0^+) + \nn\overline{\mathrm{W}}(-\x) = 0
\eeq
\item[$\bullet$] Let $\mathrm{f}$ and $\mathrm{g}$ are meromorphic functions with one cut on $\gamma$, which are solution of Eqn.~\ref{eq:H}. Then, the quantity:
\beq
\label{eq:s}(\mathrm{f}|\mathrm{g})(\x) = \mathrm{f}(\x)\mathrm{g}(\x) + \mathrm{f}(-\x)\mathrm{g}(-\x) + \frac{\nn}{2}\left(\mathrm{f}(\x)\mathrm{g}(-\x) + \mathrm{f}(-\x)\mathrm{g}(\x)\right)
\eeq
is meromorphic, even, and has no cut. Hence, it is a rational function of $\mathrm{x}^2$.
\item[$\bullet$] The vector space of solutions of Eqn.~\ref{eq:H}, over the field of rational functions of $\x^2$, has dimension 2.
\end{itemize}

\subsection*{Chebyshev functions}

A basis of functions having similar properties is known, namely the Chebyshev functions $T_{\mu}(z)$ and $U_{\mu}(z)$:
\bea
&& z = \cos\phi \nonumber \\
&& T_{\mu}(z) = \cos\mu\phi = \frac{1}{2}\left[\left(-z + i\sqrt{1 - z^2}\right)^{\mu} + \left(-z - i \sqrt{1 - z^2}\right)^{-\mu}\right] \nonumber \\
&& U_{\mu}(z) = \frac{\sin(\mu + 1)\phi}{\sin\phi} = \frac{1}{\sqrt{1 - z^2}}\,\frac{1}{2i}\left[\left(-z + i\sqrt{1 - z^2}\right)^{\mu} + \left(-z - i\sqrt{1 - z^2}\right)^{-\mu}\right] \nonumber
\eea
We prefer to use an orthogonal basis for the product of Eqn.~\ref{eq:s}. So, we set:
\beq
\widehat{T}_{\mu}(z) = \mathrm{cotan}\phi\,\sin\mu\phi = U_{\mu}(z) - T_{\mu}(z)
\eeq
These functions are analytical on the complex plane, except on one cut $]-\infty,-1]$. We have the following properties:
\bea
(T_{\mu}|T_{\mu}) & = & \sin^2\pi\mu \\
(\widehat{T}_{\mu}|\widehat{T}_{\mu}) & = & \mathrm{tan}^2\phi\,\sin^2\pi\mu \\
(T_{\mu}|\widehat{T}_{\mu}) & = & 0
\eea
and:
\bea
T_{\mu}(z + i0^+) - T_{\mu}(z - i0^-) & = & 4\sin\pi\mu\,\sin\mu(\phi - \pi) \\
\widehat{T}_{\mu}(z + i0^+) - \widehat{T}_{\mu}(z - i0^-) & = & 4\sin\pi\mu\,\mathrm{cotan}\phi\,\cos\mu(\phi - \pi)
\eea
For information, we give the asymptotics of $T_{\mu}$.
\begin{itemize}
\item[$\bullet$] When $z \rightarrow \infty$, i.e. $x \rightarrow 0$:
\beq
T_{\mu}(z) = 2^{(\mu - 1)}z^{-\mu}\left(1 - \frac{\mu}{4z^2} + \cdots\right) + 2^{-(\mu + 1)}z^{\mu}\left(1 + \frac{\mu}{4z^2} + \cdots\right)
\eeq
\item[$\bullet$] When $z \rightarrow 0$, i.e $x \rightarrow \infty$:
\bea
\label{eq:asy1}T_{\mu}(z) = \cos\left(\frac{\pi\mu}{2}\right) - \mu\sin\left(\frac{\pi\mu}{2}\right) z - \frac{\mu^2}{2}\cos\left(\frac{\pi\mu}{2}\right) z^2 + o(z^2) \\
\label{eq:asy2}\widehat{T}_{\mu}(z) = \sin\left(\frac{\pi\mu}{2}\right) z - \left(\frac{\mu^2}{2} + 2\mu\right)\cos\left(\frac{\pi\mu}{2}\right) z^2 + o(z^2)
\eea
\end{itemize}

\subsection*{Case $\gamma = [0,b]$}
\label{sec:baba}
If we perform the change of variable $z = -b/\x$ sending the segment $\gamma$ to $z(\gamma) = ]-\infty,-1]$, we can write the general solution of Eqn.~\ref{eq:H} in the form:
\bea
\overline{\mathrm{W}}(\x) & = & \frac{1}{\sin^2\pi\mu}\left[\mathrm{A}(\x^2)\cos\mu\phi + \widehat{\mathrm{A}}(\x^2)\mathrm{tan}\phi\,\sin\mu\phi\right] \\
\label{eq:blas} & = & \frac{1}{\sin^2\pi\mu}\left[\mathrm{A}(\x^2)\,T_{\mu}(z) + \widehat{\mathrm{A}}(\x^2)\left(\frac{1}{z^2} - 1\right)\,\widehat{T}_{\mu}(z)\right]
\eea
where $\mathrm{A}$ and $\widehat{\mathrm{A}}$ are rational functions:
\beq
\label{eq:AB}\mathrm{A}(\x^2) = (T_{\mu}|\mathrm{W}),\qquad \widehat{\mathrm{A}}(\x^2) = (\widehat{T}_{\mu}|\mathrm{W})
\eeq

For our initial problem, $\mathrm{A}$ and $\widehat{\mathrm{A}}$ are determined by the required analytical properties for $\mathrm{W}(\x)$. One can see on Eqn.~\ref{eq:AB} that $\mathrm{A}$ and $\widehat{\mathrm{A}}$ can have singularities only at $\x = \infty$, hence are polynomials. And we know the behavior of $\overline{\mathrm{W}}$ near $\x \rightarrow \infty$:
\beq
\overline{\mathrm{W}}(\x) = - \frac{2\mathrm{V}'(\x) - \nn \mathrm{V}'(-\x)}{4 - \nn^2} + o(1)
\eeq
Thus:
\bea
\mathrm{A}(\x^2) & = & - \left[\Big(\frac{2\mathrm{V}'(\x) - \nn \mathrm{V}'(-\x)}{4 - \nn^2}\,\Big|\,T_{\mu}(-b/\x)\Big)\right]_+ \nonumber \\
& = & - \mathrm{Even}\;\mathrm{polynomial}\;\mathrm{part}\;\left[\mathrm{V}'(\x)T_{\mu}(-b/\x)\right]
\eea
All the same:
\beq
\widehat{\mathrm{A}}(\x^2) = -\mathrm{Even}\;\mathrm{polynomial}\;\mathrm{part}\;\left[\mathrm{V}'(\x)\widehat{T}_{\mu}(-b/\x)\right]
\eeq
The corresponding density is:
\bea
\rho(\x) & = & \frac{1}{2i \pi t \sin(\pi\mu)}\left[\mathrm{A}(\x^2)\sin\mu(\phi - \pi) + \widehat{\mathrm{A}}(\x^2)\mathrm{tan}\phi\,\cos\mu(\phi - \pi)\right] \nonumber \\
& = & \frac{\sqrt{1 - \frac{\x^2}{b^2}}}{2\pi t\sin(\pi\mu)}\left[\mathrm{A}(\x^2)\,U_{\mu}(-z) + \widehat{A}(\mathrm{x}^2)T_{\mu}(-z)\right]
\eea

\medskip

For example, in the fully packed case, $\mathrm{V}'(\x) = \x - c/2$. When we use the asymptotics given in Eqns.~\ref{eq:asy1} and ~\ref{eq:asy2}, we find that $\mathrm{A}$ and $\widehat{\mathrm{A}}$ are constants:
\beq
\mathrm{A}(\x^2) = b\mu\sin\left(\frac{\pi\mu}{2}\right),\qquad \widehat{\mathrm{A}}(\x^2) = b\sin\left(\frac{\pi\mu}{2}\right)
\eeq
So, we have:
\beq
\mathrm{W}_1^{(0)}(\x) = \frac{1}{4\cos^2\left(\frac{\pi\mu}{2}\right)}\left[\x + \frac{b}{\sin\left(\frac{\pi\mu}{2}\right)}\left(\mu T_{\mu}(-b/\x) + \widehat{T}_{\mu}(-b/\x)\right)\right]
\eeq
and the density:
\beq
\rho(x) = \frac{\sqrt{b^2 - \x^2}}{4\pi t \cos\left(\frac{\pi\mu}{2}\right)}\left[\mu\,\widehat{T}_{\mu}(-\x/b) + T_{\mu}(-\x/b)\right]
\eeq
When $b$ is fixed, $t$ is the normalization such that $\int_{0}^b \rho = 1$.

\end{document}